\UseRawInputEncoding
\documentclass[11pt, a4paper]{article}
\usepackage{jcappub}
\usepackage{epsfig}
\usepackage{natbib}
\usepackage{amsmath}
\usepackage{amsfonts}
\usepackage{amssymb}
\usepackage{booktabs}
\usepackage{graphicx}
\usepackage{hyperref}
\usepackage{caption}
\usepackage{subcaption}
\usepackage{ragged2e}
\usepackage{multicol}
\usepackage{multirow}
\usepackage{comment}
\usepackage{paralist}
\usepackage[shortlabels]{enumitem}
\usepackage[normalem]{ulem}
\usepackage[dvipsnames]{xcolor}

\title{Learning from galactic rotation curves: a neural network}
\author[a]{Bihag Dave,}
\author[b, c]{and Gaurav Goswami}

\affiliation[a]{School of Engineering and Applied Science, Ahmedabad University, Commerce Six Roads, Navrangpura, Ahmedabad - 380009, India}
\affiliation[b]{Division of Mathematical and Physical Sciences, School of Arts and Sciences, Ahmedabad University, Commerce Six Roads, Navrangpura, Ahmedabad - 380009, India}
\affiliation[c]{International Centre for Space and Cosmology, Ahmedabad University, Commerce Six Roads, Navrangpura, Ahmedabad - 380009, India}

\emailAdd{bihag.d@ahduni.edu.in}
\emailAdd{gaurav.goswami@ahduni.edu.in}

\abstract{
For a galaxy, given its observed rotation curve, can one directly infer parameters of the dark matter density profile (such as dark matter particle mass $m$, scaling parameter $s$, core-to-envelope transition radius $r_t$ and NFW scale radius $r_s$), along with Baryonic parameters (such as the stellar mass-to-light ratio $\Upsilon_*$)?
In this work, using simulated rotation curves, we train neural networks, which can then be fed observed rotation curves of dark matter dominated dwarf galaxies from the SPARC catalog, to infer parameter values and their uncertainties. 
Since observed rotation curves have errors, we also explore the very important effect of noise in the training data on the inference. 
We employ two different methods to quantify uncertainties in the estimated parameters, and compare the results with those obtained using Bayesian methods.
We find that the trained neural networks can extract parameters that describe observations well for the galaxies we studied. 
}

\keywords{Artificial Neural Networks, Rotation Curves, Ultra Light Dark Matter}

\arxivnumber{2412.03547}

\begin{document}
\maketitle
\flushbottom

\section{\label{sec:introduction}Introduction}

In the last few decades, large observational data sets at both astrophysical and cosmological scales have enabled us to place ever-stringent constraints on many exciting new physics ideas such as dark matter, dark energy, inflation, etc. (see for instance \cite{Workman_2022, Planck_2018}). 
In the near-future, we expect even larger data sets with more accurate observations from various upcoming experiments like LSST, CMB-S4, DESI, etc. \cite{Ivezic_2019, Abazajian_CMBS4_Snowmass2021, DesiCollabVI_2024}. 
At the same time, in the last decade or so, advances in computer hardware, and especially parallel computing, have led to a renewed interest in machine learning techniques. In particular, deep learning using Artificial Neural Networks (ANNs), Convolutional Neural Networks (CNNs), transformers, etc. \cite{Mehta_2019, Alzubaidi_2021, Vaswani_2023}, has proved to be very useful in extracting information from data. 

These novel tools and techniques are currently being applied to a wide range of astrophysical and cosmological datasets \cite{Graff_2012, Wang_2020, Wang_ECoPANN_2020, Fluri_2021, Ho_2021, Gomez-Vargas_2021, Pal_2023, Pal_ANN-CMB_2023, Pal_CNN-CMB_2023, Shah_2024, GarciaArroyo_2024, Pal_2024, Hagimoto_2024, Artola_2024, Garuda_2024, Nerin_2024}. 
For instance, Refs.~\cite{Graff_2012, Gomez-Vargas_2021} use ANNs to compute likelihood in Bayesian inference to reduce computational time. 
On the other hand, Ref.~\cite{Wang_2020} carries out non-parametric reconstruction of the Hubble parameter as a function of redshift while  Ref.~\cite{GarciaArroyo_2024} does so for galactic rotational velocity as a function of radius.
Ref.~\cite{Wang_2020} finds that if one conducts Bayesian inference using Markov Chain Monte-Carlo (MCMC) on the reconstructed data, the resultant cosmological parameter posteriors are in agreement to those obtained from observed data. 
Deep learning has also been applied to the reconstruction of CMB B-modes\cite{Pal_2024}, as well as to reconstruct full CMB spectra from partial sky data \cite{Pal_ANN-CMB_2023, Pal_CNN-CMB_2023}.
Neural networks have also been used to estimate parameters from observational data like $H(z)$ data \cite{Pal_2023}, CMB angular power spectrum \cite{Wang_ECoPANN_2020}, CMB birefringence maps \cite{Hagimoto_2024} as well as Lyman-$\alpha$ spectra \cite{Artola_2024}.

In this work, we use neural networks to extract model parameters from galactic rotation curves from the Spitzer Photometry \& Accurate Rotation Curves (SPARC) catalog \cite{Lelli_2016}. 
Recall that rotation curves  are pivotal in tracing the mass distribution of dark matter and baryons in galaxies, and are an important test of dark matter models \cite{Salucci_2019, Profumo_2019}.
We consider dark matter to comprise of a spin-zero particle with mass $m\sim 10^{-22}\ \text{eV}$ (called Ultra-Light Dark Matter (ULDM)), whose large deBroglie wavelength leads to the formation of a flat density core surrounded by a cold dark matter-like envelope \cite{Schive_2014} (see also \cite{Ferreira_2021, Hui_2021, UBDM_book_2023}).
We thus have the following five free parameters: mass of the dark matter particle $m\ (\text{eV})$, along with galaxy specific parameters such as the scaling parameter $s$, which characterizes the dark matter core, core-to-envelope transition radius $r_t\ (\text{kpc})$ and NFW scale radius $r_s\ (\text{kpc})$ which characterize the surrounding halo. The effect of Baryons is parameterized by the stellar mass-to-light ratio $\Upsilon_*\ (M_\odot/L_\odot)$, which tunes contribution from the stellar disk (see section~\ref{sec:model_and_data} for details). 

For a model with the above parameters, we ask the following: for a chosen galaxy, given the observed rotation curve [i.e. observed values of velocities $V_{obs}(r)$ for some finite number ($N_{obs}$) of radius values along with their uncertainties $\sigma(r)$], what can we say about the values and uncertainties of parameters $m$, $s$, $r_t$, $r_s$ and $\Upsilon_*$?

The usual approach to answer this question involves Bayesian inference, where given some prior distribution of parameters and a likelihood function, one can obtain the posterior distribution of parameters using Bayes' theorem.
MCMC methods \cite{McKay_book_2023} are then used to sample from this posterior which in-turn gives the best-fit parameters along with confidence intervals (see \cite{Bernal_2017, Delgado_2022, Khelashvili_2023, Banares_Hernandez_2023} for some recent work on constraining ULDM parameters using rotation curves). 
In the context of our problem, given the $N_{obs}$ values of rotational velocity, for the case of uniform priors, this problem is equivalent to the problem of finding regions in the five dimensional parameter space in which the likelihood function is large. 

Since the five parameters are estimated from $N_{obs}$ values of rotational velocity, it is interesting to ask whether there could be a well defined function from $\mathbb{R}^{N_{obs}}$ to $\mathbb{R}^5$ which, when fed the rotation curve (i.e. a point in $\mathbb{R}^{N_{obs}}$), gives the ``best fit" parameters (i.e. a point in $\mathbb{R}^5$). 

We explore whether, using simulated rotation curves, one can train a neural network to approximate this function. 
Typically, $N_{obs}\sim \mathcal{O}(15)$ for the galaxies we consider, while the number of parameters we want to infer is $5$ (or $10$ if uncertainties are also inferred). 
If the neural network has $2$ hidden layers with $200$ neurons each, it will have $\sim (40-50)\times 10^3$ internal adjustable parameters (called weights and biases).
To fix these internal parameters, we need to train the neural network. 
For training, we use simulated rotation curves whose parameter values are already known. 
We generate training data for $7$ dwarf galaxies from the SPARC catalog \cite{Lelli_2016} and train a different neural network for each galaxy. 
The size of the training data, i.e., the number of known pairs of rotation curves and parameters in our work is $\sim 10^5$. 
The details of neural networks used and their architectures are discussed in Sections~\ref{sec:ANN_basics} and~\ref{sec:architecture} respectively. 

To test our trained neural networks, we use the observed rotation curves for the $7$ galaxies as input and infer parameter values in section~\ref{sec:noiseless_case}.
Then, in section~\ref{sec:noisy_case}, we explore the effect of noise in the training data on the performance of the neural network during parameter inference, and find that including noise improves point-estimates of parameters when confronted with observed rotation curves, i.e., the rotation curve obtained from these parameter values agree well with the observed rotation curves. 
In section~\ref{sec:uncertainties}, we also utilize two different ways of obtaining uncertainties in the model parameters: during inference (as carried out in \cite{Wang_ECoPANN_2020}), or during training (following the work in \cite{Pal_2023}). 
Finally, we compare the parameter point-estimates and uncertainties obtained using our approach to those obtained using MCMC in section~\ref{sec:mcmc_comparison}.
We conclude in section~\ref{sec:discussion}.

\section{\label{sec:rot_curves_ANN}Rotation curves and artificial neural networks}

\subsection{Model and data\label{sec:model_and_data}}

Galactic rotation curves, i.e. orbital velocity of stars and gas as a function of distance from the centres of galaxies are an important probe of the matter (visible and dark) distribution in said galaxies~\cite{Salucci_2019}. 

The total gravitational potential of the galaxy includes contribution from both baryonic (disk, bulge, gas) and dark matter components, allowing one to split the total velocity~\cite{Lelli_2016}:  

\begin{equation}\label{eq:vel_components}
    V_{obs} = \sqrt{V_{DM}^2 + V_{g}|V_g| + \Upsilon_d V_{d}|V_d| + \Upsilon_b V_{b}|V_b|}\ ,
\end{equation} 
where $V_d$, $V_b$, and $V_g$ are contributions from the stellar disk, bulge and gas components, while $V_{DM}$ is the dark matter contribution.
Contributions from the stellar disk and bulge can be further tuned by $\Upsilon_d$ and $\Upsilon_b$, i.e. the disk and bulge mass-to-light ratios respectively, which are free parameters. 
Baryonic velocities, i.e. $V_d$, $V_g$, and $V_b$ can be obtained by fitting relevant density profiles to observed surface brightness profiles \cite{Lelli_2016}.
For galaxies without a bulge, $V_b = 0$ at all radius values, and $\Upsilon_* \equiv \Upsilon_d$ is the only free parameter. 

In this work, we consider dark matter to comprise of ultralight spin-zero scalars, with $m\sim 10^{-22}\ \text{eV}$. 
Due to the large deBroglie wavelength ($\lambda_{dB}\sim \mathcal{O}(\text{kpc})$), simulations suggest that dark matter halos in the ULDM paradigm have a core-halo structure where, the inner regions of the halo are described by flat density cores \cite{Schive_2014}. 
These cores are stationary state solutions of the Schrödinger-Poisson system of equations.
In the outer regions, beyond a transition radius, ULDM behaves like CDM and the corresponding density profile can be described by the well known Navarro-Frenk-White (NFW) profile \cite{Navarro_1996}. 
Hence, for a galactic halo, the total dark matter density profile can be written as
\begin{equation}\label{eq:uldm_nfw}
    \rho_{DM}(r) = \rho_{ULDM}\Theta(r_t - r) + \rho_{NFW}\Theta(r - r_t)\ ,
\end{equation}
where $r_t$ is the transition radius. 
In the absence of self-interactions, instead of solving for the stationary state solution, $\rho_{ULDM}$ can also be described by the following fitting function \cite{Schive_2014}

\begin{equation}\label{eq:schive_profile}
    \rho_{ULDM}(r) \simeq \frac{0.019\times (m/10^{-22}\ \text{eV})^{-2}(r_c/\text{kpc})^{-4}}{\left[1 + 0.091\times (r/r_c)^2\right]^8}\ M_\odot/\text{pc}^3\ ,
\end{equation}
where $r_c$ is defined as the radius at which the density becomes half its central value, and is given by 

\begin{equation}\label{eq:core_radius}
    r_c = 0.8242\left(\frac{s}{10^4}\right)\left(\frac{m}{10^{-22}\ \text{eV}}\right)^{-1}\ \text{kpc}\ .
\end{equation}
Note that the free parameters here are the ULDM particle mass $m$ and the scale parameter $s$. The scale parameter allows one to describe solitonic solutions of different masses and radii \cite{Chakrabarti_2022, Dave_2023}.

The NFW density profile, obtained form CDM-only simulations \cite{Navarro_1996} is given by 
\begin{equation}\label{eq:nfw}
    \rho_{NFW}(r) = \frac{\rho_s}{\frac{r}{r_s}\left(1 + \frac{r}{r_s}\right)^2}\ M_\odot/pc^3.
\end{equation}
Here $\rho_s$ and $r_s$ are halo-specific parameters. 
Since we impose continuity at the transition radius, i.e. $\rho_{ULDM}(r_t) = \rho_{NFW}(r_t)$, one can eliminate $\rho_s$ and describe the NFW part of the profile using only $r_t$ and $r_s$.

The circular velocity of a test particle moving under the influence of the spherically symmetric density profile in eq.~(\ref{eq:uldm_nfw}), is simply 

\begin{equation}\label{eq:circ_vel}
    v(r) = \sqrt{\frac{GM(r)}{r}} = \sqrt{\frac{4\pi G\int_0^r\rho_{DM}(r')r'^2dr'}{r}}\ .
\end{equation}

Hence, the free parameters that characterize the rotation curve are: mass of the ULDM particle $m$, the scaling parameter $s$, the radius at which ULDM transitions to NFW $r_t$, the scale radius of the NFW profile $r_s$ and the stellar mass-to-light ratio $\Upsilon_*$, i.e. 
\begin{equation}\label{eq:params}
    {\bf P} = \left(m, s, r_t, r_s, \Upsilon_*\right)\ .
\end{equation}

\subsubsection{Observed Rotation Curves}\label{sec:RCs}

In this work, we utilize observed rotation curves from the SPARC catalog which hosts
high quality HI/H$\alpha$ rotation curves for 175 galaxies \cite{Lelli_2016}. 
The SPARC catalog has been utilized previously to constrain ULDM parameters \cite{Bernal_2017, Bar_2022, Delgado_2022, Khelashvili_2023} as well as models of modified gravity \cite{de_Almeida_2018, Bhatia_2024} and obtain bounds on the cosmological constant \cite{Benisty_2024}.

Since ULDM affects dark matter distribution in the inner regions of galaxies, one must look at galaxies where baryons are not the dominant component even at small radii, or dark matter dominated galaxies. 
To study ULDM, authors in Ref.~\cite{Delgado_2022} chose $17$ dark matter dominated dwarf galaxies from the SPARC catalog with well-defined inner regions.
In this work, we choose a subset of $7$ galaxies from the sample of $17$. 
It is important to note that, along with observed rotation curves, the SPARC catalog provides values for $V_d$ and $V_g$ by fitting relevant stellar density profiles. We shall utilize these values directly in eq.~(\ref{eq:vel_components}), while allowing $\Upsilon_*$ to vary. 

At this point, as discussed in section~\ref{sec:introduction}, we note that we are trying to approximate a function that can take an observed rotation curve as input and infer model parameters in eq.~(\ref{eq:params}) as well as their uncertainties as output. 
To understand how a neural network can do this, we must briefly discuss the ingredients involved in defining and training an artificial neural network in the following sub-section.

\subsection{Artificial neural networks}\label{sec:ANNs}

\subsubsection{Basics of neural networks}\label{sec:ANN_basics}

Neural networks (NNs) are an important tool in supervised machine learning, that can approximate complex relationships between some input ${\bf x}$ and output ${\bf y}$. 
This is done by looking at a set of examples called `training data' consisting of known pairs of inputs (also called features) and corresponding outputs (also called targets). 
Note that the output ${\bf y}$ can either be a vector of continuous values (in case of regression), or categories from a finite set (in case of classification). 
For a set of $I$ known input-output pairs $\{{\bf x}_i, {\bf y}_i\}_{i = 1}^I$, a feed-forward neural network is simply a function ${\bf f}$ of the input ${\bf x}$ parameterized by ${\bf \Omega}$, 

\begin{equation}\label{eq:ANN_func}
    {\bf f} = {\bf f}({\bf x}; {\bf \Omega})\ .
\end{equation}
Here, ${\bf \Omega}$ are the internal adjustable parameters (IAPs) of the neural network. 
To understand how to construct such a function, we first look at the fundamental building block of a neural network, the neuron. 
For an input vector ${\bf x}\in \mathbb{R}^{N_0}$, the output of a neuron is defined as 

\begin{equation}\label{eq:neuron}
    v = a({\boldsymbol \omega}\cdot {\bf x} + \beta)\ .
\end{equation}
Here, components of ${\boldsymbol \omega}$ are called weights and $\beta$ is called bias. 
The function $a(z)$ is called the activation function, which imparts a non-linearity to the transformed input. 
The choice of the activation function depends on the kind of problem at hand. 
For instance, in the case of classification problems where the required output is discreet, the sigmoid function, $a(z) = (1+e^{-z})^{-1}$ is useful since the output is contained between $0$ and $1$. For regression problems, where the required output is continuous, the rectified linear unit (ReLU), $a(z) = \mathrm{max}(z, 0)$ can be used. 

One can also define a layer (often called a hidden layer) of $N$ neurons, where the input for each neuron is the same albeit with different weights and biases. 
The output of the layer can be written as a vector of size $N$, where each component is given by eq.~(\ref{eq:neuron}) with different weights and biases, 
\begin{equation}
    {\bf v} =  {\bf a}\left({\boldsymbol \omega}{\bf x} + {\boldsymbol \beta}\right)\ .
\end{equation}
In this case, ${\boldsymbol \omega}$ denotes a matrix of size $N\times N_0$ while $\beta$ is a $N\times 1$ column matrix. 
Also note that activation function vector ${\bf a}$ is applied element-wise in the above equation. 

A neural network can have multiple such layers, where the output of one layer acts as the input for the next one; in particular when for all hidden layers, the output of each neuron in the current layer acts as an input to every neuron in the next layer, it is called a fully connected neural network. 
For a neural network with $L$ layers, we use the index $j$ to keep track of which layer we are talking about, where $1 \leq j \leq L$. The number of neurons in the $j$th layer is denoted by $N_j$.

In this case, output of the $j$th layer with $N_j$ neurons, is defined to be ${\bf v}_j$,

\begin{equation}\label{eq:layer}
  {\bf v}_j = {\bf a}\left({\boldsymbol \omega_j}{\bf v}_{j-1} + {\boldsymbol \beta}_j\right)\ . 
\end{equation}
Here, $j = 1, 2, ..., L$, while ${\boldsymbol \omega}_j$ and ${\boldsymbol \beta}_j$ are the weights and biases for the $j$th layer. 
For ease of notation, one can denote weights and biases for all layers by $\boldsymbol{\Omega}$. 

The final output of a neural network, can be written as a composition of multiple functions ${\bf f}({\bf x}, {\bf \Omega}) = {\bf v}_L({\bf v}_{L-1}(...{\bf v}_1({\bf x}, {\bf \Omega})))$, where $L$ is the total number of layers in the neural network \cite{Strang_book_2018}. 

It must be noted that one can use $j = 0$ to denote the input layer, which is just the input vector ${\bf x}$. For ANNs, no transformations occur at this layer.
One can also define $j = L+1$ as the output layer, whose size will be the same as ${\bf y}$. 
Unlike the input layer, going from $j=L$ to $j = L+1$ does involve a transformation similar to eq.~(\ref{eq:neuron}), but with $a(z) = z$. 

The universal approximation theorem \cite{Hornik_1990} shows that a neural network with a single hidden layer can approximate any non-linear function with a finite number of neurons.
However, it is often found that utilizing multiple layers, i.e. deep neural networks are easier to train and generalize better than shallow ones \cite{Prince_book_2023}. 
This is not understood well and is an active area of research \cite{Mhaskar_2017, Meir_2023}.

\begin{figure}[ht]
    \centering
    \includegraphics[width = 0.7\textwidth]{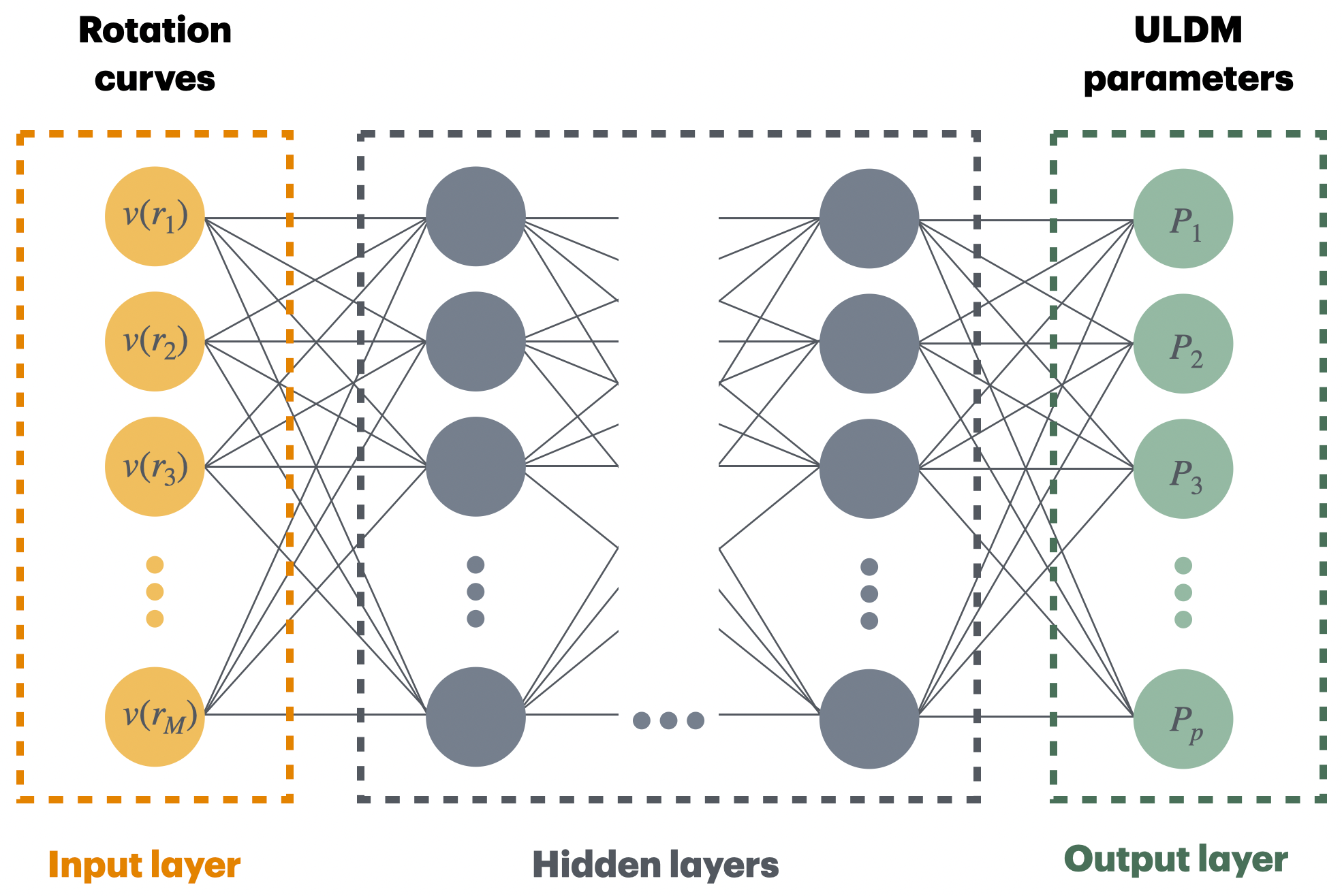}
    \caption{\justifying A schematic of the neural network that we want to construct, where given a rotation curve of dimension $M$ for a galaxy as input, one can obtain the corresponding $p$ ULDM parameters. In our work, while the number of parameters to predict will be $5$ (eq.~(\ref{eq:params})), the number of output neurons $p = 5$ or $p = 10$ depending on the loss-function used.}
    \label{fig:ANN_cartoon}
\end{figure}

In our case, the neural network of interest is shown in figure~\ref{fig:ANN_cartoon}, where the velocities for $N_{obs}$ radius values are in the input layer, while the model parameters in eq.~(\ref{eq:params}) are in the output layer, and we allow for more than $1$ hidden layers. 
It is worth noting, that for the case of a heteroscedastic loss function, as we shall see in section~\ref{sec:hetero_uncertainty}, the size of the output layer will be doubled. 
This is because we shall require the uncertainties in the parameter values to be learned during training itself. 

It is also important to note that the number of observed data points in the rotation curves of different galaxies will not be the same, implying that the size of the input layer will be different for each galaxy. 
Further, the range of radius values for which there are observed velocities will also be different for each galaxy. 
Due to this, one must define a different neural network for every galaxy in our sample, i.e., we train a total of $7$ neural networks for every case in section~\ref{sec:MSE_predictions} and~\ref{sec:uncertainties}.

\subsubsection{Training a neural network}\label{sec:ANNs_training}

The goal of training a neural network is to find the optimum values of the IAPs or weights and biases ${\bf \Omega}$, such that for an input from the training data, the neural network output ${\bf f}$ (also called predicted or inferred output) is close to the target output ${\bf y}$ (also called ground truth). 
To quantify this closeness, one employs a loss function $\mathcal{L}({\boldsymbol \Omega})$, which assigns a real number to each pair of predicted and target output. 
Training thus involves finding a set of ${\boldsymbol \Omega}$ such that $\mathcal{L}$ is minimized. 
Well-known examples of loss functions are the mean-squared error (regression tasks) and binary cross-entropy (classification tasks). 
In this paper, we shall consider two different loss functions: (a) mean-squared error (MSE) in eq.~(\ref{eq:MSE_loss}) and (b) heteroscedastic loss in eq.~(\ref{eq:hetero_loss}).  
The mean-squared-error loss is simply the proportional to the Euclidean distance between the predictions and the target values, averaged over the number of samples, given by

\begin{equation}\label{eq:MSE_loss}
    \mathcal{L} = \frac{1}{np}\sum_{i=1}^n|{\bf y}_i - {\bf \hat{y}_i}|^2\ ,
\end{equation}
where ${\bf y}_i$ is the output corresponding to the $i$th input ${\bf x}_i$, while ${\bf \hat{y}}_i \equiv {\bf f({\bf x}_i, {\bf \Omega})}$ is the prediction made by the neural network for the same input. 
$p$ is the size of the output vector. 

The goal during training is to find the global minimum of the loss function in the $\boldsymbol{\Omega}$-space \footnote{This can be a high-dimensional space, since the number of IAPs range from a few thousand to tens of millions.}. 
The usual way involves utilizing a gradient descent algorithm which requires two ingredients: (a) an efficient way to calculate the gradient of the loss function w.r.t $\boldsymbol{\Omega}$, (b) a rule to update the weights and biases in the direction of the steepest descent. 

The former can be computed using the backpropagation algorithm \cite{Rumelhart_1986} which computes the gradient backwards from the last layer, using the chain rule and layered structure of a neural network to avoid redundant calculations. 
Once the gradient is calculated, it is scaled by a step size (also called the learning rate) and the IAPs are updated in the opposite direction. 
There are numerous algorithms to carry out the update, like stochastic gradient descent, nesterov, adaptive moment estimation (Adam), etc. (see \cite{Prince_book_2023} for a detailed discussion on backpropagation and gradient descent methods). 

Usually, during training the above procedure must be carried out many times, i.e. the same training data is passed through neural network and the parameters are updated multiple times to reduce the loss.  
Training is complete when the loss converges to a minimum. The true test of a neural network is how it deals with data it was not trained on, or unseen data. 
A well-trained neural network should generalizes well, i.e. it should make accurate predictions even for the inputs that are not present in the training data. 
If the training of a neural network goes on for too long, it can memorize the training data and start to perform worse on unseen data. This is called overfitting \cite{Prince_book_2023}, and one must stop the training before this.

We would like to point to our reader Refs.\cite{Strang_book_2018, Mehta_2019, Prince_book_2023} for excellent pedagogical discussions on the various aspects of machine learning and neural networks.
For our work, the choice of the gradient descent algorithm, learning rate, and other parameters related to the optimization of the loss function for our problem are discussed in section~\ref{sec:architecture}.

\subsection{Generating simulated rotation curves\label{sec:simulated_RCs}}

Neural networks usually require a sufficiently large number of known pairs of inputs ${\bf x}$ and target outputs ${\bf y}$ to train on. 
In this case it will be a set of velocities from a rotation curve as input and the parameter vector ${\bf P}$ that can generate the rotation curve. 
However, for a fixed galaxy, we only have a single set of observed velocities, i.e. one rotation curve.
Hence, to successfully train a network to learn the relationship between rotation curves and parameters, we have to rely on a set of simulated rotation curves as training data.  
Consider a galaxy with an observed rotation curve between $R_{min}$ and $R_{max}$ consisting of $N_{obs}$ data points (i.e. $N_{obs}$ values of radii for which there are observed velocities). 
To generate $I$ simulated rotation curves for this galaxy, we follow the procedure below: 

\begin{enumerate}
    \item First, we define a uniform distribution for each parameter.
    While any random combination of parameters can form a velocity curve, not all of them will be visually similar to the observed one. 
    Hence, choosing sensible ranges for the uniform distributions for each galaxy is important. 
    We do this by examining how the numerically generated curves vary with each parameter in comparison to the observed rotation curve and choose the lower and upper limits accordingly.
    \item We make a random draw from each parameter distribution, which will give a parameter vector {\bf P} (see eq.~(\ref{eq:params})) in the $5$D parameter space.
    \item We use $\{m, s\}$ to obtain the ULDM density profile from eq.~(\ref{eq:schive_profile}), and then use $\{r_t, r_s\}$ to obtain the profile of the NFW skirt.
    Finally, given the total density profile, we calculate the dark matter velocity profile $V_{DM}$ using eq.~(\ref{eq:circ_vel}). 
    \item Using the randomly drawn mass-to-light ratio $\Upsilon_*$ along with the fixed $V_d$ and $V_g$ values provided by the SPARC catalog, we obtain the baryonic contribution to the velocity curve. 
    \item We finally construct the full velocity curve between $R_{min}$ and $R_{max}$ for all the $N_{obs}$ radius values using eq.~(\ref{eq:vel_components}). 
\end{enumerate}

The ranges of the uniform distribution for each parameter are shown in Table~\ref{tab:param_space} for all galaxies. 
We employ the above procedure for $I = 5\times 10^5$ randomly chosen parameter vectors ${\bf P}$ from the above-mentioned ranges to obtain simulated rotation curves $v(r)\in \mathbb{R}^{N_{obs}}$.
These simulated rotation curves will serve as the training inputs in figure~\ref{fig:ANN_cartoon}, while ${\bf P}$ will be the target parameters that the neural networks attempts to predict for every input rotation curve.

\begin{table}[ht]
\resizebox{\columnwidth}{!}{
\begin{tabular}{@{}cccccc@{}}
\toprule
\multirow{2}{*}{\textbf{Galaxy}} & \multicolumn{5}{c}{\textbf{Parameter ranges}} \\ \cmidrule(l){2-6}
& $m$ ($10^{-23}$\ eV) & $s\ (10^3)$ & $r_t$ (kpc) & $r_s$ (kpc) & $\Upsilon_* (M_\odot/L_\odot)$ \\ \cmidrule(r){1-6}
DDO 154 & $\left[1, 10\right]$ & $\left[3, 9\right]$ & $\left[1, 5.99\right]$ & $\left[1, 15\right]$ & $\left[0.3, 0.8\right]$ \\ \cmidrule(l){1-6}
ESO444-G084 & $\left[1, 10\right]$ & $\left[2, 9\right]$ & $\left[1, 4.44\right]$ & $\left[1, 15\right]$ & $\left[0.3, 0.8\right]$\\ \cmidrule(l){1-6}
UGC 5721 & $\left[1, 10\right]$ & $\left[1.5, 5\right]$ & $\left[1, 6.74\right]$ & $\left[1, 15\right]$ & $\left[0.3, 0.8\right]$ \\ \cmidrule(l){1-6}
UGC 5764 & $\left[1, 10\right]$ & $\left[2, 9\right]$ & $\left[1, 3.62\right]$ & $\left[1, 15\right]$ & $\left[0.3, 0.8\right]$\\ \cmidrule(l){1-6}
UGC 7524 & $\left[1, 10\right]$ & $\left[1, 9\right]$ & $\left[1, 10.69\right]$ & $\left[1, 15\right]$ & $\left[0.3, 0.8\right]$\\ \cmidrule(l){1-6}
UGC 7603 & $\left[1, 10\right]$ & $\left[2, 7\right]$ & $\left[1, 4.11\right]$ & $\left[1, 15\right]$ & $\left[0.3, 0.8\right]$\\ \cmidrule(l){1-6}
UGC A444 & $\left[1, 10\right]$ & $\left[2, 9\right]$ & $\left[1, 2.55\right]$ & $\left[1, 15\right]$ & $\left[0.3, 0.8\right]$\\
\bottomrule
\end{tabular}
}
\caption{\justifying Uniform ranges for all parameters. Note that ULDM mass is chosen to be such that the size of the core is $\sim \mathcal{O}(1\ \text{kpc})$. 
Similarly, requiring that ULDM describes the inner regions for all galaxies, the lower limit for the transition radius is $r_t \geq 1\ \text{kpc}$. The upper limit is fixed to be the largest radius bin for which there is an observation and hence is galaxy specific.}
\label{tab:param_space}
\end{table}

\subsection{\label{sec:preprocessing}Pre-processing}

Pre-processing of training data usually involves converting it to a more usable type or to normalize the input data such that all inputs have a similar range of values \cite{Eisert_2022}. 
This is required to ensure that no component of the input is considered to be more important if it has a larger absolute value or variation.
It also prevents the gradient descent algorithms from taking too small or too large steps based on the absolute value of the inputs. 

Before proceeding, we split our simulated data into three parts. For each galaxy considered in our analysis, we reserve $0.8I$ ($4\times 10^5$) examples for training, and $0.1I$ ($5\times 10^4$) examples each for validation and testing. 
The neural network will update its weights and biases only based on the examples in the training set. 
Hence, the validation set acts as unseen data for the neural network and will only be used to monitor if the network is overfitting. 
It can also be used to measure performance across various hyperparameters values. 
Finally, test data also acts as unseen data and will be used to characterize the performance of the final neural network once it has been trained. A well trained neural network will have a similar loss value across all three datasets. 
Note that for the remainder of this paper, training set or data will refer to the $80\%$ subsample of the simulated dataset. 

While there are various techniques to scale the components of the input, we use z-score normalization (also called standardization) on our input.
For each radius bin for which we have a velocity value, we subtract the velocity from the mean and divide by the standard deviation of the training data, 
\begin{equation}
    \Tilde{{\bf v}}_l = \frac{{\bf v}_l - \mu_l}{\sigma_l}\ .
\end{equation}
Here $\mu_l$ and $\sigma_l$ are the mean and standard deviation over the $l$th radius value of the training data (validation and test data are not included in this calculation). The boldface here implies a column vector the size of the training data, i.e. $4\times 10^5$. 
We then use the mean and standard deviation of the training set itself to scale the validation and test sets, as well as the observed rotation curve before feeding them to the neural network. 

We also scale output parameters to ensure that each component has values which are close to $\mathcal{O}(1)$. This leads to scaling the output parameter vector as: 

\begin{equation}
    {\bf P} = \{m/10^{-23},\  s/10^3,\  r_t,\  r_s,\  10\Upsilon_*\}\ .
\end{equation}
The output parameters (${\bf P}$) that the neural network predicts therefore need to be scaled back to familiar units during final predictions.

\subsection{\label{sec:architecture}Neural network architecture}

It is easy to see from the discussion in section~\ref{sec:ANN_basics} that a larger number of neurons per layer as well as a larger number of layers in a neural network will enable it to approximate a complex function better. 
However, the performance of the network also depends on other parameters like batch size, learning rate of the optimizer, choice of the optimization algorithm and activation function as well as the loss function used, etc. 
These parameters that characterize a neural network are called hyperparameters and choosing their optimal values for a neural network is a difficult task. This is because they must be chosen empirically, i.e. by training the neural network multiple times with various combinations of hyperparameters and choosing those which perform best on unseen data. 

Often a full grid search in the hyperparameter space is computationally expensive, which has lead to the use to some other methods like random search or genetic algorithms \cite{GarciaArroyo_2024}. 
We employ a grid search for only the base architecture of the neural network, i.e. number of layers and number of neurons per layer, in a small grid of parameters. 
The rest of the hyperparameters are chosen by trial and error across various training runs.

The fixed hyperparameters, except for the base architecture of the network, are the following: 
\begin{enumerate}
    \item Activation function: For every neuron in the hidden layers, we implement the ReLU (Rectified linear unit) activation function, given by $a(z) = \mathrm{max}(z, 0)$. 
    \item Loss function: We use the mean-squared-error (MSE) loss given by eq.~(\ref{eq:MSE_loss}). Note that we shall also implement a different loss function in section~\ref{sec:hetero_uncertainty} given by eq.~(\ref{eq:hetero_loss}). 
    \item Optimization algorithm: We use a momentum-based stochastic gradient descent algorithm called ADAM \cite{Kingma_ADAM_2017}. Parameters of the algorithm save for the learning rate, are kept at their default values.
    \item Learning rate: The step-size used for updating IAPs; we fix it to $10^{-4}$. 
    \item Batch size: For a stochastic gradient descent method, IAPs are updated based on loss calculated for a small subset of size $\mathcal{B}$ of the total training set. 
    This small subset is called a batch (or a mini-batch), and is randomly drawn from the training set without replacement. 
    We use $\mathcal{B} = 32$, implying $4\times 10^5/32 = 12500$ updates (or iterations) of the internal adjustable parameters before all samples from the training set are fed to the neural network once. 
    \item No. of epochs: When the entire training set passes through the neural network once, it is called an epoch.
    Usually, multiple epochs are required to adequately train a neural network. 
    Note that, if the training goes on for too many epochs, the neural network can memorize the training set, which leads to a poor performance on unseen data, i.e. validation and test sets (this is called overfitting). 
    To prevent this, we also keep an eye on the validation loss while training, and find that by $250$ epochs, the validation loss stops decreasing appreciably or starts to increase. 
    \item Dropout: To prevent overfitting we also utilize dropout regularization, where for every sample from the training set, the output of each neuron is set to zero with probability $d$, so that each sample is passed through a different `thinned' neural network \cite{Srivastava_2014}. 
    We set $d=0.2$ for every hidden layer in the network. 
\end{enumerate}

To find the optimum architecture of the neural network, we first fix the above hyperparameters at the values mentioned. We then construct a grid that consists of the number of hidden layers $L \in [1, 2, 3, 4]$ and neurons per layer $N_j \in [100, 150, 200]$. 
For a galaxy, for each combination of $L$ and $N_j$ we train a neural network for 250 epochs.
The performance of these trained networks is then evaluated using the loss on validation data, (recall that the weights and biases are not updated based on validation data) and we choose the architecture for which the validation loss is the lowest.
Note that this grid search is carried out with noise-induced simulated rotation curves (see section~\ref{sec:noisy_case} for details) with the mean-squared-error loss function, while the obtained optimal architecture is also utilized for the case of noiseless training data as well as the case with a heteroscedastic loss function (as we shall discuss in section~\ref{sec:hetero_uncertainty}).

We perform this grid search for $4$ galaxies, and find that validation loss was lowest for networks with at least $2$ hidden layers, while the number of neurons per layer varied. 
It is worth noting that the difference between validation loss for the best performing architectures for a given galaxy was very small ($\mathcal{O}(10^{-3})$). 
Therefore for purposes of this paper, we fix the number of hidden layers to $2$ and neurons per layer to $200$ for all 7 galaxies in the sample.

Finally, using the index notation discussed in section~\ref{sec:ANN_basics}, the neural network architecture will comprise of an input layer ($j=0$), whose size is the number of observed radius values for a particular galaxy, $2$ hidden layers ($j = 1,2$) with $N_j = 200$ each and an output layer $j = 3$ with $N_3 = 5$ outputs corresponding to a vector given by eq.~(\ref{eq:params}) in parameter space.

\section{\label{sec:MSE_predictions}Inferring model parameters using mean-squared-error loss function}

Once the simulated rotation curves are obtained and the architecture is finalized, we can now train our neural networks.
In this section, we train two neural networks for each galaxy in the sample: one without noise included in the inputs of the training set (i.e. simulated rotation curves) and one with noise included. 

For both neural networks, we use the mean-squared-error loss defined in eq.~(\ref{eq:MSE_loss}) during training, while all other hyperparameters including the architecture are fixed to what was discussed in section~\ref{sec:architecture}.

\subsection{\label{sec:noiseless_case}Noiseless Case}

Let us start with the simpler case, where simulated rotation curves without any noise are used for training. 
We follow the pre-processing steps in section~\ref{sec:preprocessing} and then train $7$ different neural networks, each for a galaxy in our sample of DM dominated dwarf galaxies. 
The neural networks are trained for $250$ epochs while both training and validation loss are monitored.
The loss as a function of epochs is plotted in Appendix~\ref{app:loss_epochs} in Figs.~\ref{fig:loss_1} and~\ref{fig:loss_2} for all galaxies.
We then test the performance of the network by calculating the loss on the test data.
This is done to ensure that test loss is not too different from training loss, i.e. the neural network does just as well on unseen data as it does on training data. 

For the final test, the trained neural networks are given the observed rotation curves - one for each neural network - as input. 
Note that only the central values of the observed rotation curves are fed to the network. 
The neural network, using its trained weights and biases predicts a parameter vector $\bf{\hat{P}}$. 

Since we do not have target parameters for the observed rotation curves to check how good these predictions are, we must utilize another metric to discern if the predicted parameters are good. 
One way to do so is to use the predicted parameter vector ${\bf \hat{P}}$ to construct a rotation curve using eqs.~(\ref{eq:uldm_nfw}) - (\ref{eq:circ_vel}), which can then be compared to the observed one using a $\chi^2_{red}$ (reduced $\chi^2$) value. 

We note the following results, using the example of UGC 5721, which carry over to other galaxies as well: 

\begin{itemize}
    \item To make sure performance on test data is good, along with calculating loss, one can also compare constructed rotation curves using the target parameters to those obtained from the predicted parameters. We do this for three randomly selected examples from the test data in figure~\ref{fig:test_noiseless}. 
    It can be seen that, while the rotation curves don't match exactly (since the loss is not zero for parameter predictions), the predicted curves look visually similar and are close to the target rotation curves. 
    \begin{figure}[h]
    \centering
    \includegraphics[width = 0.7\textwidth]{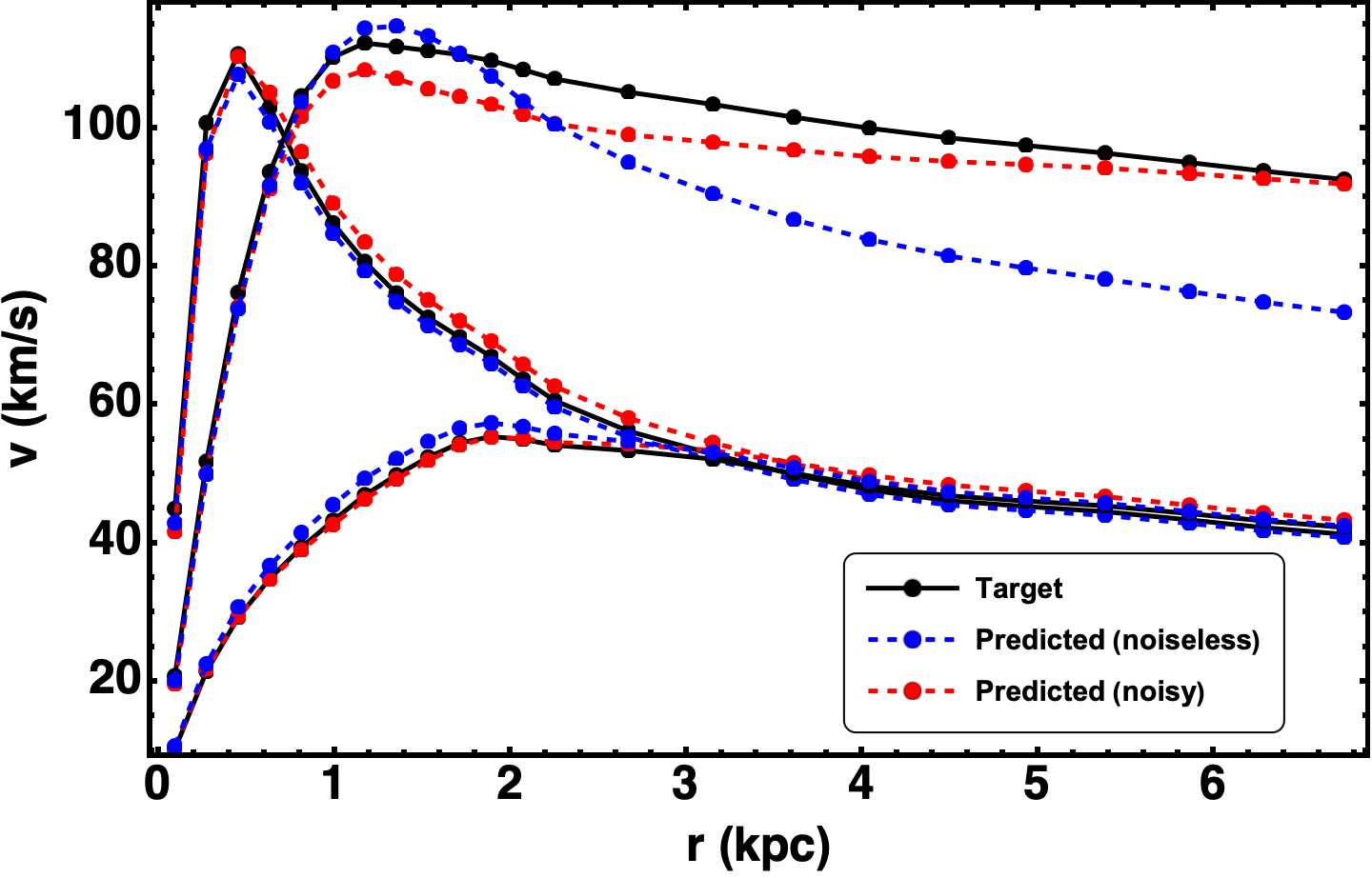}
    \caption{Target (black) and predicted rotation curves (blue for the noiseless case and red for the noisy case) constructed from randomly selected target parameters from the test dataset and the corresponding predicted parameters that the neural network infers respectively.}
    \label{fig:test_noiseless}
    \end{figure}
    
    \item However, when observed rotation curves are given as input, the predicted parameters are not able to reconstruct the observed rotation curve with a low $\chi^2_{red}$, which can be seen by the blue curve in figure~\ref{fig:UGC05721_compare}. One reason for this could be that even without the error bars, central values of the observed velocities do not form a smooth rotation curve (see for instance, the dip in the observed velocity at $\sim 1\ \text{kpc}$ in figure~\ref{fig:UGC05721_compare}). 
    Since the dark matter component for simulated data is a smooth curve, these bumps are not learned by the neural network.
    This can be seen in other galaxies as well, where in figure~\ref{fig:MSE_comparison}, parameters for observed velocities (without uncertainties) which are smoother are predicted better by the corresponding neural network. 

    \begin{figure}[h]
    \centering
    \includegraphics[width = 0.7\textwidth]{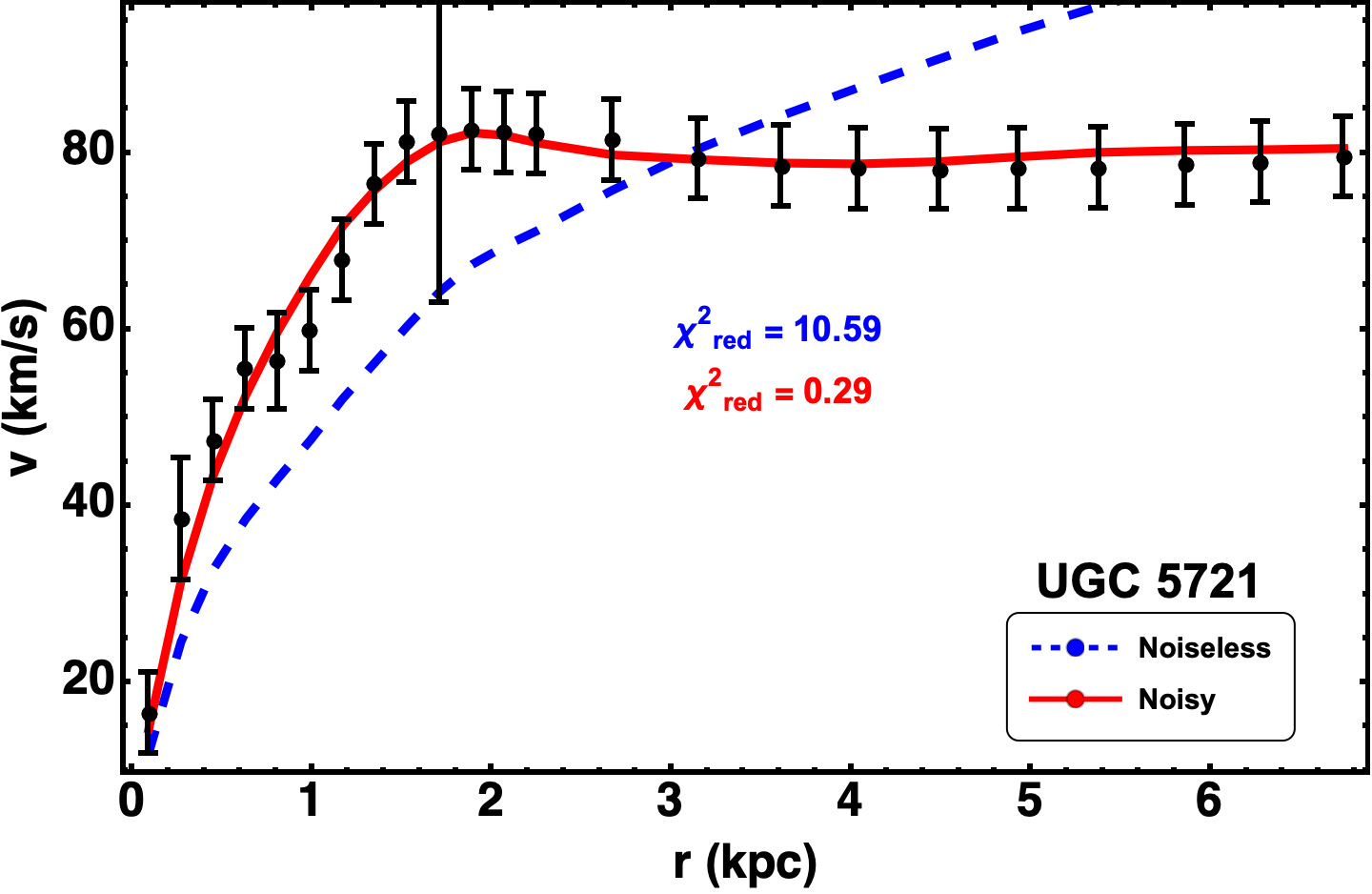}
    \caption{Performance of the neural networks trained on MSE loss with (red) and without (blue) noisy inputs, for UGC 5721.}
    \label{fig:UGC05721_compare}
\end{figure}
\end{itemize}

\subsection{Noisy case}\label{sec:noisy_case}

\begin{figure*}
\begin{tabular}{ccc}
\includegraphics[width=0.3\textwidth]{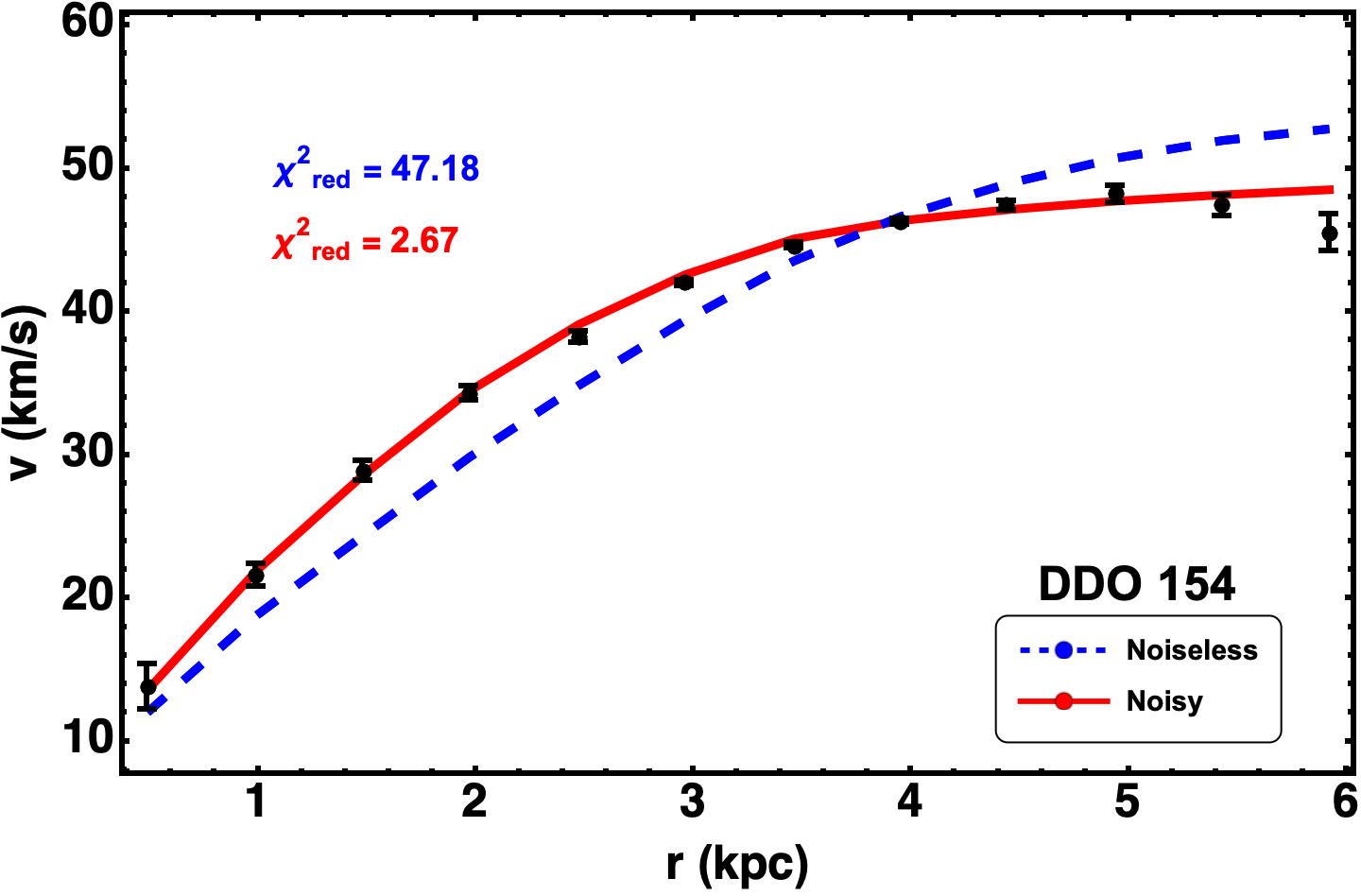} & 
\includegraphics[width=0.3\textwidth]{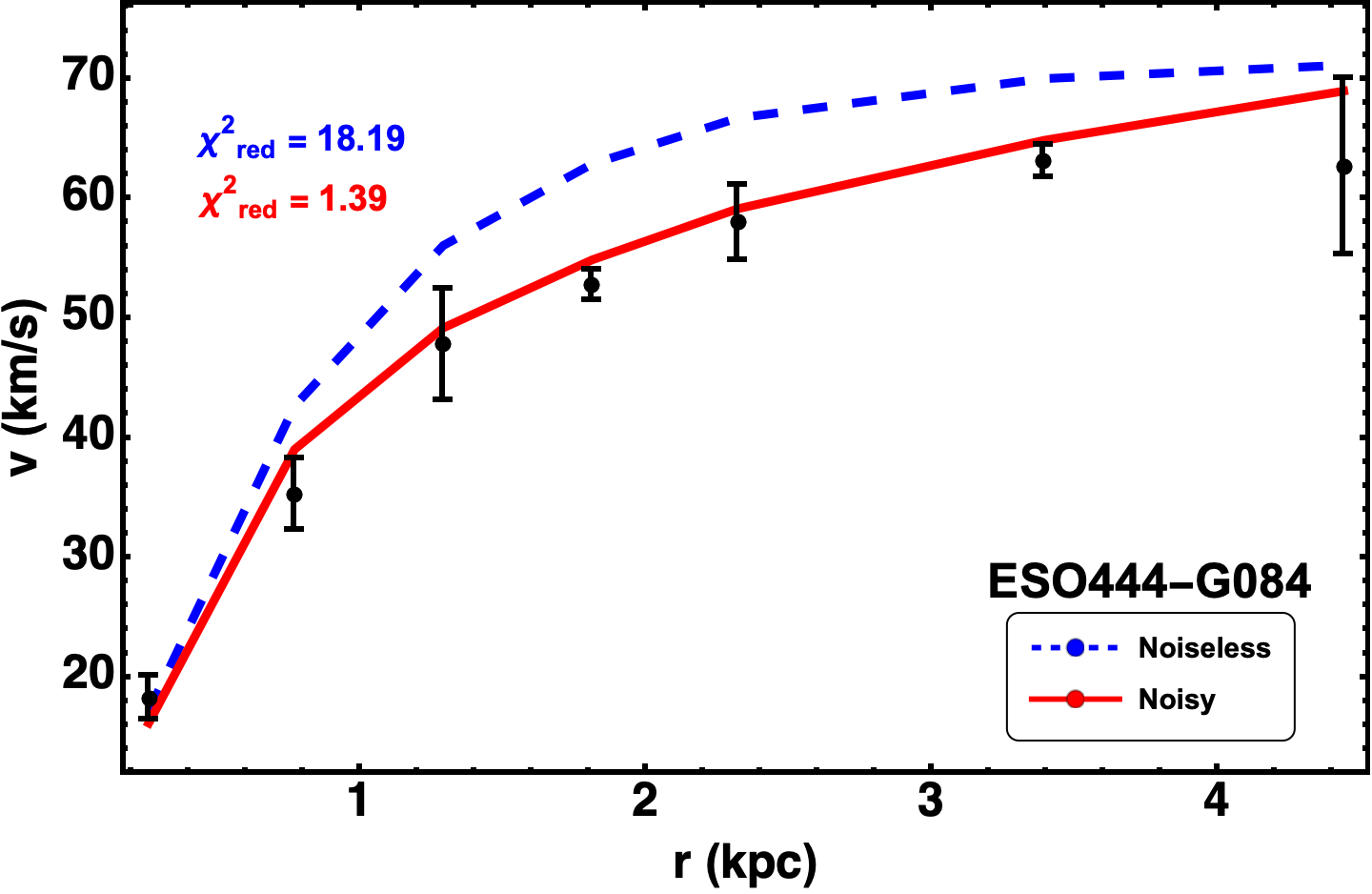} &
\includegraphics[width=0.3\textwidth]{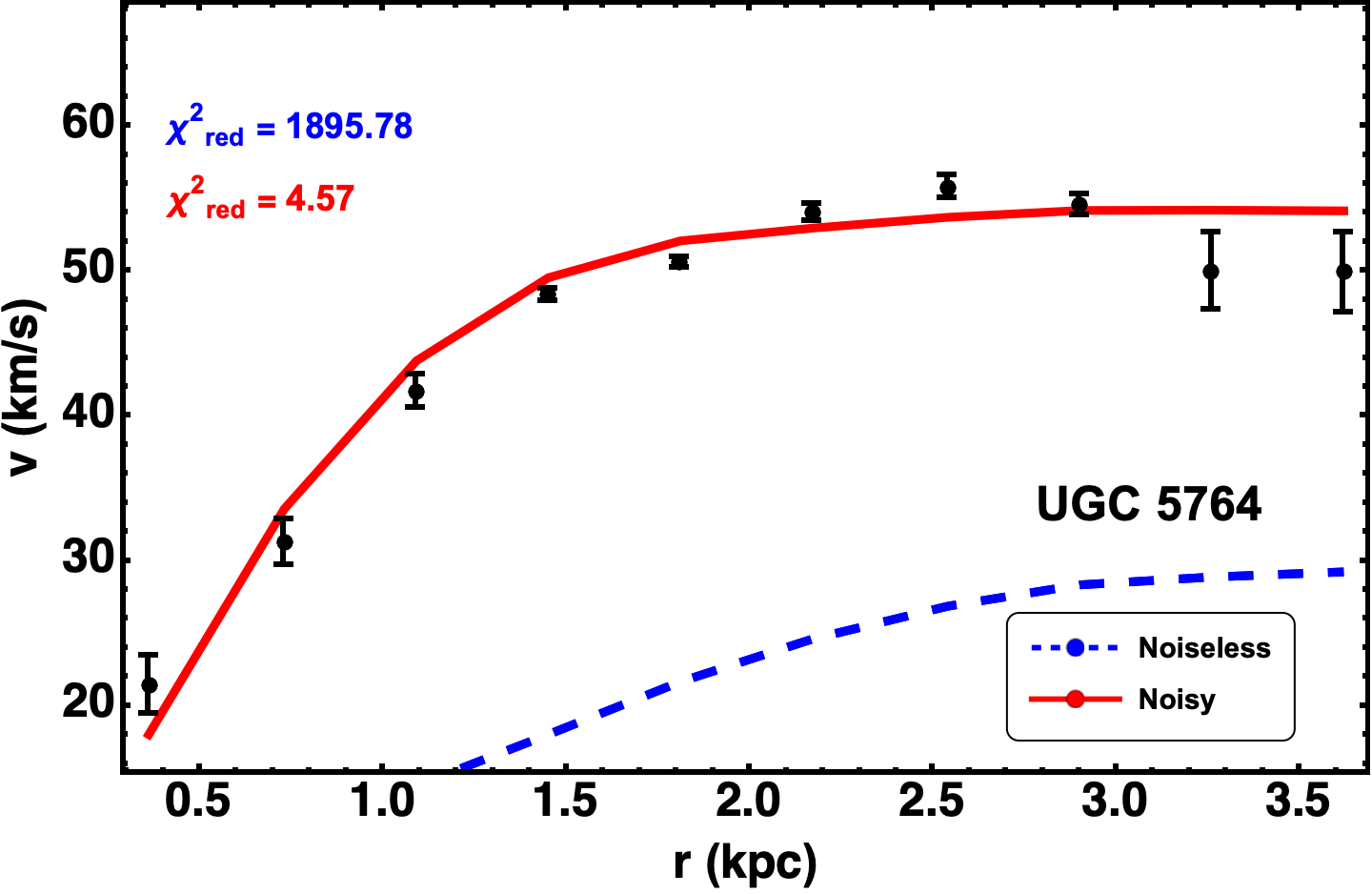} \\
(a) DDO154 & (b) ESO 444-G084 & (c) UGC 5764 \\[6pt]
\includegraphics[width=0.3\textwidth]{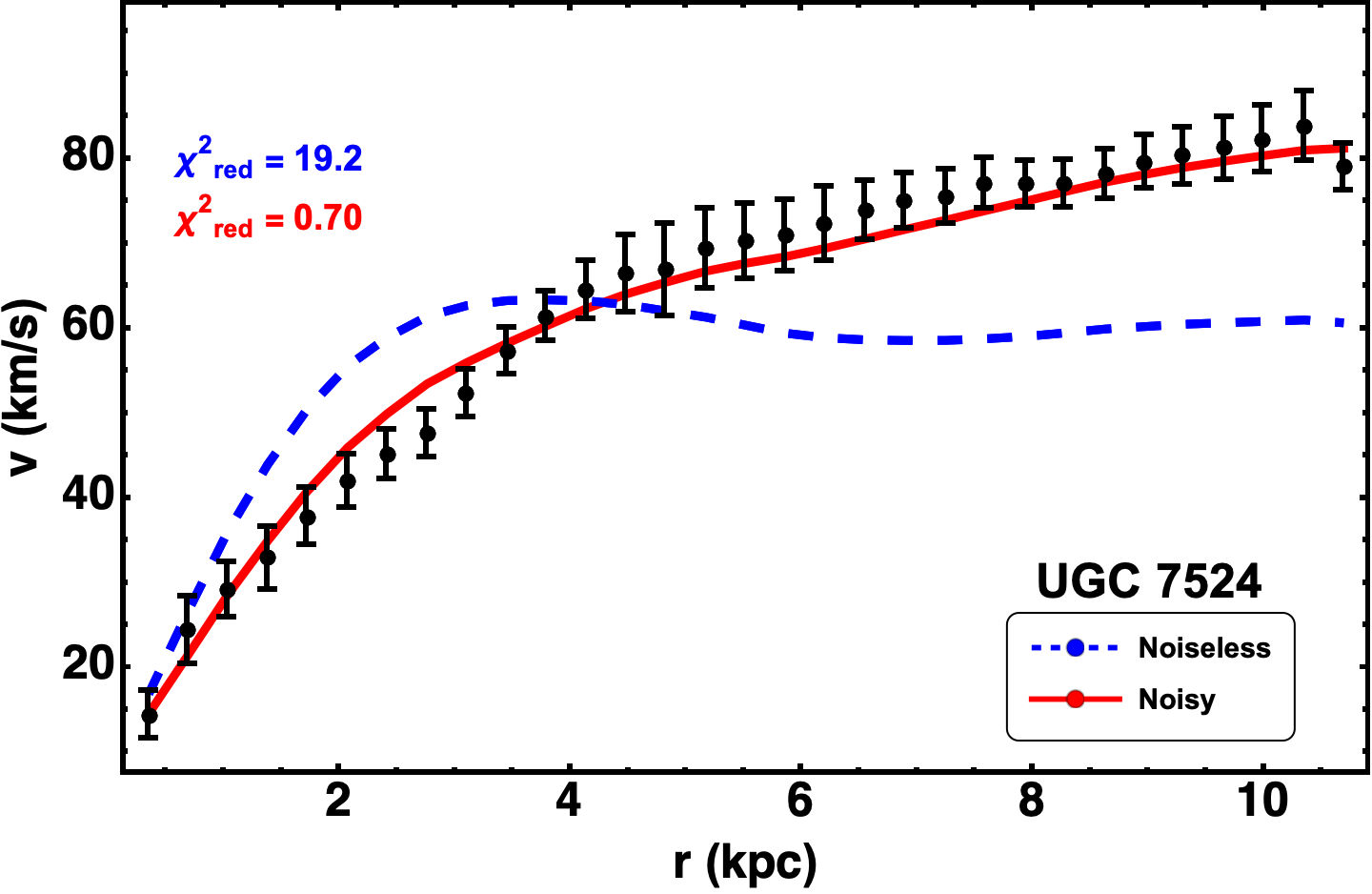} &   \includegraphics[width=0.3\textwidth]{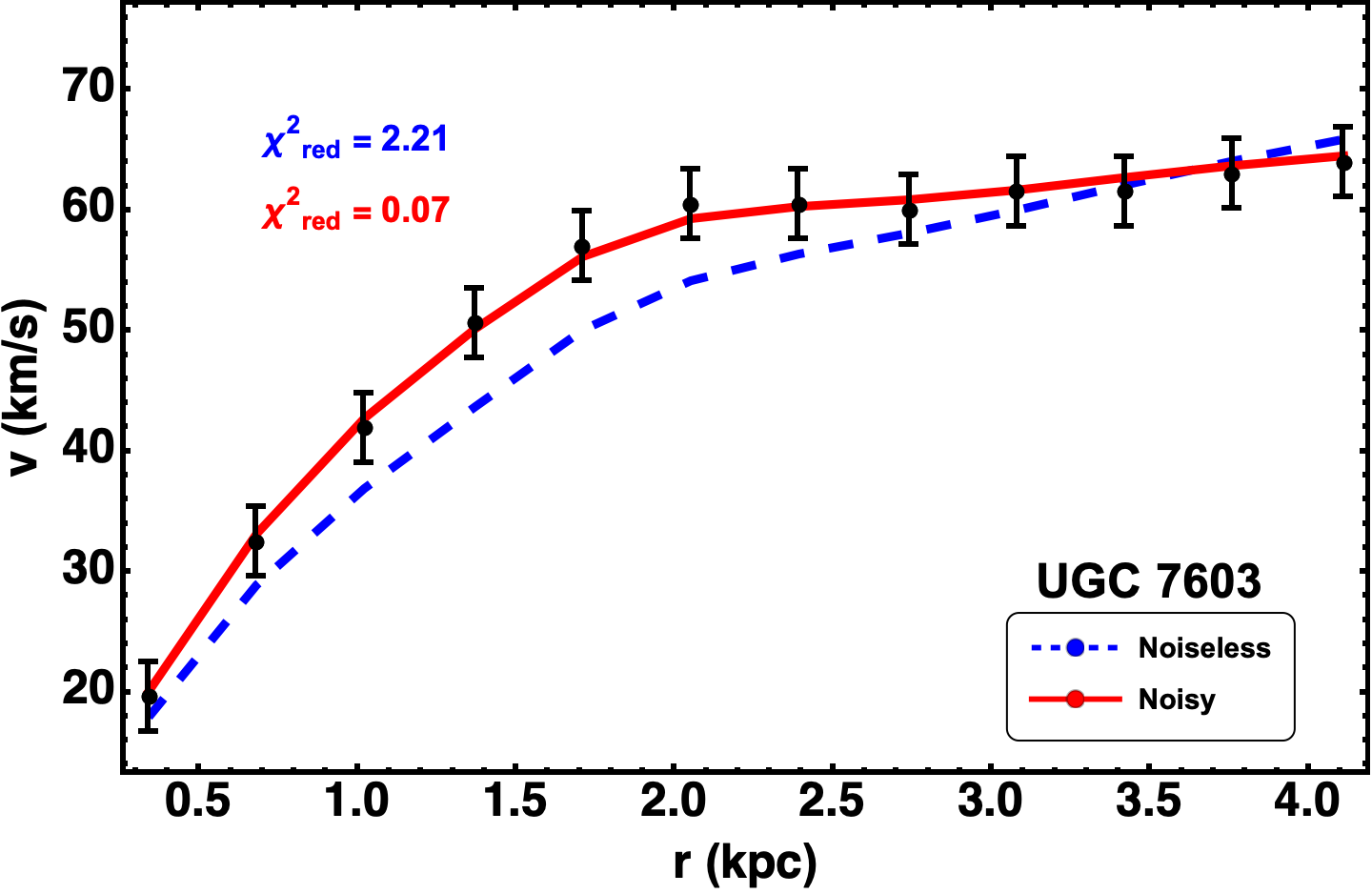} &
\includegraphics[width=0.3\textwidth]{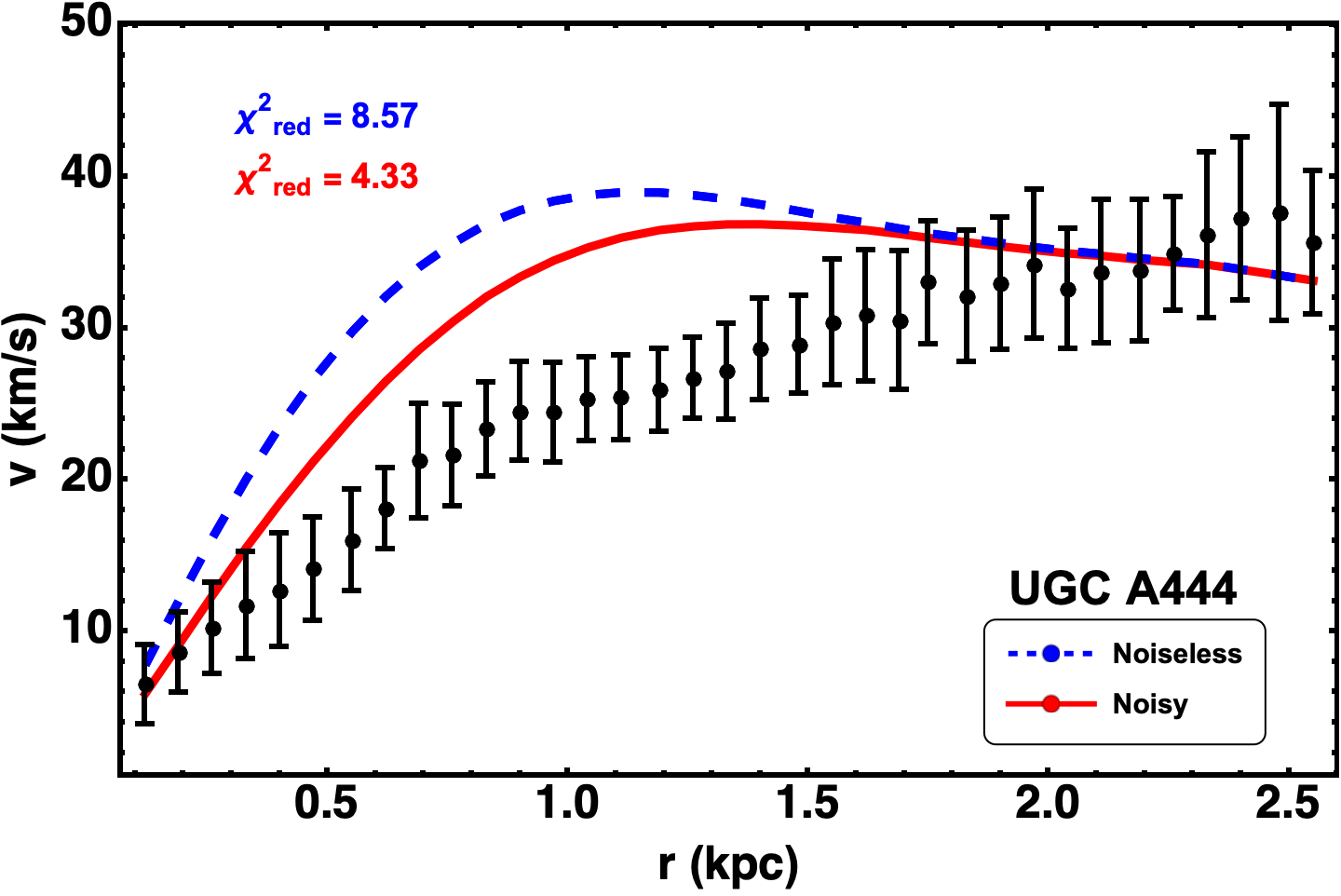} \\
(d) UGC 7603 & (e) UGC 7524 & (f) UGCA 444
\end{tabular}
\caption{\justifying Curves corresponding to the predicted parameters for the neural networks trained on noisy (red) and noiseless (blue) input data with MSE loss function. Note that only the central value of the observed rotation curves were given as input.}
\label{fig:MSE_comparison}
\end{figure*}

In this section, we consider a neural network with the same hyperparameters and training time as in the previous section, but with noise included in simulated rotation curves. 
We shall see how this improves performance of the network when confronted with observed rotation curves.

The SPARC catalog provides (for all galaxies) central values for the observed velocity at different radius values along with the uncertainty at that bin, i.e. $v(r) = v_{obs}(r) \pm \sigma(r)$. 
Hence, at some radius bin $r_i$ we denote the observed velocity as $v_{obs}(r_i) \pm \sigma(r_i)$. 
This noise, when incorporated during training, could possibly lead to better predictions \cite{Wang_2020, Pal_2023}. 
As we shall see in this section, that is precisely the case. 
 
In every simulated rotation curve, for each radius bin $r_i$, we add a random draw from the distribution $\mathcal{N}(0, \sigma(r_i)^2)$ as noise to the simulated velocity $v_{sim}(r_i)$.
Note that the target outputs, i.e. the known parameters, are not noisy and will not be affected by this process. 
The usual pre-processing steps are carried out as discussed in section~\ref{sec:preprocessing}.

It is interesting to note that training with noisy inputs is equivalent to Tikhonov regularization \cite{Bishop_1995}, a strategy that is also employed in \cite{Wang_ECoPANN_2020} for cosmological parameter estimation. 
However, unlike the procedure in the above papers, we do not add noise after every epoch, and choose to include noise at the pre-processing stage itself. 

We train neural networks for $250$ epochs with the same hyperparameters except that inputs will now be noisy. In this case, validation loss does converge by $250$ epochs, after which it starts to increase, signaling overfitting. Similar to the previous section, using the example of UGC 5721, we note the following results: 

\begin{itemize}
    \item After training, we evaluate the performance of the neural network on test data, for which we know the target parameters. We find that the test loss is similar to the training and validation loss implying no overfitting. 
    It is worth noting that due to the regularization effect of the input noise, the MSE loss for all data sets is higher compared to the noiseless case. 
    However, when rotation curves corresponding to target and predicted parameters are constructed, we find good agreement between the two as seen in figure~\ref{fig:test_noiseless}. 

    \item On the other hand, unlike the noiseless case, we find that the performance on the observed rotation curve has improved drastically. We feed as input, the central values of the observed velocities and construct the rotation curve from the output parameters predicted. The corresponding curve and $\chi^2_{red}$ are shown in figure~\ref{fig:UGC05721_compare} in red. Note that the $\chi^2_{red}$ value is far smaller compared to the noiseless case.
\end{itemize}

Similar results are obtained for all the other galaxies in our sample, as seen in figure~\ref{fig:MSE_comparison}.
In conclusion, neural networks trained with noisy training data perform much better than those trained without noise. 
In the next section we shall discuss how one can also obtain the uncertainties associated with the parameters along with the point-estimates.

\section{Estimating uncertainties in model parameters}{\label{sec:uncertainties}}

As we discussed in the previous section, observed rotation curves have errors associated with every velocity observation. 
Given this uncertainty, it would be useful to obtain uncertainties in the parameter predictions as well.  

It is important to note that in machine learning uncertainty is usually separated into \cite{Kendall_HS_2017}: (a) Epistemic uncertainty viz., the uncertainty associated with the choice of the model and its weights, which can be reduced by using more training data and (b) Aleatoric uncertainty viz., the uncertainty that is inherent to the data itself (like errors in measurement or observations) and cannot be reduced by increasing the size of the training set.

In this section, following recent work \cite{Wang_ECoPANN_2020, Ho_2021, Pal_2023} we try to account for the aleatoric uncertainty in multiple ways, for neural networks trained on noisy inputs. 
First we explore how multiple realizations of the observed rotation curves can enable us to obtain multiple estimates of parameters in section~\ref{sec:multiple_realizations}. 
These point-estimates can be used to construct `joint' and `marginalized' distributions for the predicted parameters. 
In section~\ref{sec:hetero_uncertainty} we also explore the use of a heteroscedastic loss function to implicitly learn the uncertainty during training itself.

\subsection{Using multiple realizations of the observations}\label{sec:multiple_realizations}

A straightforward way to obtain uncertainties in parameter predictions - assuming that errors in observations are Gaussian - is to draw multiple samples from the Gaussian $\mathcal{N}(v_{obs}(r), \sigma_r^2)$ for each value of radius $r$ observed. 

Thus, for every galaxy in our sample, we now have multiple realizations of the observed rotation curves for that galaxy. 
We can use these realizations as inputs to the trained neural network corresponding to each galaxy. 
Each realization will lead to a slightly different parameter prediction ${\bf \hat{P}}$. 
We generate $1000$ realizations and obtain $1000$ predictions for each galaxy in the sample. 

For every galaxy, we then consider the $50\%$ quantile (i.e., median) for each parameter as the point-estimate and $16\%$, $84\%$ quantiles (corresponding to a $1\sigma$ region of a $1$D normal distribution) as the uncertainties. 
This definition will help us in comparing our results to those obatined using MCMC in section~\ref{sec:mcmc_comparison}.
Note that a similar procedure was used in \cite{Wang_ECoPANN_2020} to estimate cosmological parameters from CMB observations.
The resultant parameters and their uncertainties are shown in Table~\ref{tab:preds_hetero}. 
We also plot the rotation curves constructed from predictions and obtain the goodness of fit to the observations using $\chi^2_{red}$ (reduced $\chi^2$).
The rotation curves are shown in figure~\ref{fig:all_comparison} by the purple dashed curves. 
It is clear that similar to the case with a single realization of the observed rotation curve, the rotation curves corresponding to the median value of parameters fit observations just as well, or even better, especially in the case of DDO 154 and UGCA 444.

\begin{table}
\centering
\resizebox{\linewidth}{!}{
\begin{tabular}{c|c|ccccc}
\toprule
\multirow{2}{*}{\textbf{Galaxy}} & \multirow{2}{*}{\textbf{Method}} & \multicolumn{5}{c}{\textbf{Predictions}} \\ \cmidrule(l){3-7}
& & $m$ ($10^{-23}$ eV) &  scale $s$ & $r_t$ (kpc) & $r_s$ (kpc) & $\Upsilon_*$ ($M_\odot/L_\odot$)\\ \cmidrule(r){1-7}

\multirow{3}{*}{DDO 154} & MSE & $1.96_{-0.16}^{+0.12}$ & $5380.39_{-103.57}^{+101.64}$ & $3.63_{-0.69}^{+0.85}$ & $6.37_{-0.57}^{+0.63}$ & $0.61_{-0.08}^{+0.05}$ \\ [0.2cm]

& Hetero & $1.95_{-0.30}^{+0.30}$ & $5350.71_{-292.16}^{+292.16}$ & $3.50_{-0.87}^{+0.87}$ & $6.87_{-3.90}^{+3.90}$ & $0.61_{-0.12}^{+0.12}$ \\ [0.2cm]

& MCMC & $1.79_{-0.10}^{+0.16}$ & $5328.85_{-90.48}^{+136.65}$ & $3.25_{-0.83}^{+1.61}$ & $4.01_{-1.56}^{+3.56}$ & $0.70_{-0.14}^{+0.08}$ \\ [0.1cm] \cmidrule(r){1-7}

\multirow{3}{*}{ESO 444-G084} & MSE & $5.06_{-0.98}^{+0.69}$ & $4954.22_{-157.20}^{+123.29}$ & $1.26_{-0.08}^{+0.2}$ & $9.91_{-1.06}^{+0.72}$ & $0.58_{-0.01}^{+0.01}$ \\ [0.2cm]

& Hetero & $5.26_{-1.07}^{+1.07}$ & $5055.20_{-363.85}^{+363.85}$ & $1.26_{-0.27}^{+0.27}$ & $9.53_{-3.38}^{+3.38}$ & $0.56_{-0.14}^{+0.14}$ \\ [0.2cm]

& MCMC & $5.29_{-1.02}^{+0.94}$ & $5181.16_{-324.19}^{+257.23}$ & $1.14_{-0.11}^{+0.21}$ & $10.39_{-3.63}^{+3.15}$ & $0.58_{-0.18}^{+0.16}$ \\ [0.1cm]  \cmidrule(r){1-7}

\multirow{3}{*}{UGC 5721} & MSE & $2.28_{-0.30}^{+0.37}$ & $3042.17_{-127.19}^{+155.87}$ & $2.44_{-0.34}^{+0.57}$ & $7.74_{-1.76}^{+0.94}$ & $0.65_{-0.04}^{+0.03}$ \\ [0.2cm]

& Hetero & $2.21_{-0.34}^{+0.34}$ & $3052.34_{-163.26}^{+163.26}$ & $2.55_{-0.72}^{+0.72}$ & $7.12_{-4.04}^{+4.04}$ & $0.66_{-0.13}^{+0.13}$ \\ [0.2cm]

& MCMC & $2.12_{-0.22}^{+0.30}$ & $3064.23_{-99.85}^{+124.44}$ & $2.56_{-0.41}^{+0.36}$ & $7.27_{-4.74}^{+5.20}$ & $0.70_{-0.13}^{+0.07}$ \\ [0.1cm]  \cmidrule(r){1-7}

\multirow{3}{*}{UGC 5764} & MSE & $3.97_{-0.42}^{+0.35}$ & $4658.97_{-119.27}^{+122.23}$ & $1.92_{-0.32}^{+0.64}$ & $4.18_{-1.00}^{+1.83}$ & $0.56_{-0.01}^{+0.01}$ \\ [0.2cm]

& Hetero & $3.87_{-0.43}^{+0.43}$ & $4769.11_{-192.48}^{+192.48}$ & $2.00_{-0.51}^{+0.51}$ & $4.75_{-3.74}^{+3.74}$ & $0.57_{-0.14}^{+0.14}$ \\ [0.2cm]

& MCMC & $4.25_{-0.35}^{+0.43}$ & $4879.48_{-106.92}^{+123.52}$ & $1.40_{-0.30}^{+0.28}$ & $1.42_{-0.32}^{+0.90}$ & $0.56_{-0.18}^{+0.17}$ \\ [0.1cm]  \cmidrule(r){1-7}

\multirow{3}{*}{UGC 7524} & MSE & $1.90_{-0.43}^{+0.62}$ & $4403.68_{-256.26}^{+376.14}$ & $3.02_{-0.75}^{+1.30}$ & $10.35_{-1.70}^{+1.29}$ & $0.57_{-0.03}^{+0.03}$ \\ [0.2cm]

& Hetero & $1.55_{-0.52}^{+0.52}$ & $4370.11_{-422.96}^{+422.96}$ & $3.02_{-1.43}^{+1.43}$ & $9.56_{-2.86}^{+2.86}$ & $0.57_{-0.14}^{+0.14}$ \\ [0.2cm]

& MCMC & $1.36_{-0.28}^{+0.70}$ & $4196.24_{-315.63}^{+609.70}$ & $2.48_{-0.89}^{+1.22}$ & $10.35_{-3.22}^{+2.92}$ & $0.64_{-0.17}^{+0.12}$ \\ [0.1cm] \cmidrule(r){1-7}

\multirow{3}{*}{UGC 7603} & MSE & $2.59_{-0.27}^{+0.47}$ & $4128.01_{-177.46}^{+326.71}$ & $2.29_{-0.42}^{+0.47}$ & $7.35_{-0.36}^{+0.68}$ & $0.54_{-0.04}^{+0.05}$ \\ [0.2cm]

& Hetero & $2.61_{-0.65}^{+0.65}$ & $4243.85_{-368.33}^{+368.33}$ & $2.20_{-0.77}^{+0.77}$ & $7.43_{-4.04}^{+4.04}$ & $0.56_{-0.14}^{+0.14}$ \\ [0.2cm]

& MCMC & $2.48_{-0.42}^{+0.62}$ & $4186.56_{-221.71}^{+345.33}$ & $2.18_{-0.83}^{+0.74}$ & $6.03_{-3.56}^{+5.78}$ & $0.57_{-0.17}^{+0.15}$ \\ [0.1cm] \cmidrule(r){1-7}

\multirow{3}{*}{UGCA 444} & MSE & $7.96_{-0.95}^{+1.05}$ & $7178.33_{-510.66}^{+384.62}$ & $1.74_{-0.01}^{+0.01}$ & $7.93_{-0.02}^{+0.02}$ & $0.55_{-0.001}^{+0.001}$ \\ [0.2cm]

& Hetero & $6.84_{-1.61}^{+1.61}$ & $7958.19_{-742.22}^{+742.22}$ & $1.47_{-0.35}^{+0.35}$ & $8.23_{-3.91}^{+3.91}$ & $0.57_{-0.14}^{+0.14}$ \\ [0.2cm]

& MCMC & $6.90_{-1.40}^{+1.13}$ & $8467.32_{-640.01}^{+386.14}$ & $1.34_{-0.22}^{+0.47}$ & $8.57_{-4.56}^{+4.38}$ & $0.57_{-0.18}^{+0.16}$ \\
\bottomrule
\end{tabular}
}
\caption{\justifying Parameters and their uncertainties obtained by feeding observed rotation curves as input to neural networks trained using: (a) the multiple realization method using MSE loss function (section~\ref{sec:multiple_realizations}), (b) a heteroscedastic loss function (section~\ref{sec:hetero_uncertainty}) and (c) using MCMC (section~\ref{sec:mcmc_comparison}).}
\label{tab:preds_hetero}
\end{table}

The method of multiple realizations that we have discussed here gives us a sequence of parameters which can then be used to construct joint and marginal distributions. 
One could then compare these distributions to the posteriors obtained from a likelihood-based MCMC sampling approach.
In section~\ref{sec:mcmc_comparison} we do this for UGC 5721 and compare the contours with those obtained using the approach described in this section.

\subsection{\label{sec:hetero_uncertainty}Employing heteroscedastic loss function}

Another way to account for aleatoric uncertainty involves changing the loss function used for training the neural network. 
Consider the case where, for every output corresponding to the parameter prediction, one assigns an additional output that represents the Gaussian variance in that output. 
This implies that instead of learning a point estimate of the parameter, the neural network now learns the mean and variance of the Gaussian distribution to which the parameter belongs. 
Hence, in the output layer, for every $k$th parameter in ${\bf P}$, we assign an additional output to be its variance $\sigma_k^2$. 
For our neural network, we shall now have an output layer $j = 3$ with $N_3 = 10$ outputs, with the first five outputs corresponding to the parameter vector ${\bf \hat{P}}$ and the last five to the uncertainties of the parameters. 

Here, it is important to note that we do not have known values for $\sigma_k^2$ in the output of the simulated training set. To learn these uncertainties, one must introduce an uncertainty term in the loss function which is given by (for $n$ samples),
\begin{equation}\label{eq:hetero_loss}
    \mathcal{L}_{HS} = \frac{1}{n}\sum_{i=1}^{n}\left[\frac{1}{2p}\sum_{k=1}^p\left(e^{-s_{ik}}(y_{ik} - \hat{y}_{ik})^2 + s_{ik} \right)\right]\ ,
\end{equation}
where $y_{ik}$ is the $k$th parameter for the $i$th sample in the training set, while $\hat{y}_{ik} \equiv f_{k}({\bf x}_i, {\boldsymbol \Omega})$ is the neural network prediction for the same. This is called heteroscedastic loss \cite{Kendall_HS_2017}. 
Similar loss functions have been utilized in parameter estimation recently in the context of cosmological parameters~\cite{Pal_2023, Fluri_2019} as well as gravitational wave parameters~\cite{Andres_Carcasona_2023}. 

Here, the actual outputs corresponding to the uncertainties have been redefined to $s_{ik} \equiv \log{\hat{\sigma}_{ik}^2}$ where $\hat{\sigma}_{ik}^2$ is the predicted variance for the predicted parameter $\hat{y}_{ik}$, ensuring a more stable loss \cite{Kendall_HS_2017}.
The first term (i.e. the squared difference between the predicted and target value) in the parenthesis is weighed by $\exp{(-s_{ik})}$, which penalizes too small values of $s_{ik}$ by increasing the loss. On the other hand, the second term ensures that too large values of $s_{ik}$ are also penalized.

An important caveat to note here is that this loss function assumes that the parameters are independent random variables drawn from a Gaussian distribution.

\subsubsection{\label{sec:hetero_inference}Training and inference using heteroscedastic loss}

As with the previous section, we shall first look at the UGC 5721 galaxy as an example. 
Utilizing the same architecture and the hyperparameter values as the previous case with noisy input data, we train the neural network for $250$ epochs. We find that the validation loss has converged by this epoch. 

\begin{figure*}[ht]
\centering
\begin{tabular}{cccc}
\includegraphics[width=0.3\textwidth]{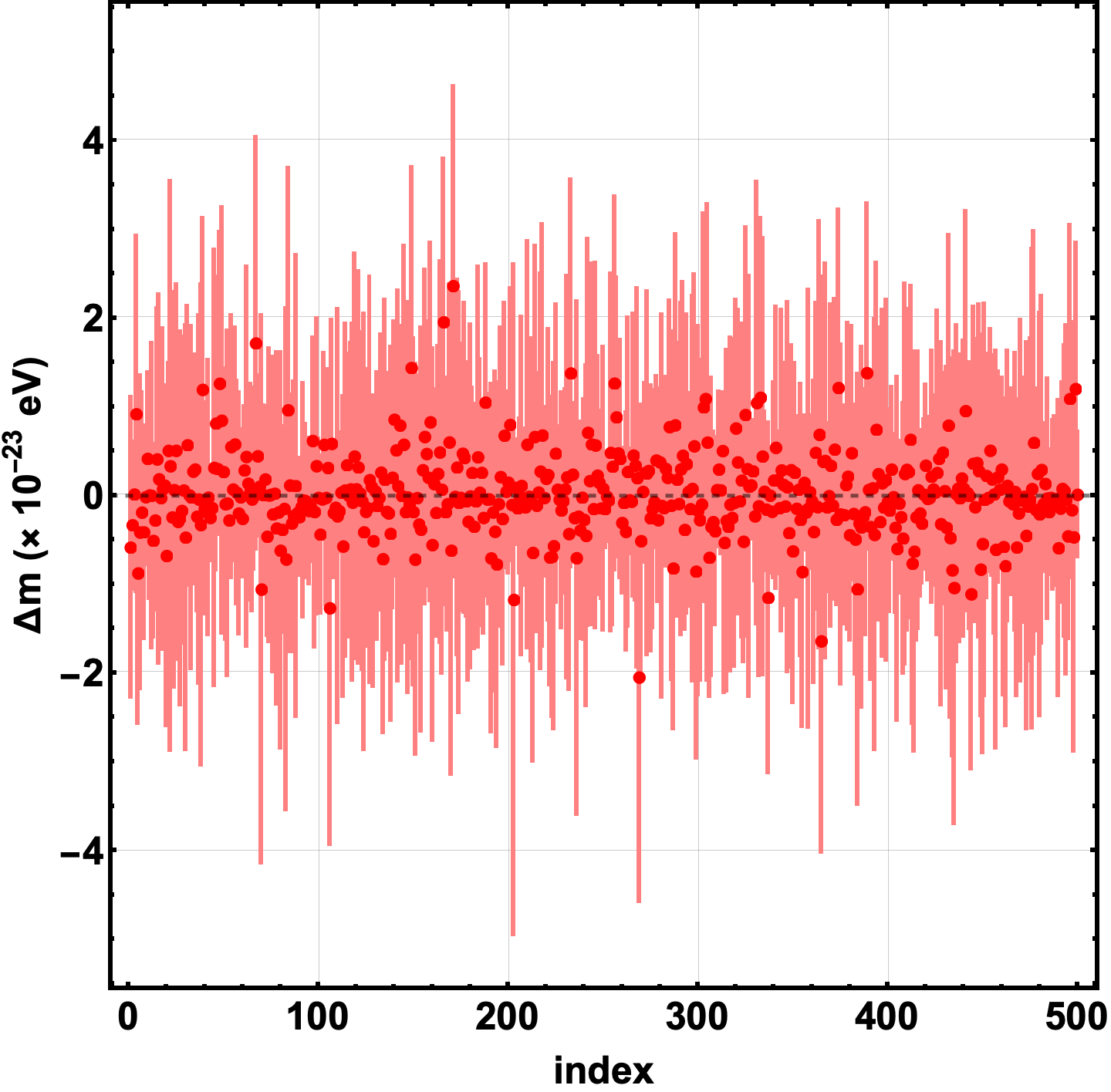} &
\includegraphics[width=0.3\textwidth]{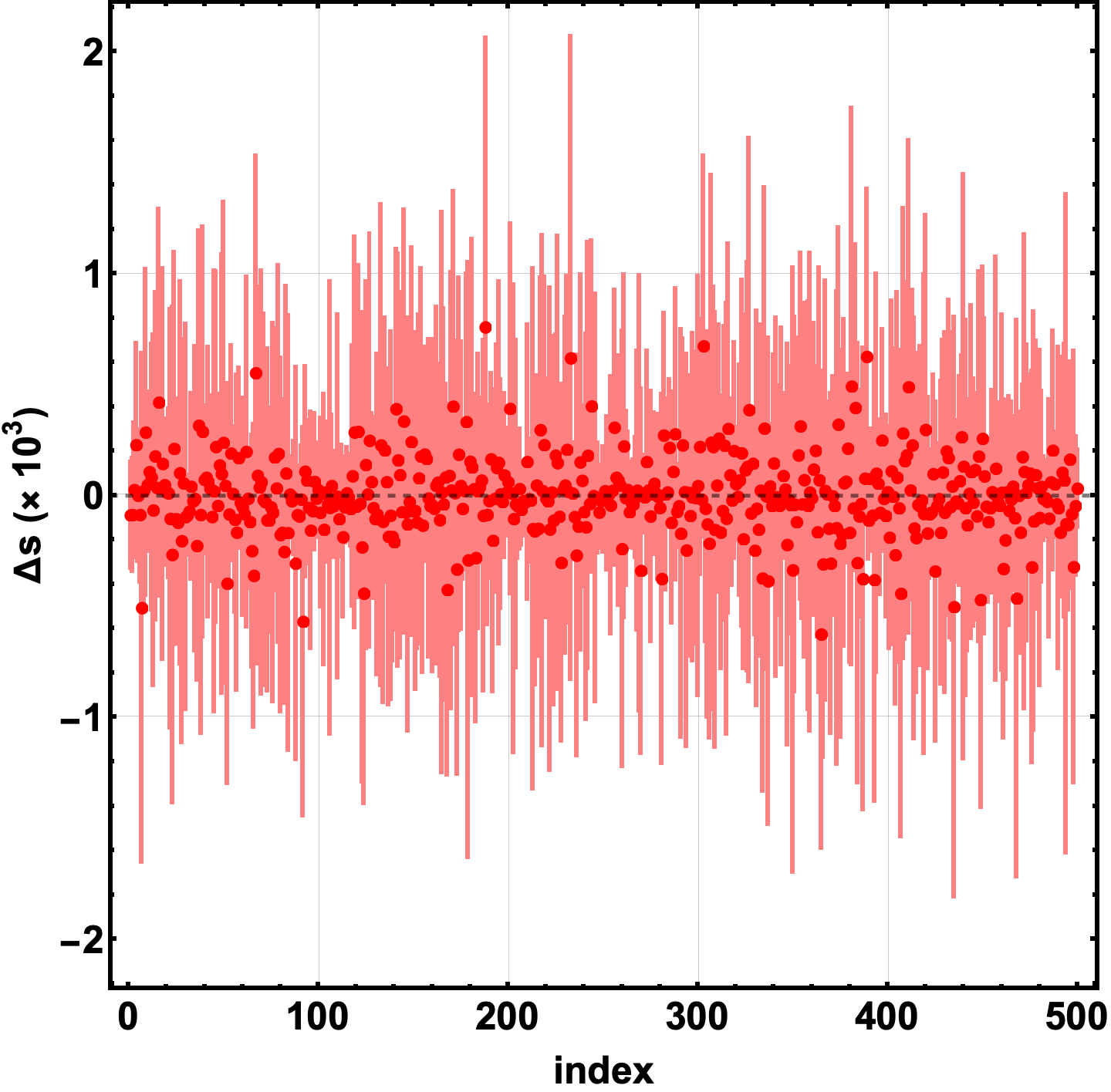} &
\includegraphics[width=0.3\textwidth]{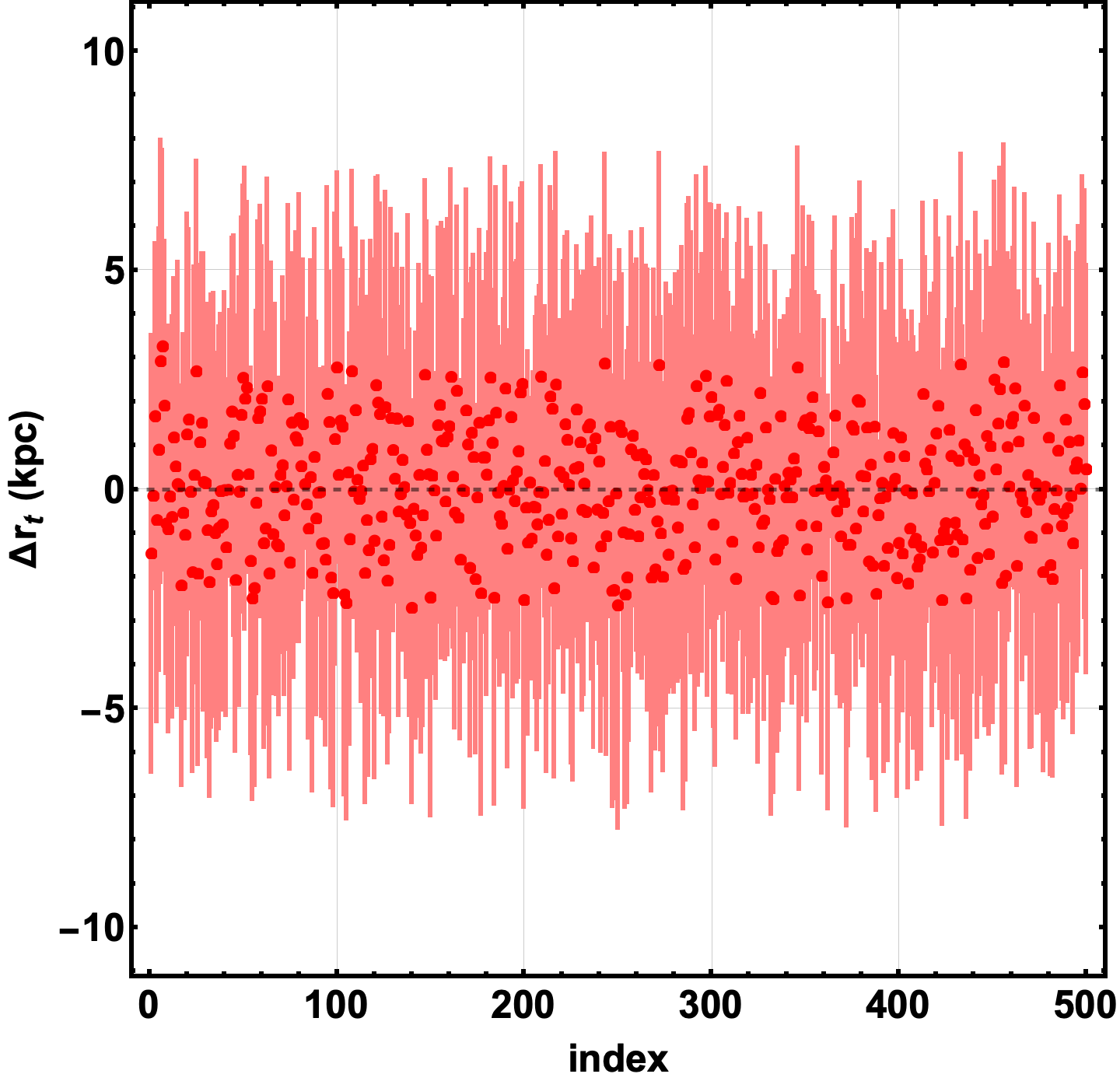} \\
(a) ULDM mass, $m$ & (b) scale parameter, $s$ & (c) transition radius, $r_t$  \\[6pt]
\end{tabular}
\begin{tabular}{cccc}
\includegraphics[width=0.3\textwidth]{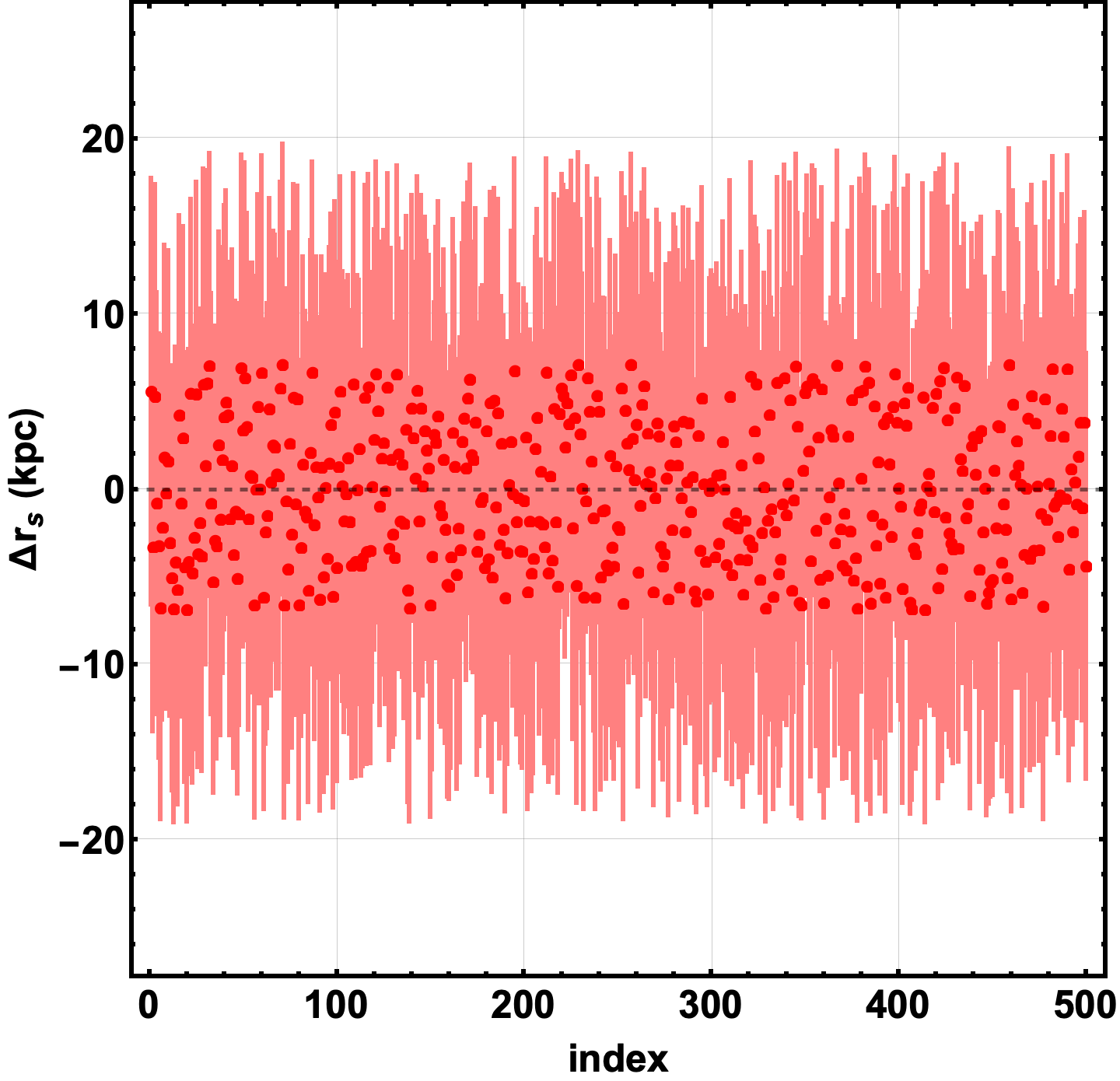} &
\includegraphics[width=0.3\textwidth]{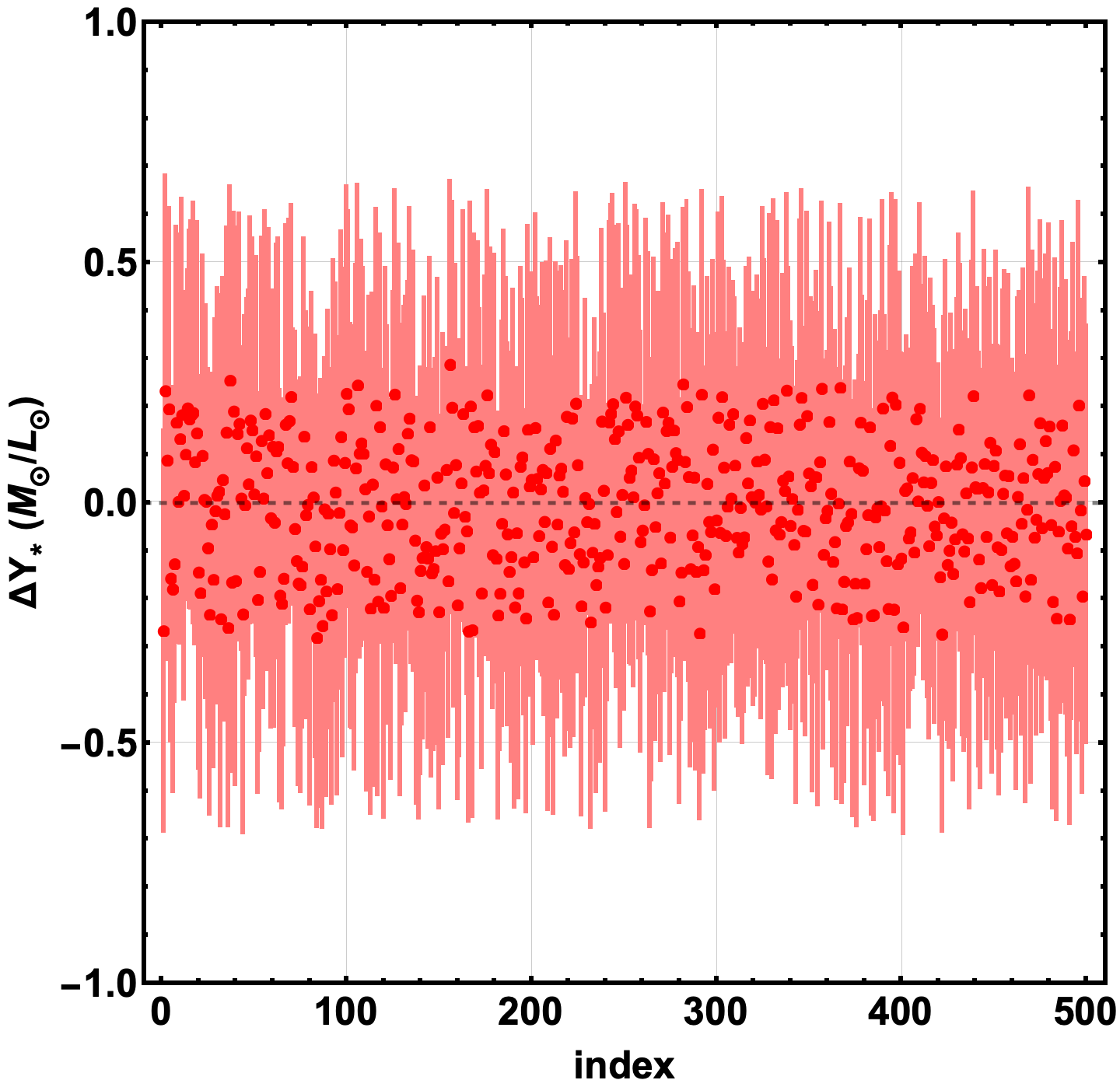} \\
(d) scale radius, $r_s$ & (e) stellar mass-to-light ratio, $\Upsilon_*$ \\[6pt]
\end{tabular}
\caption{\justifying Performance of the neural network for UGC 5721 trained using a heteroscedastic loss function on test data. The red dots are the difference between inferred and target parameter values, e.g. $\Delta m = m_{\text{true}} - m_{\text{inferred}}$, while the pink lines correspond to three times the inferred uncertainty of prediction.}
\label{fig:hetero_uncert}
\end{figure*}

Here, to test on unseen data, since the neural networks also predict uncertainties along with the point estimates of the parameters, we employ a different method than earlier. 
For $500$ samples from test data, we plot the difference between the predicted value and target value for each parameter. We also plot $3\hat{\sigma}_p$ for each sample. 
The plot is shown in figure~\ref{fig:hetero_uncert}, where the pink lines are $3\hat{\sigma}_p$ values while the red dots are differences between predicted and target parameters. 
Similar to \cite{Pal_2023}, for most samples, the difference between predicted and target value of each parameter lies within three times the uncertainty values.
Hence, these uncertainties appear to capture the ability of the neural network (with the chosen hyperparameters) to learn the parameters from the training data. 

Note that larger differences in $r_t$, $r_s$ and $\Upsilon_*$ and correspondingly larger uncertainties. 
The larger uncertainties in the $r_s$ and $\Upsilon_*$ are expected, since rotation curves of dark matter dominated dwarf galaxies are not sensitive to these parameters.
For the case of $r_s$, since the inner regions are described by a soliton core, i.e., we force $r_t\geq 1\ \text{kpc}$ and rotation curves extend only till $\mathcal{O}(5)\ \text{kpc}$ for most galaxies in our sample, a change in the NFW scale radius does not alter the rotation curve significantly. 
Similarly, because dark matter dominated galaxies have small baryonic contribution to begin with, changing $\Upsilon_*$ does not affect the rotation curve much.
We shall see in section~\ref{sec:mcmc_comparison}, how this quantification of the uncertainty compares with the uncertainty in parameters obtained from MCMC methods.

For the final test, we feed the central values of the rotation curve to the trained neural network as input, giving us an output parameter vector as well as the uncertainty associated with each parameter. 
The parameters and uncertainties are shown in Table~\ref{tab:preds_hetero}. 
Notice again, the large uncertainties (i.e., the same order of magnitude as the inferred values) obtained for $r_s$ and $\Upsilon_*$. 

\begin{figure}[ht]
    \centering
    \includegraphics[width=0.7\textwidth]{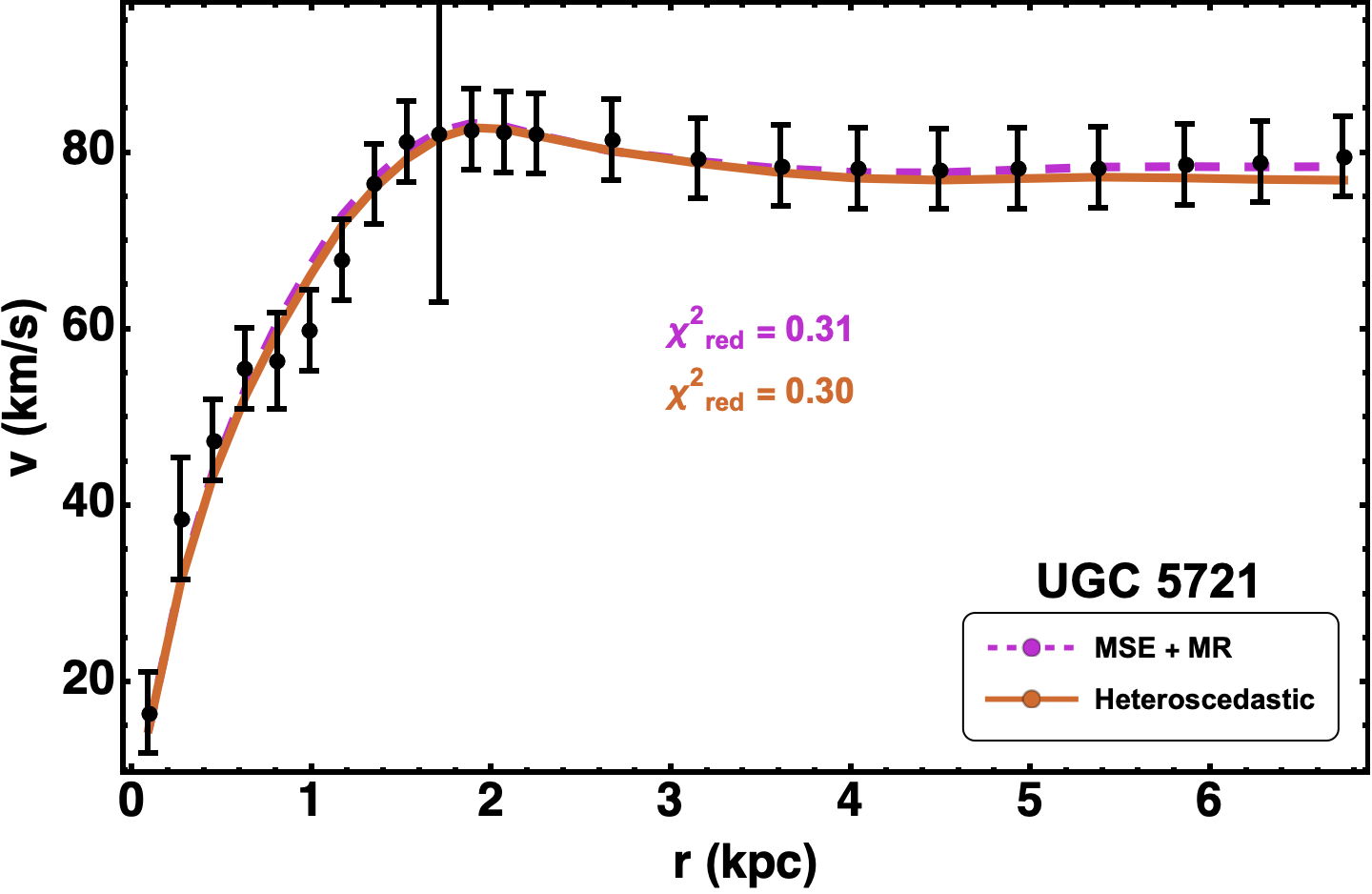}
    \caption{\justifying Rotation curves corresponding to the predicted parameters for the neural network trained for UGC 5271 on noisy input data with: (a) MSE loss and mean of multiple parameter predictions (purple dashed) and (b) heteroscedastic loss (orange).}
    \label{fig:UGC05721_hetero}
\end{figure}

The rotation curve corresponding to the above inferred parameters agrees well with observations for UGC 5721 which can be seen in figure~\ref{fig:UGC05721_hetero} by the orange curve.
We reiterate that only the central value of observed rotation curve is given to the neural network during inference, hence, there is only one prediction for the parameters and their uncertainties.
Then network has learned the uncertainty implicitly from the noisy training data. 

\begin{figure*}
\begin{tabular}{ccc}
\includegraphics[width=0.3\textwidth]{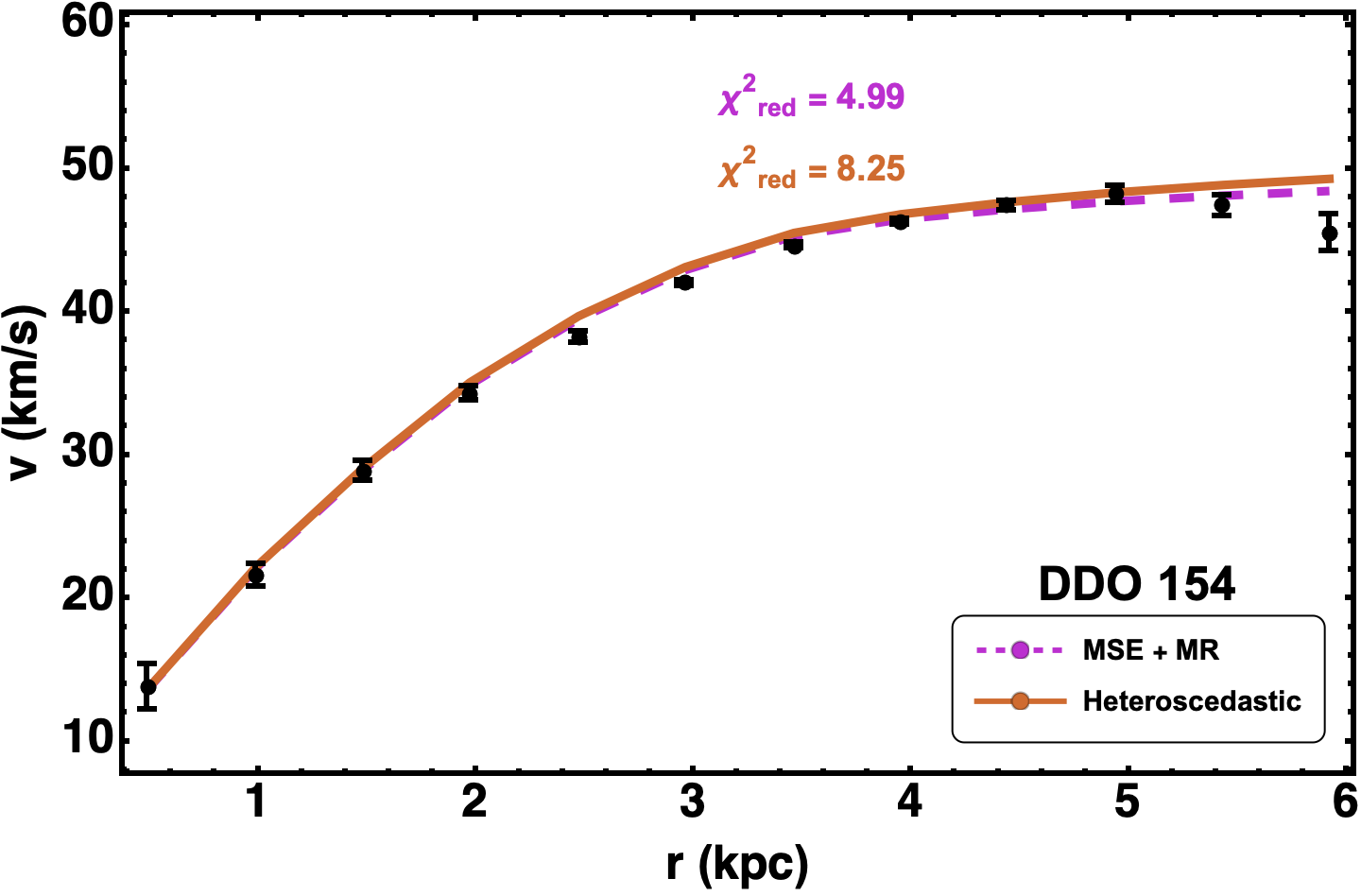} & 
\includegraphics[width=0.3\textwidth]{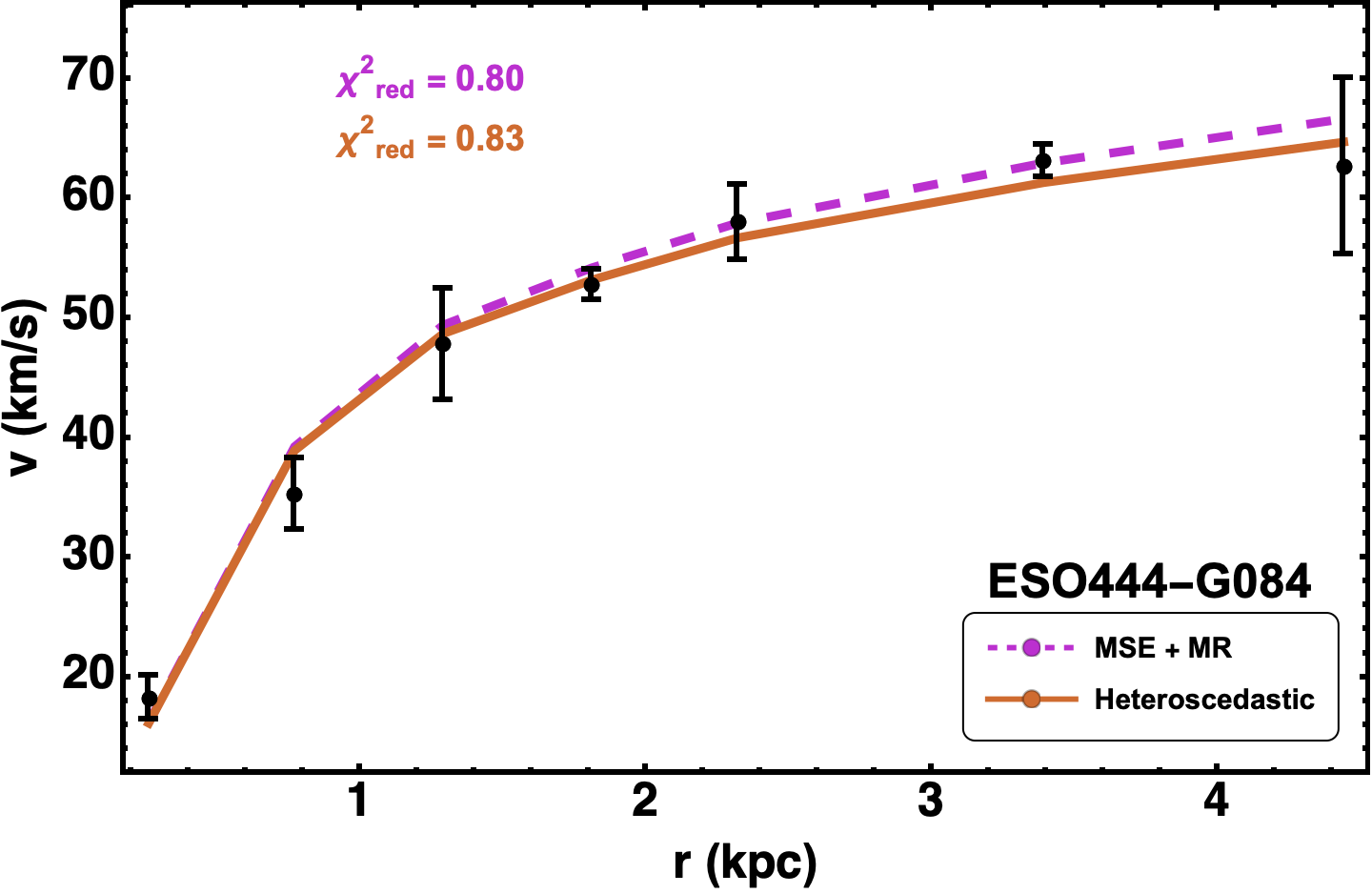} &
\includegraphics[width=0.3\textwidth]{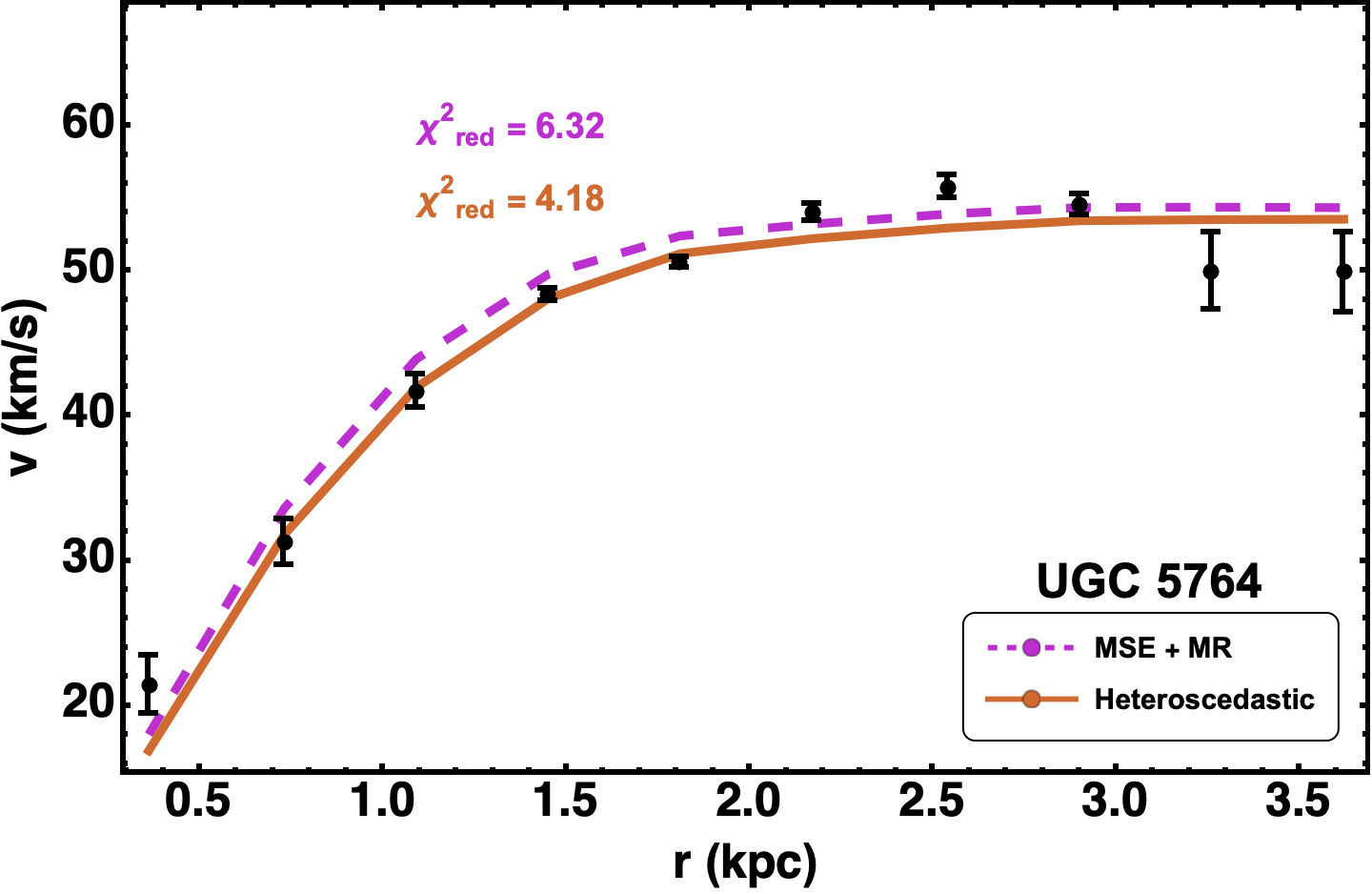} \\
(a) DDO154 & (b) ESO 444-G084 & (c) UGC 5764 \\[6pt]
\includegraphics[width=0.3\textwidth]{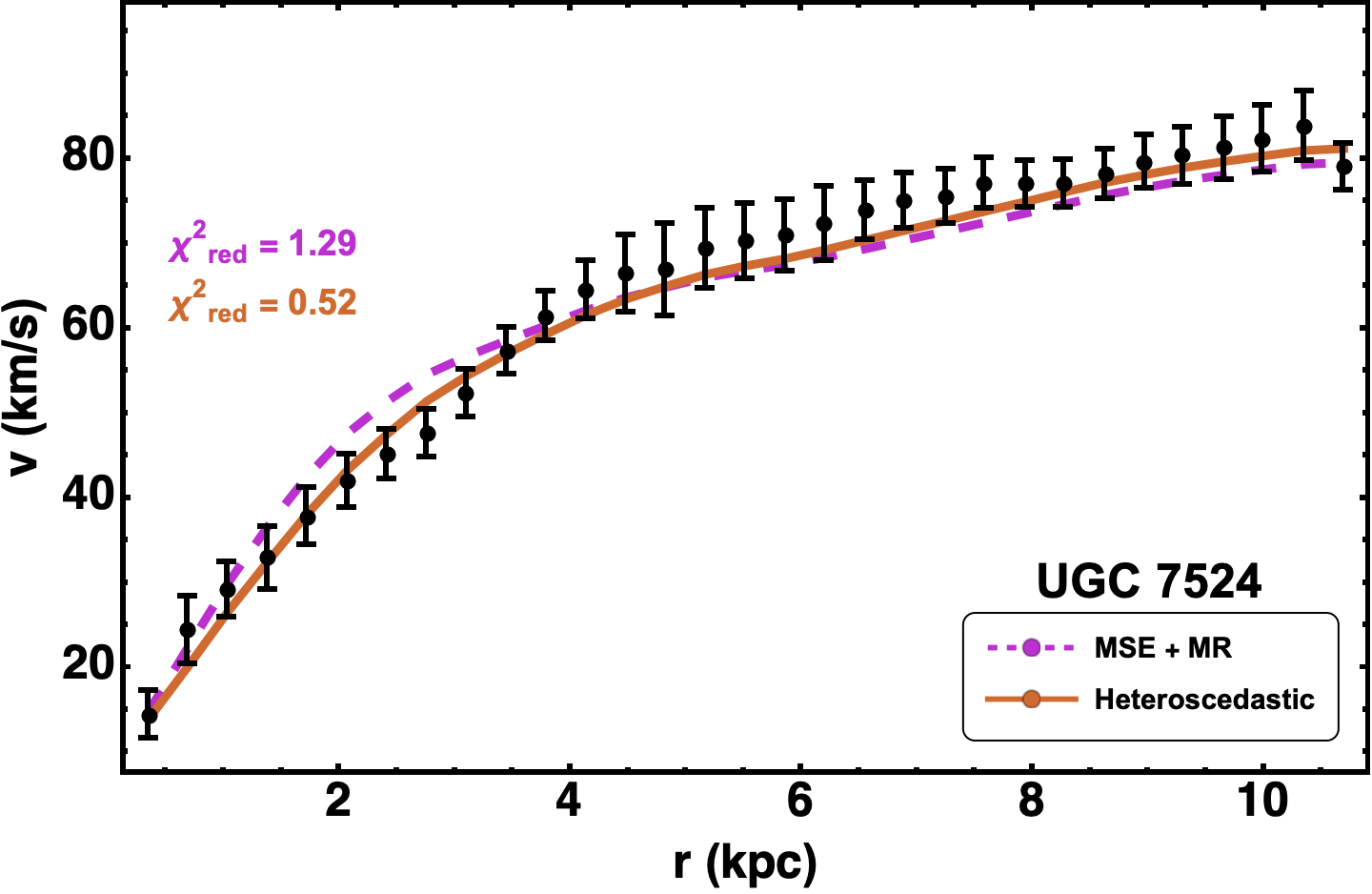} &   \includegraphics[width=0.3\textwidth]{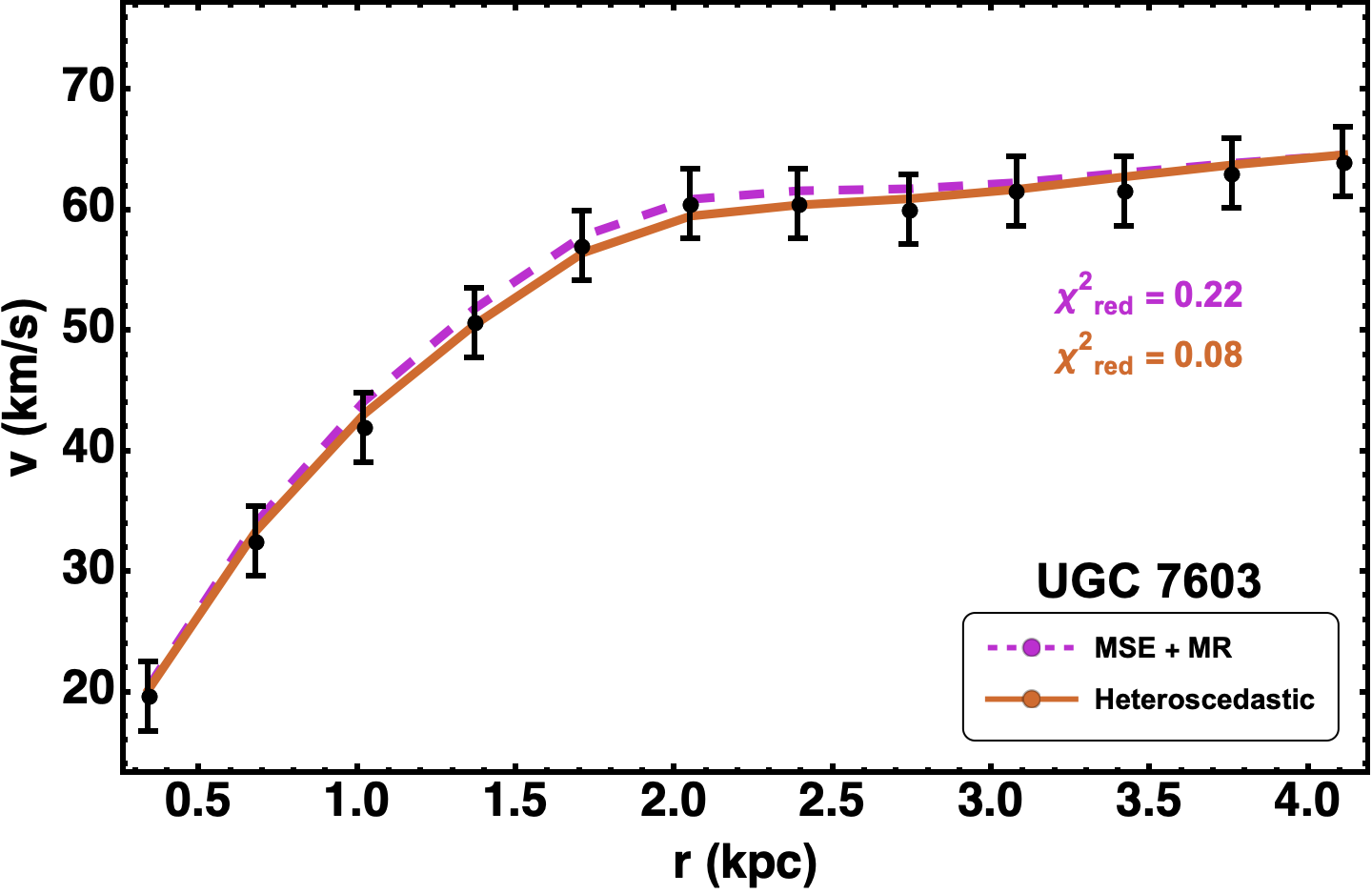} &
\includegraphics[width=0.3\textwidth]{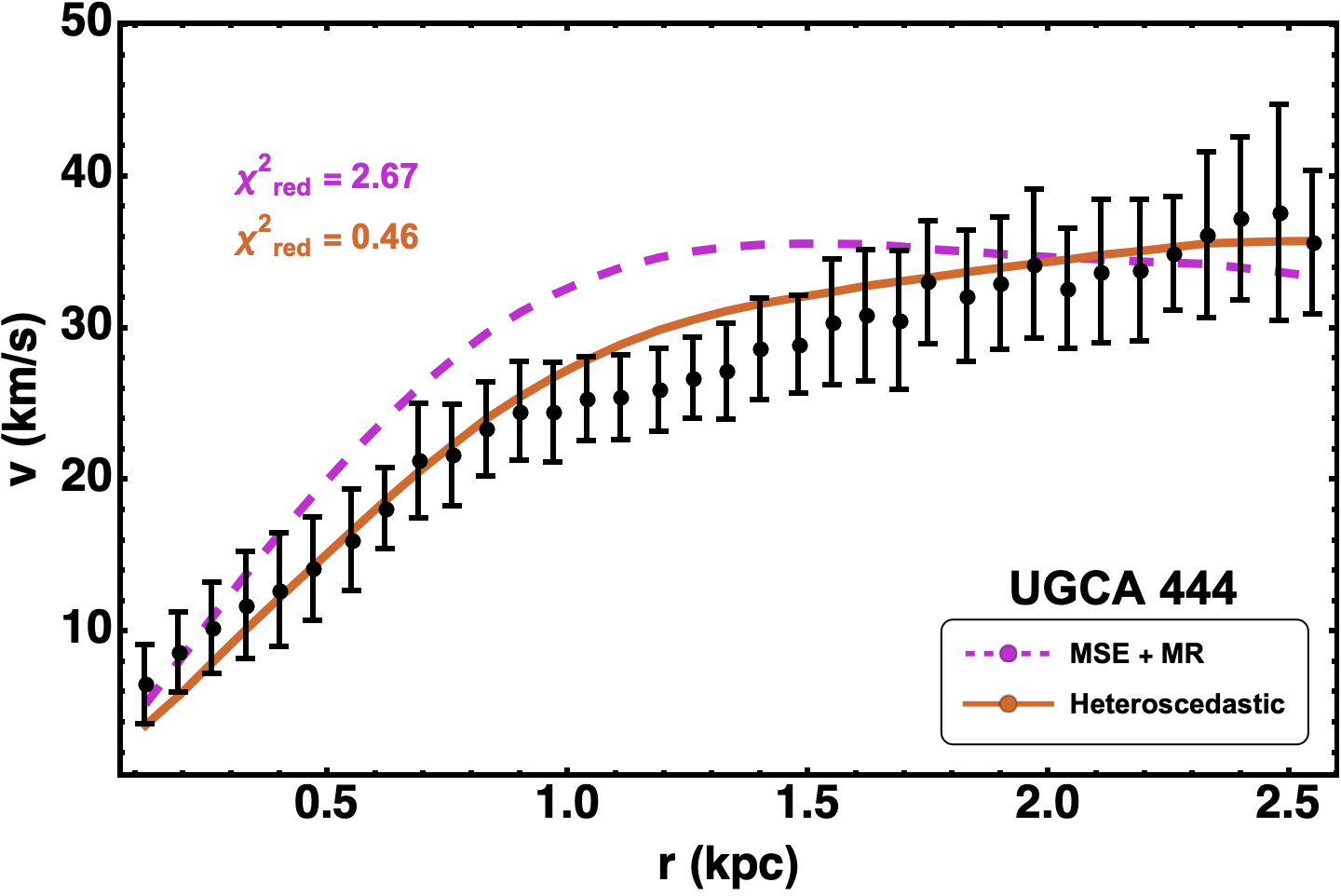} \\
(d) UGC 7603 & (e) UGC 7524 & (f) UGCA 444
\end{tabular}
\caption{\justifying Rotation curves corresponding to the predicted parameters for each of the three cases studied:(a) MSE loss function with noisy input data with multiple realizations of the observed rotation curves from section~\ref{sec:multiple_realizations} (purple dashed curves) , and (b)  Heteroscedastic loss function with noisy inputs from section~\ref{sec:hetero_uncertainty} (orange curves). The reduced $\chi^2$ are also shown for each case.}
\label{fig:all_comparison}
\end{figure*}

The story remains the same for all galaxies in the samples, rotation curves for which are shown in figure~\ref{fig:all_comparison}, while the parameter values and their uncertainties are listed in Table~\ref{tab:preds_hetero}.
Notice the larger uncertainties associated with $r_s$ and $\Upsilon_*$ for the other galaxies as well, which is remarkably consistent with the expectation that the rotation curves for DM dominated galaxies with $r_t \geq 1\ \text{kpc}$ are not very sensitive to $r_s$ and $\Upsilon_*$ values. 
This implies that uncertainty in the heteroscedastic loss function captures, in some sense, the effect a parameter will have on a rotation curve. 
This is interesting since the neural network has only approximated the functional dependence between parameters and rotation curves by looking at various examples. 

Therefore we see that without having to define a likelihood or plotting a posterior distribution, one can use heteroscedastic loss function obtain accurate parameter predictions as well as uncertainties that capture, to some extent, the sensitivity of the data to the parameters.

\section{Comparison with MCMC}\label{sec:mcmc_comparison}

In this section, we compare the multiple realizations approach of obtaining a sequence of parameters as well as the uncertainties obtained using the heteroscedastic loss function with a likelihood-based MCMC approach. 

Recall from section~\ref{sec:multiple_realizations}, that for a fixed galaxy, we sample from the Gaussian $\mathcal{N}(v_{obs}(r), \sigma_r^2)$ for all radius values to generate multiple realizations of the observed rotation curve. 
One can think of this as sampling curves from the vicinity of the observed rotation curve while being consistent with observations. 
For a sufficiently accurate neural network, one can then think of the multiple parameter predictions (using the multiple realizations as input) as a `chain' of parameters that satisfy the observed rotation curve of that galaxy. 
One can then plot 2D and 1D projections of the parameter values for this chain in a corner plot to visualize how the parameters are distributed. 

The above procedure appears qualitatively similar to the approach used in Bayesian inference using MCMC.
Indeed, authors in \cite{Wang_ECoPANN_2020} employed such a method to obtain chains of parameters and found that the corresponding 2D contour plots along with 1D projections using ANNs agreed well with the 2D joint distributions and 1D marginalized distribution obtained using MCMC sampling. 

\begin{figure}
    \includegraphics[width=\textwidth]{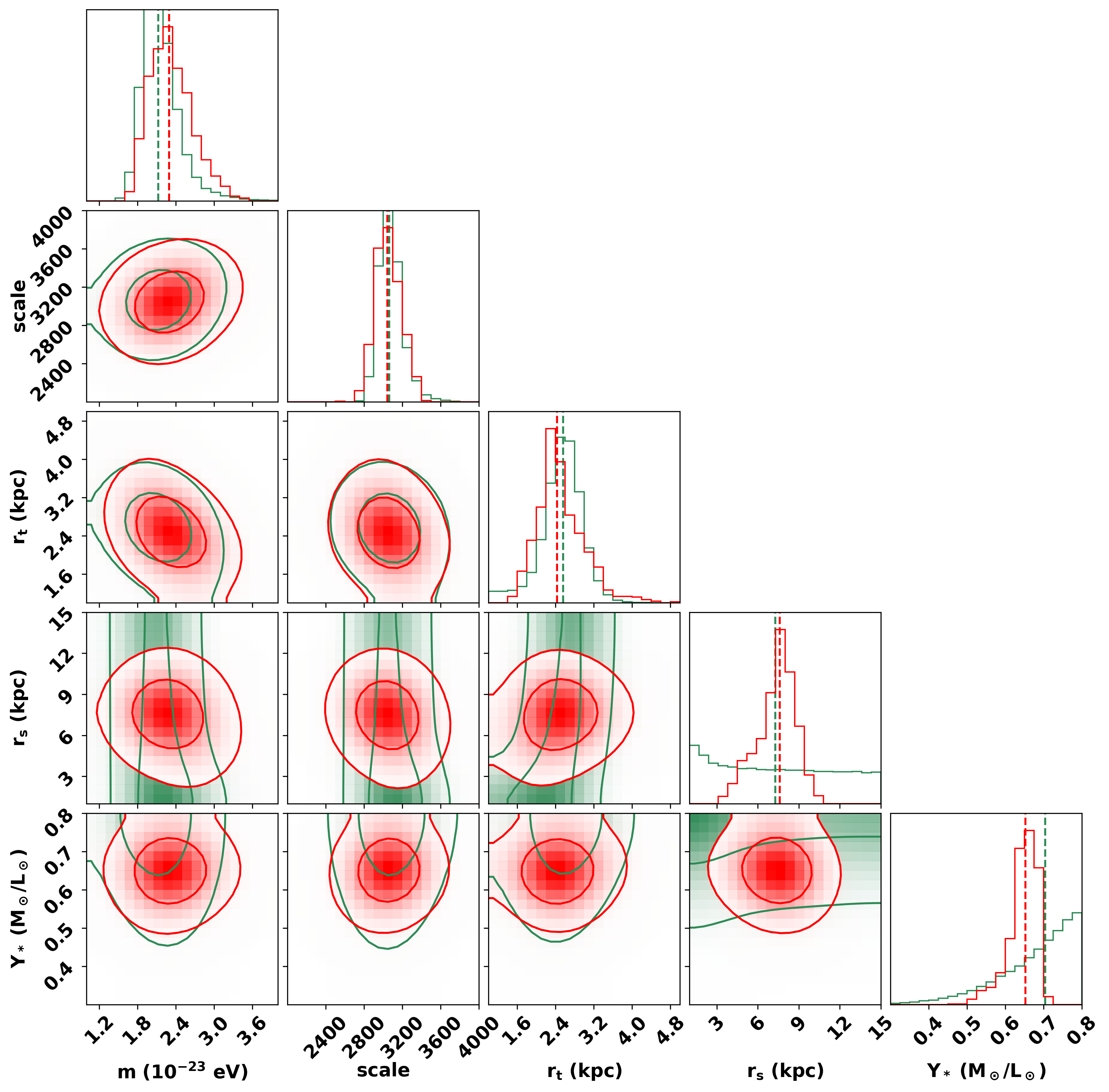}
    \caption{\justifying $2$D and $1$D distributions of parameters obtained using our approach as described in section~\ref{sec:multiple_realizations} (in red), plotted over joint and marginal distributions obtained using MCMC for UGC 5721 (in green). The vertical dashed lines (red and green) in the $1$D histograms are the $50\%$ quantile values for both approaches. The $2$D contours show $\sim 39.3\%$ and $86.4\%$ confidence regions.}
    \label{fig:ANN_MCMC_contour_plot}
\end{figure}

Therefore, we proceed to use the `chain' of $1000$ inferred parameters, to obtain the median and $1\sigma$ confidence interval for each parameter for every galaxy in our sample as shown by the purple points with error bars in figure~\ref{fig:MSE_hetero_MCMC}. 
One can also use the chain of parameters to plot contours of the 2D joint and 1D marginalized distributions in a corner plot. 
Contours obtained for the case of UGC 5721, as an example, are shown in red in figure~\ref{fig:ANN_MCMC_contour_plot}. 

To compare with Bayesian inference, we sample the posterior using an ensemble sampler incorporated in the emcee\footnote{\href{https://emcee.readthedocs.io/en/v2.2.1/}{https://emcee.readthedocs.io/en/v2.2.1/}} python package, for all galaxies in the sample. 
We use the same model described in section~\ref{sec:model_and_data}, where the theory rotation curve is parameterized by $5$ parameters in eq.~(\ref{eq:params}).
We also use the same uniform priors as those defined in Table~\ref{tab:param_space}.
Here, the log-likelihood is written as
\begin{equation}
    \mathrm{ln}\mathcal{L}_{\text{MCMC}} = -\frac{1}{2}\sum_i\left(\frac{v_{obs}(r_i) - v_{th}(r_i; {\bf P})}{\sigma_i}\right)^2\ .
\end{equation}
We then run the emcee sampler with 48 walkers (at different random initial points in the parameter space) for $20,000$ steps, out of which the initial $10,000$ are removed to account for the burn-in period. 
The median and $1\sigma$ confidence intervals are shown in figure~\ref{fig:MSE_hetero_MCMC} by the green points with error bars for all parameters for every galaxy in our sample (they are also listed in Table~\ref{tab:preds_hetero}).  
These chains can also be used to plot $2$D and $1$D posterior distributions, which are shown for the case of UGC 5721 in figure~\ref{fig:ANN_MCMC_contour_plot} by the green contours.

Note that, for the neural network trained using the heteroscedastic loss function, one does not get multiple parameter estimates.
Here, the uncertainty is learned implicitly during training, and hence one cannot plot the joint distributions.
We therefore plot the point-estimate and uncertainties obtained using this method in figure~\ref{fig:MSE_hetero_MCMC}, denoted by the orange points with error bars. 

We note the following observations:
\begin{itemize}
    \item The median values obtained from the multiple realizations approach as well as the point-estimates obtained from neural networks trained using heteroscedastic loss function lie within $1\sigma$ intervals obtained from the MCMC approach for most parameters and galaxies.
    
    \item As seen from the case of UGC 5721 in figure~\ref{fig:ANN_MCMC_contour_plot}, it is also clear that contours obtained using the ANN approach seem to capture correct correlations between some parameters, for instance, the positive correlation between $m$ and $s$, the negative correlation between $r_t$ and $m$ as well as $r_t$ and $s$.
    
    \item Uncertainties obtained from all three methods for $m$, $s$ and $r_t$ are also similar.
    However, the multiple realizations approach severely under-estimates the uncertainties for $r_s$ and $\Upsilon_*$, which, as we discussed in section~\ref{sec:hetero_uncertainty}, do not affect the rotation curves significantly. 
    This is captured correctly by the neural networks trained using heteroscedastic loss, which gives large uncertainties for $r_s$ and $\Upsilon_*$ for all galaxies in figure~\ref{fig:MSE_hetero_MCMC}, agreeing well with the MCMC approach. 
\end{itemize}

\subsection{Uncertainties obtained using heteroscedastic loss function}\label{sec:issues_with_hetero}

At this stage it is important to note that, one must be cautious in the interpretation of uncertainties in the field of machine learning and in parameter estimation problems from cosmology and astrophysics~\cite{Dvorkin_2022}. 
For instance, the term `uncertainty' is used to describe the $\sigma_{ik}$ value obtained from eq.~(\ref{eq:hetero_loss}) predicted by an ANN using the heteroscedastic loss function, as well as the $1\sigma$ value obtained from an MCMC chain. These two uncertainties may not be equivalent. 
This can be understood by noting that in the MCMC approach, the uncertainty propagates from the data to the parameter posteriors via the likelihood function, given a distribution for the errors associated with observations (in this case, Gaussian). 
On the other hand, the uncertainties predicted using the heteroscedastic loss function characterize, to an extent, the difference between predicted and target parameter values, based on the input simulated rotation curves. 

It is therefore interesting to note that the uncertainties obtained using both approaches are similar, as seen in figure~\ref{fig:MSE_hetero_MCMC}; in particular for the parameters that do not affect rotation curves much and thus have larger uncertainties, i.e., NFW scale radius $r_s$ and stellar mass-to-light ratio $\Upsilon_*$. 
It is worth noting that heteroscedastic loss functions have been used previously to obtain uncertainties associated with predictions~\cite{Fluri_2019, Andres_Carcasona_2023, Pal_CNN-CMB_2023, Pal_2023}.
In particular, in~\cite{Pal_2023} the authors utilized this loss function for parameter estimation of cosmological parameters from $H(z)$ data.
For this case, the values of the uncertainties obtained using the heteroscedastic loss function were also found to be similar to those obtained using the MCMC approach. 
This suggests that the heteroscedastic loss function needs to be explored further parameter estimation problems utilizing neural networks in their pipelines. 

In this work, while we have focused on exploring an alternative, complementary approach to parameter estimation using neural networks, we leave the exploration of the connection between uncertainties obtained in ANN and MCMC approaches for future work.

\section{\label{sec:discussion}Discussion and conclusion}

We live in the era of data-driven cosmology and astrophysics, where ever larger and more accurate data sets are becoming available \cite{Amin_whitepaper_Snowmass2021}. 
Extracting information from such data sets is essential to constrain models of new physics.
While the usual method for doing this involves a likelihood-based approach using Markov Chain Monte Carlo (MCMC), with advancement in computer hardware, it is worthwhile to explore machine learning and in particular, deep learning to develop novel and complementary approaches to tackle the same problem.

With that in mind, in this work, we have explored the use of artificial neural networks in learning model parameters from observed galactic rotation curves.
Neural networks are powerful tools that can be used to approximate a wide-range of complex functions \cite{Hornik_1990}, which is particularly useful when the relationship between some input vector ${\bf x}$ and output vector ${\bf y}$ is highly non-linear and complex.

In this work, unlike a likelihood-based approach, we train neural networks with rotation curves as input and the model parameters as output using a large sample of simulated rotation curves whose parameter values are already known.
\footnote{There are various other approaches that can be used here to carry out parameter estimation along with uncertainty quantification using neural networks including Bayesian neural networks (BNNs) and normalizing flows (see \cite{Hortua_2020, Hortua_2023, Stachurski_2023, Kolmus_2024} for recent work).}. 
By looking at samples in the training data, the neural network 
updates its internal adjustable parameters (called weights and biases) to approximate a function $f: \mathbb{R}^{N_{obs}}\rightarrow\mathbb{R}^P$ where $N_{obs}$ is the number of observed velocities in a rotation curve and $P = 5$ is the number of parameters, which are:  
ULDM particle mass $m$, the scale parameter $s$ associated with the mass of the core, transition radius $r_t$, i.e. the radius at which the core profile transitions to a NFW profile, the scale radius of the NFW profile $r_s$, and $\Upsilon_*$, the stellar mass-to-light ratio which tunes the baryonic contribution to the rotation curves (See section~\ref{sec:model_and_data} for a detailed discussion on the model and dataset used).  

\begin{figure}[h!]
\centering
\begin{tabular}{cc}
\includegraphics[width=0.45\textwidth]{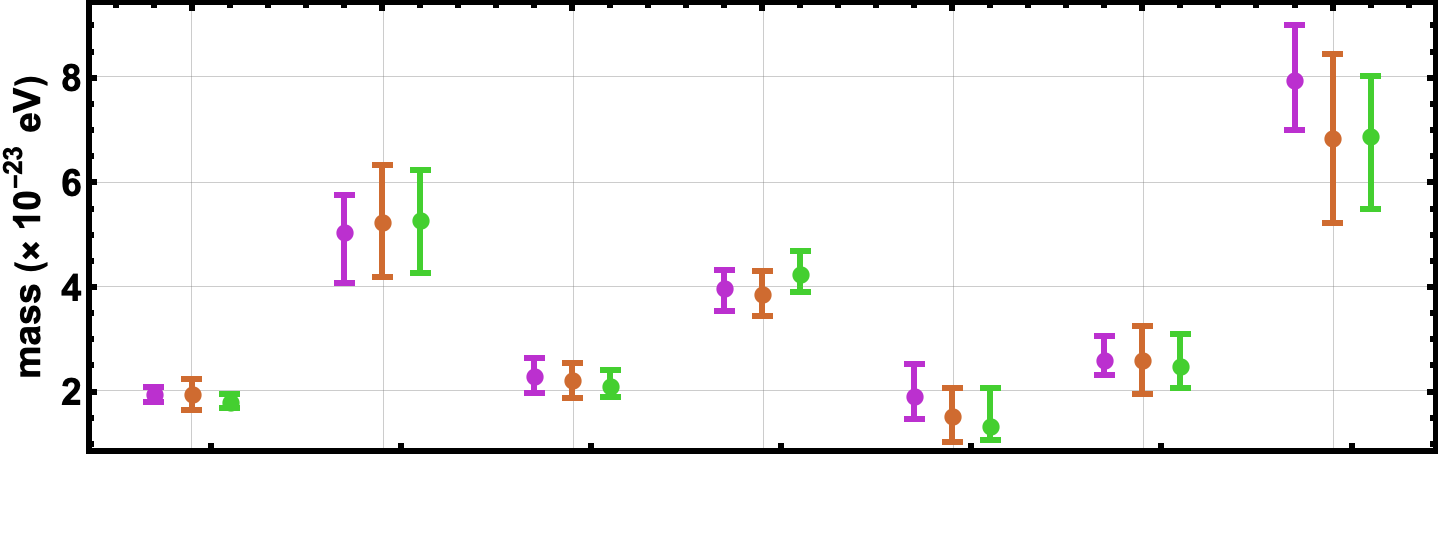} &
~ \\
(a) ULDM mass, $m$ & ~ \\[6pt]
\includegraphics[width=0.45\textwidth]{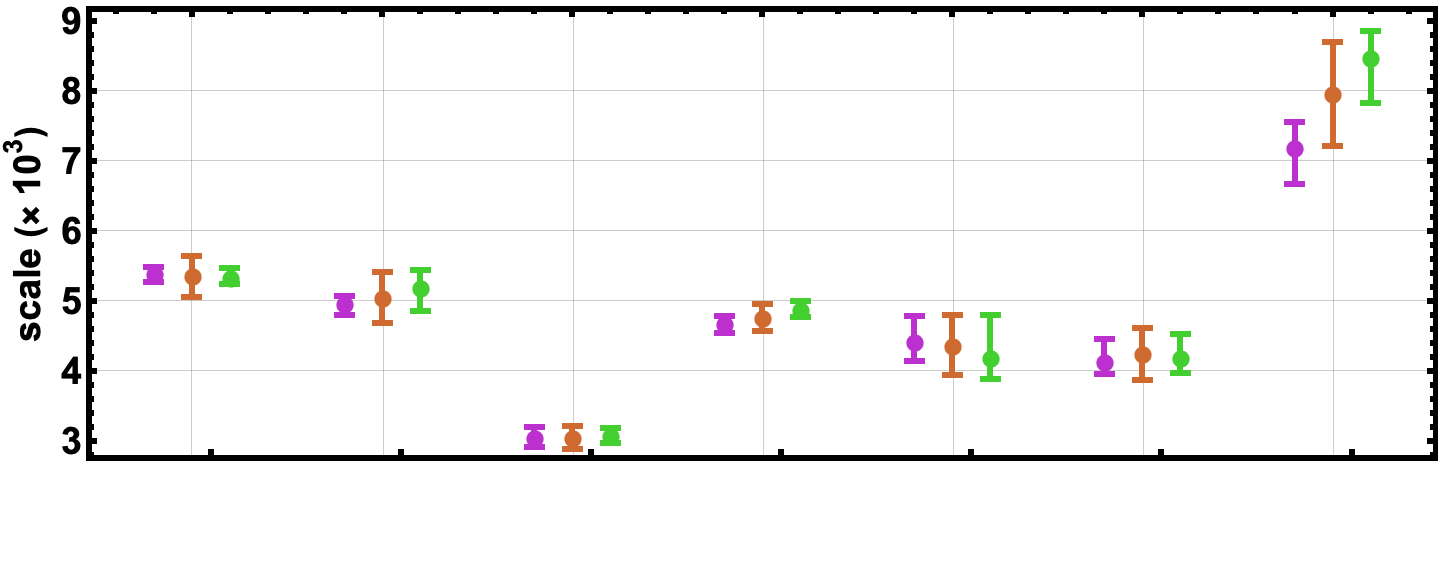} &
\includegraphics[width=0.45\textwidth]{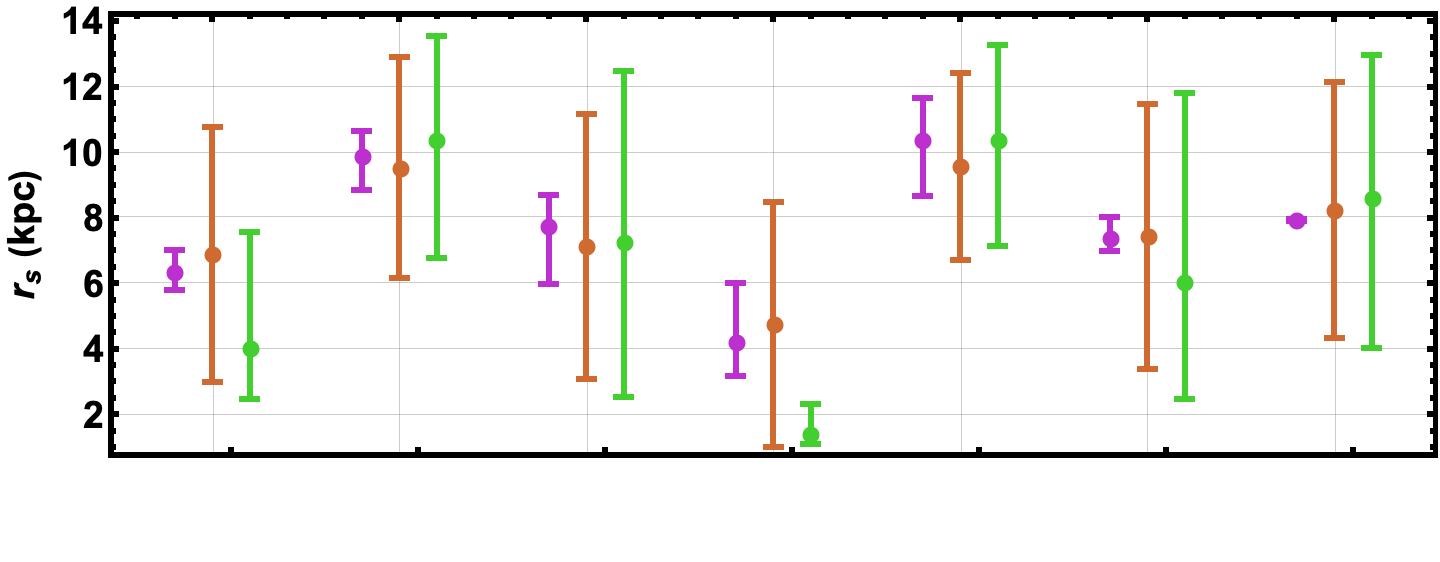}  \\
(b) scale, $s$ & (d) scale radius, $r_s$ \\[6pt]
\includegraphics[width=0.45\textwidth]{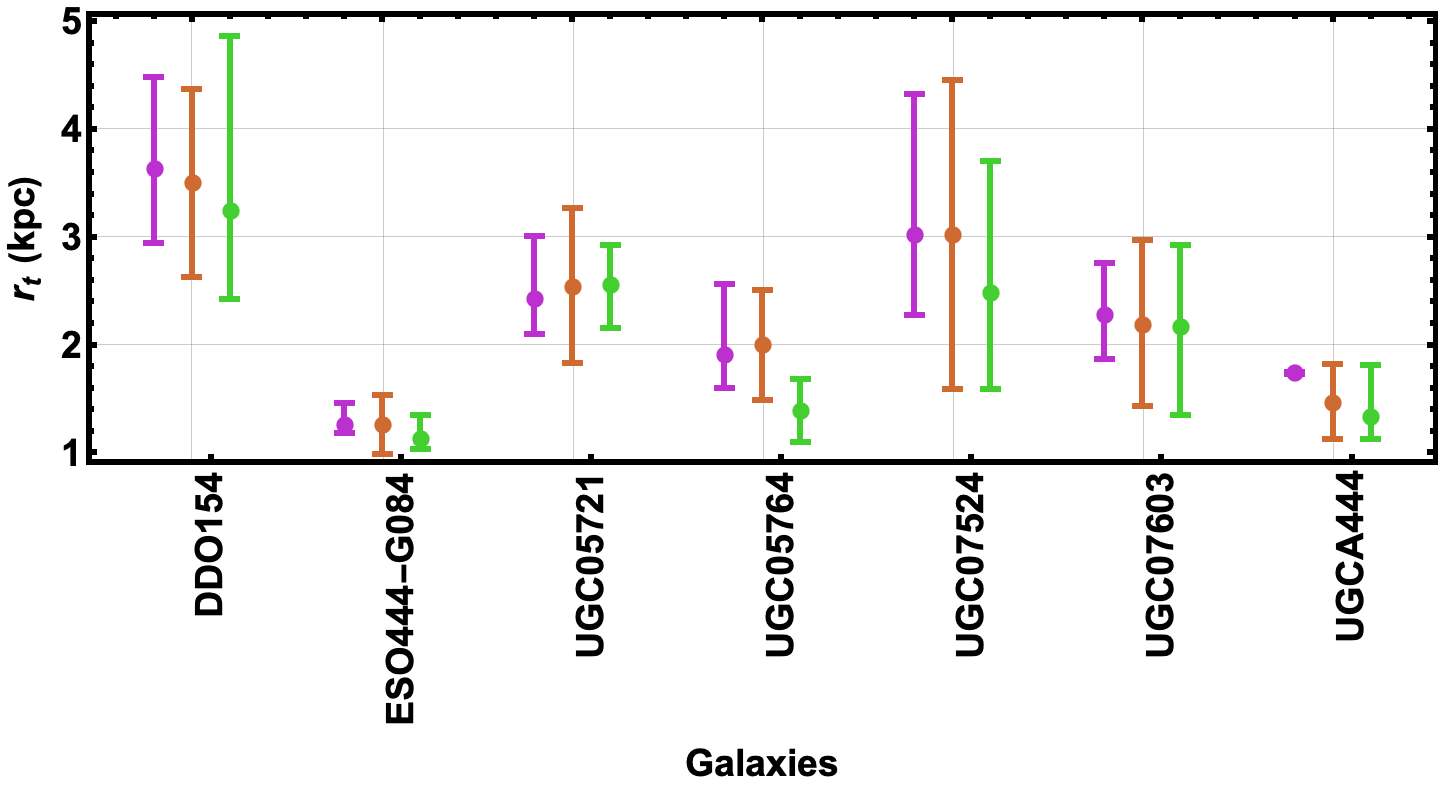} & \includegraphics[width=0.45\textwidth]{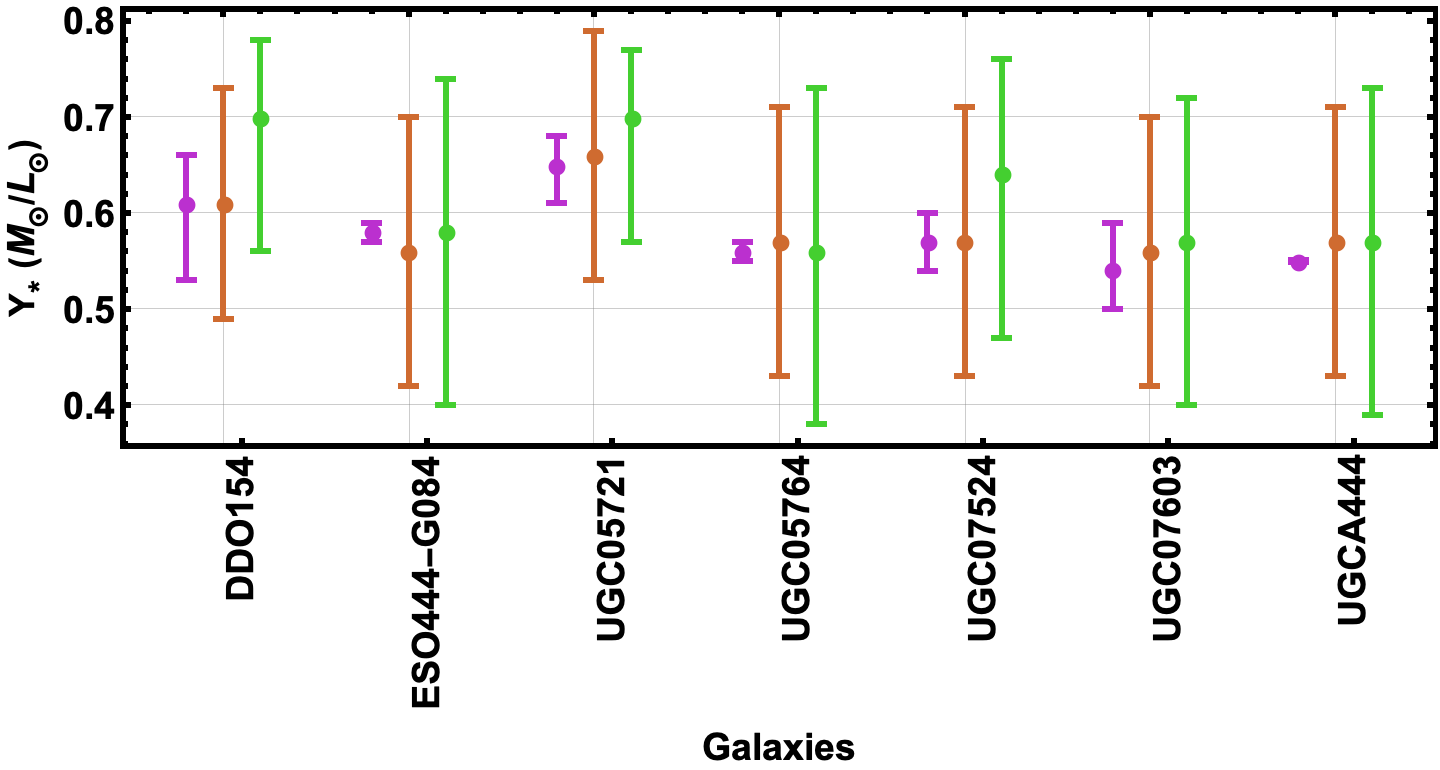} \\
(c) transition radius, $r_t$ & (e) stellar mass-to-light ratio, $\Upsilon_*$\\
\end{tabular}
\caption{\justifying Inferred parameters and their uncertainties using multiple realization method from section~\ref{sec:multiple_realizations} (purple), heteroscedastic loss function from section~\ref{sec:hetero_uncertainty} (orange) and MCMC runs using emcee in section~\ref{sec:mcmc_comparison} (green).}
\label{fig:MSE_hetero_MCMC}
\end{figure}

The observed rotation curve is then given as input to the trained neural networks and we get inferred point-estimates of parameters as output. 
Here is a short summary of our findings:

\begin{enumerate}
    \item Neural networks trained on noisy simulated rotation curves perform much better than the noiseless case when confronted with observed rotation curves. 
    The noise included during training appears to capture the fact that observed rotation curves are not ``smooth'' functions (see figure~\ref{fig:MSE_comparison} and the discussion around it).

    \item To quantify the uncertainty associated with the inferred parameters based on the uncertainty in the observations, we generated multiple realizations of the observations as described in section~\ref{sec:multiple_realizations}, which lead to multiple parameter estimates, resembling a chain of parameters. 
    The median values obtained using this method lead to rotation curves which agree well with observed rotation curves, as shown in figure~\ref{fig:all_comparison} by the purple curves.

    \item However, comparison with parameter estimation using MCMC in section~\ref{sec:mcmc_comparison} suggests that while the multiple realizations method can capture some of the correlations between a few parameters, namely $m$, $s$ and $r_t$, it struggles when confronted with hard-to-constrain parameters like $r_s$ and $\Upsilon_*$ which don't affect the rotation curves in our sample significantly (see Figs.~\ref{fig:ANN_MCMC_contour_plot} and~\ref{fig:MSE_hetero_MCMC}).

    \item Instead of the mean-squared-error (MSE) loss function, using a heteroscedastic loss function in eq.~(\ref{eq:hetero_loss}) to train the network enables us to learn the uncertainty associated with each parameter during training itself. 
    When confronted with observed rotation curves, the ANNs perform just as well as the networks trained using MSE loss function (see figure\ref{fig:all_comparison}).

    \item For the case of neural networks trained using the heteroscedastic loss function, while one cannot obtain $2$D and $1$D distribution of parameters, the ANNs predict higher uncertainties for the $r_s$ and $\Upsilon_*$ parameters compared to the multiple realizations approach. 
    Large uncertainties for $r_s$ and $\Upsilon_*$ are expected since the rotation curves are not very sensitive to a change in these parameters, which is also reflected in the large uncertainties obtained using the MCMC approach as shown in figure~\ref{fig:MSE_hetero_MCMC} in green.
\end{enumerate}

Over the last few years, there has been a lot of interest in utilizing neural networks for parameter estimation problems in astrophysics and cosmology (see section~3.2 of~\cite{CosmoVerse_2025} for a recent summary).
Many of these methods utilize neural networks as a part of a larger likelihood-free inference pipeline~\cite{Fluri_2021, Wang_ECoPANN_2020, Andres_Carcasona_2023, Pal_2023, Kolmus_2024, Hagimoto_2024, Artola_2024}.
In this work, we have explored a simple scenario where neural networks are trained to directly output parameter estimates given rotation curves as input, while also comparing different methods of quantifying uncertainties associated with each parameter to those obtained using the traditional Bayesian approach.
We have found that, given the chosen architecture and hyperparameters, neural networks can prove to be a useful tool in obtaining parameter estimates that can describe observed rotation curves well, i.e. with a small $\chi^2_{red}$.
Our work, along with other recent work \cite{Wang_2020, Wang_ECoPANN_2020, Pal_2023, Hagimoto_2024, Artola_2024} demonstrates that, the use of neural networks for this class of problems can be a useful complementary approach to standard likelihood-based approaches.
However, we would like to note that, (a) uncertainties obtained using the heteroscedastic loss function may not be equivalent to the familiar Bayesian posterior (as we have discussed in section~\ref{sec:issues_with_hetero}), and (b) the total time taken for the generation of training samples and training the neural network is much longer than the time taken to run an MCMC sampler. This implies that the ANN approach that we have incorporated in this work cannot surpass or replace the traditional MCMC approach yet.  
It is also worth noting that, the simple neural network approach we have considered in section~\ref{sec:hetero_inference} agrees well with the recent results obtained in~\cite{Pal_2023}, where the authors found that model parameters as well as their uncertainties obtained using an ANN trained with a heteroscedastic loss function compared well with the MCMC approach.

Before closing we would like to note some caveats and future prospects: 
\begin{inparaenum}[(a)]
 \item
 The true values of each parameter are assumed to be independently and normally distributed in the heteroscedastic loss function, which may not be the case. 
 \item
 We have not carried out a full hyperparameter exploration to obtain an optimal neural network. It is likely that there exists a better choice of hyperparameters that perform better by giving better estimates of the parameter values and the associated uncertainties for this dataset as well as decrease the total amount of time taken during the training phase. 
 \item
 Information about the radius values for which velocities (observed or simulated) are obtained has not been used in this analysis, since the input vector is just a list of velocities. Therefore, the neural networks we have trained cannot differentiate between two rotation curves with the same velocities but different observed radii. 
 An interesting direction would be to train a neural network capable of handling input rotation curves from multiple galaxies with a varied range of radii and velocities to infer a single parameter value for the fundamental parameter of our model, i.e., ULDM particle mass $m$, while inferring different galaxy specific parameters.
 We leave this exploration to future work.
 \item In this work, we deal with the vanilla fuzzy dark matter model, i.e., a scalar field with negligible self-interactions.
 In our recent work \cite{Dave_2023} we have demonstrated that repulsive self-interactions can satisfy observed rotation curves while also satisfying an empirical core-halo mass relation, unlike the case of no self-interactions \cite{Bar_2022}.
 It could then be interesting to include the effect of $\lambda\varphi^4$ kind of self-interactions for the ultralight scalar field, and infer $\lambda$ along with the parameters we considered in this work.  
\end{inparaenum}

\begin{acknowledgments}

We thank Jayanti Prasad, Susanta Tewari, Amit Nanavati and Kishan Malviya for helpful discussions. BD also thanks Isha Mahuvakar and Jai Chaudhari for their help with tensorflow.
This work is supported by the Department of Science and Technology, Government of India under Anusandhan National Research Foundation (formerly Science and Engineering Research Board) — State University Research Excellence (SUR/2022/005391). 
A part of the computational work in this research was carried out on the Param Shavak High Performance Computing System provided by the Gujarat Council on Science and Technology to Ahmedabad University. 
\end{acknowledgments}

\appendix

\section{Monitoring the loss function}\label{app:loss_epochs}

In this section, we plot the loss functions for each of the three neural networks, i.e., with noiseless and noisy training data for MSE loss function and noisy training data for heteroscedastic loss function - trained for each of the $7$ galaxies in our sample. 
\begin{figure*}
\begin{tabular}{ccc}
\includegraphics[width=0.28\linewidth]{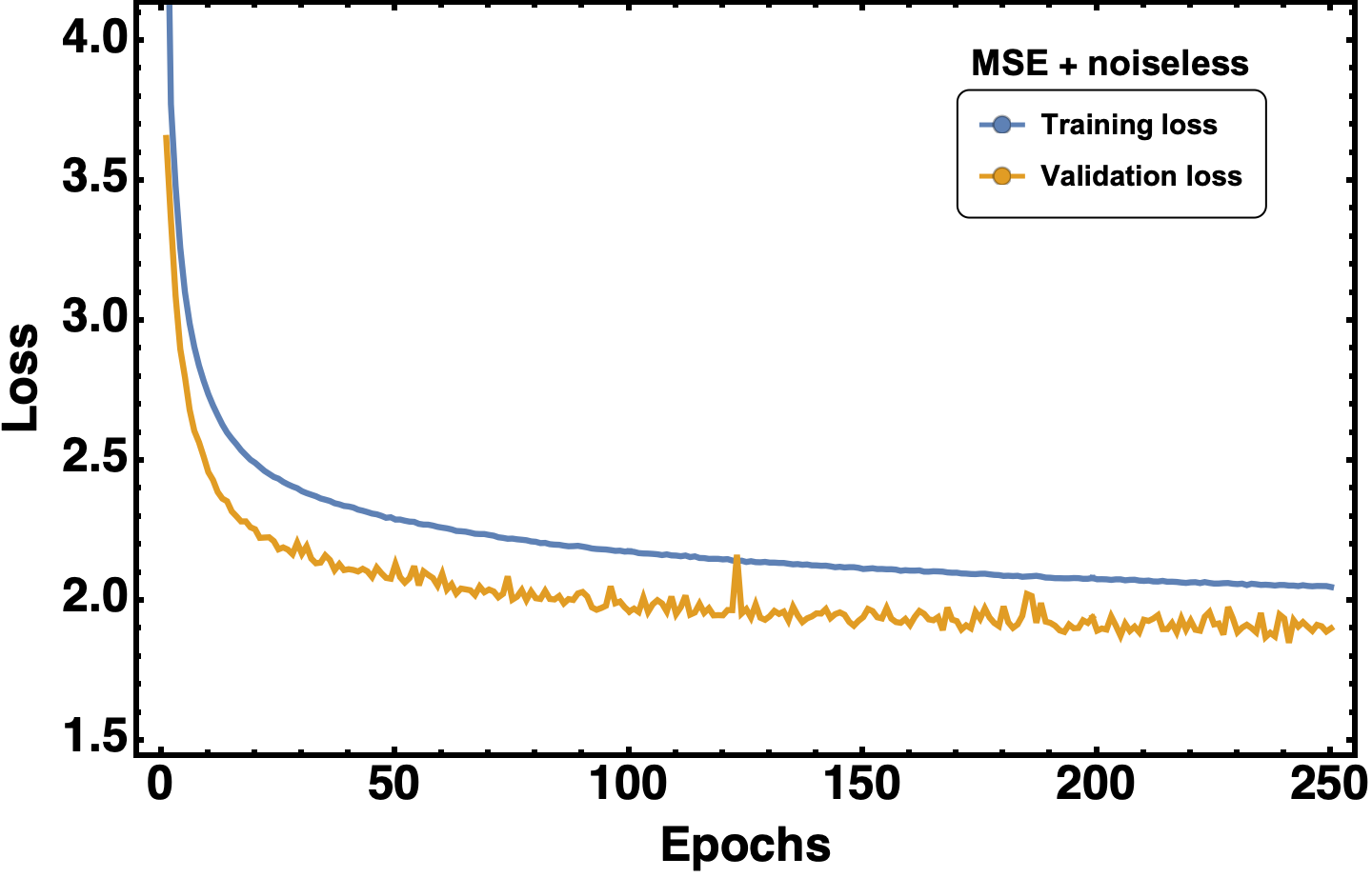} &   \includegraphics[width=0.28\linewidth]{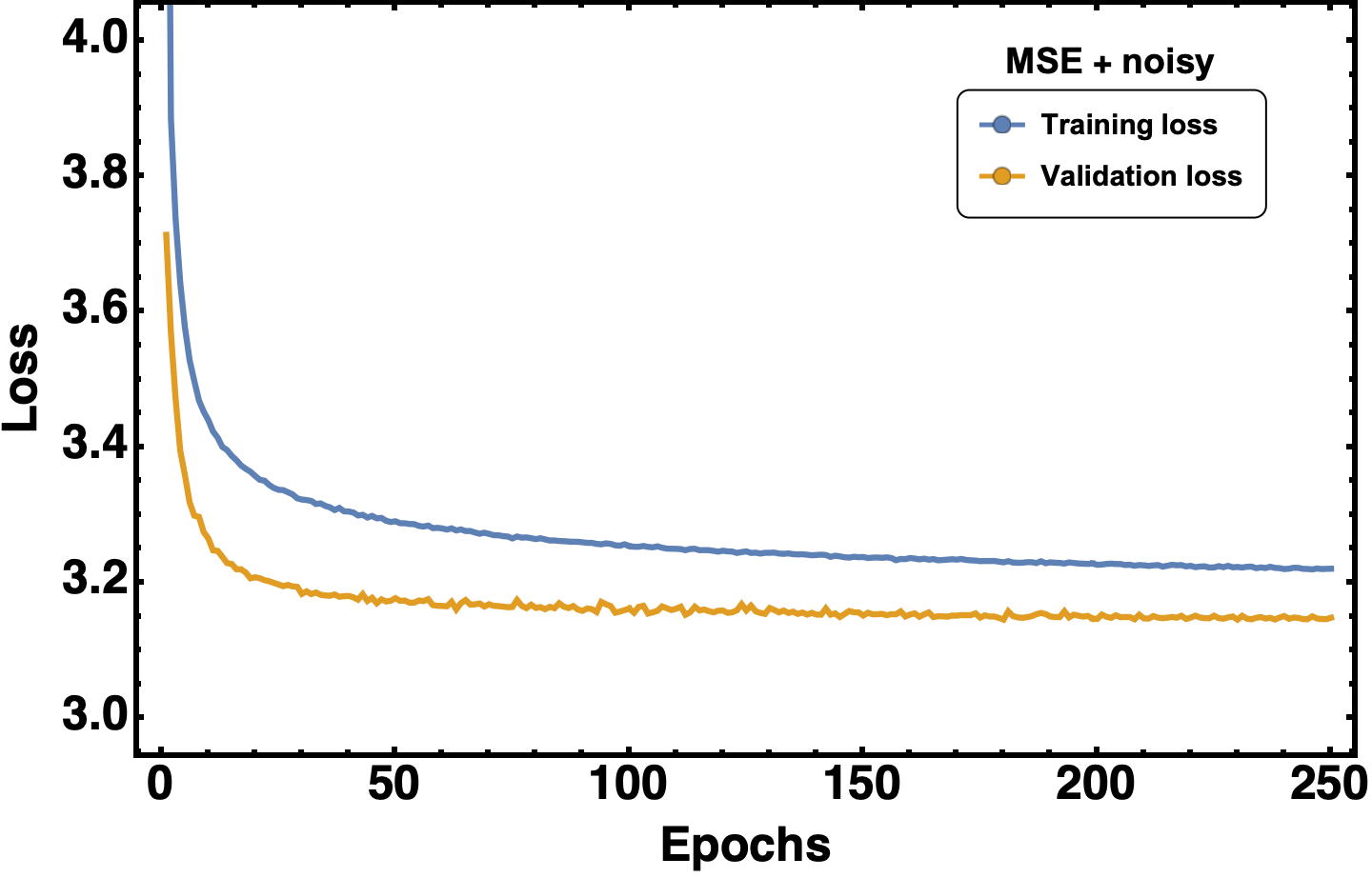} &
\includegraphics[width=0.28\linewidth]{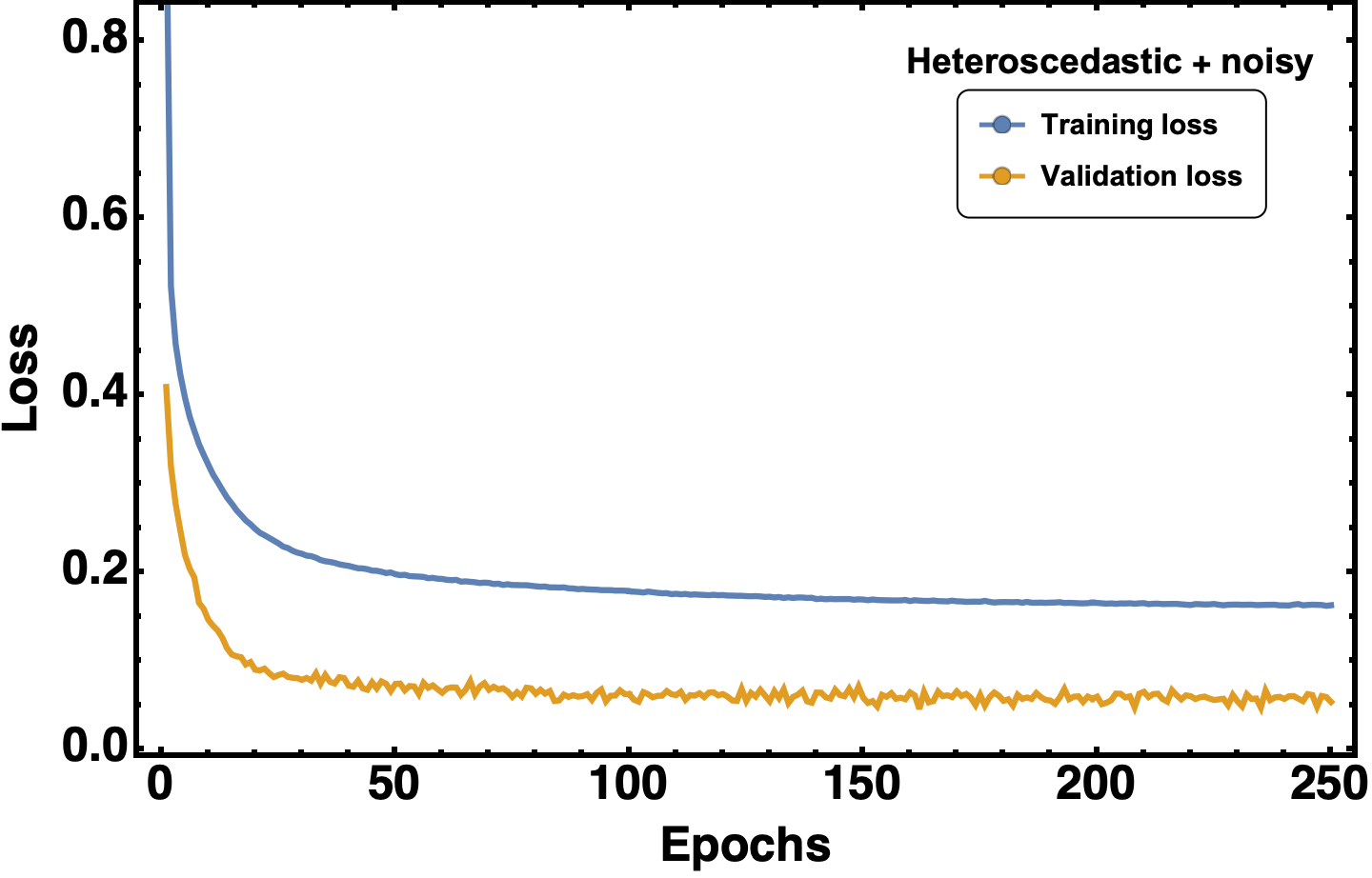} \\
& (a) DDO 154 & \\[5pt]
\includegraphics[width=0.28\linewidth]{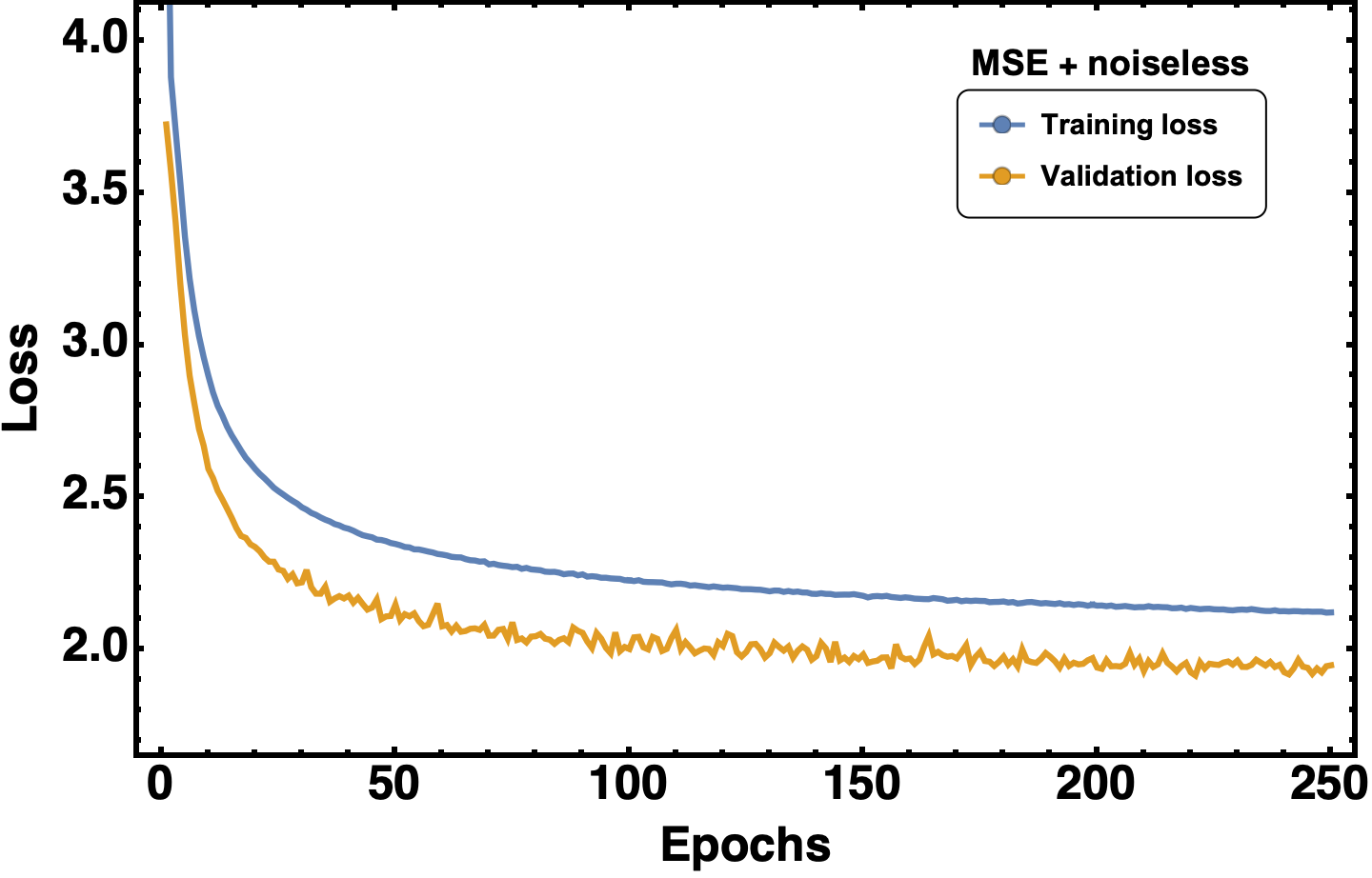} &   \includegraphics[width=0.28\linewidth]{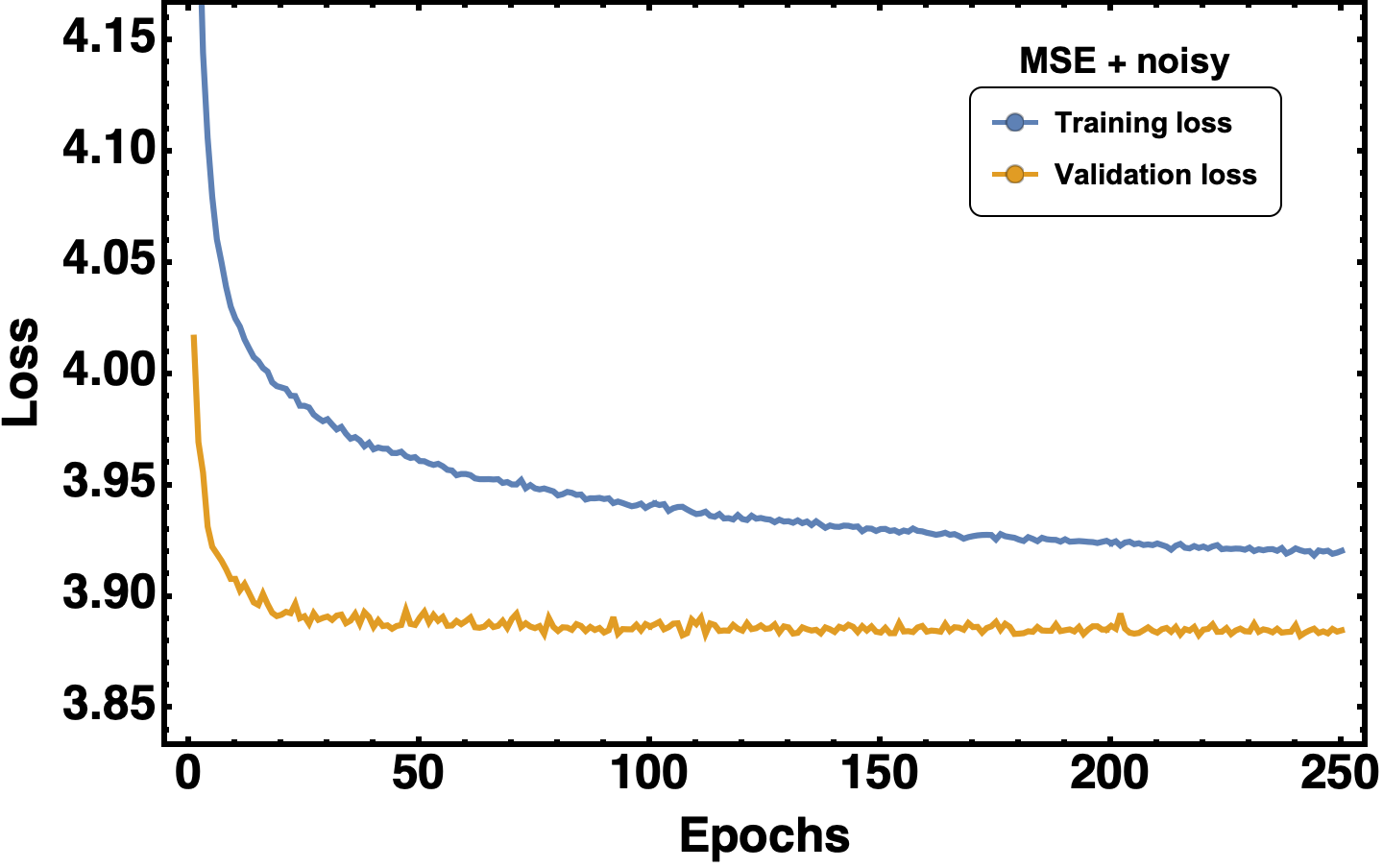} &
\includegraphics[width=0.28\linewidth]{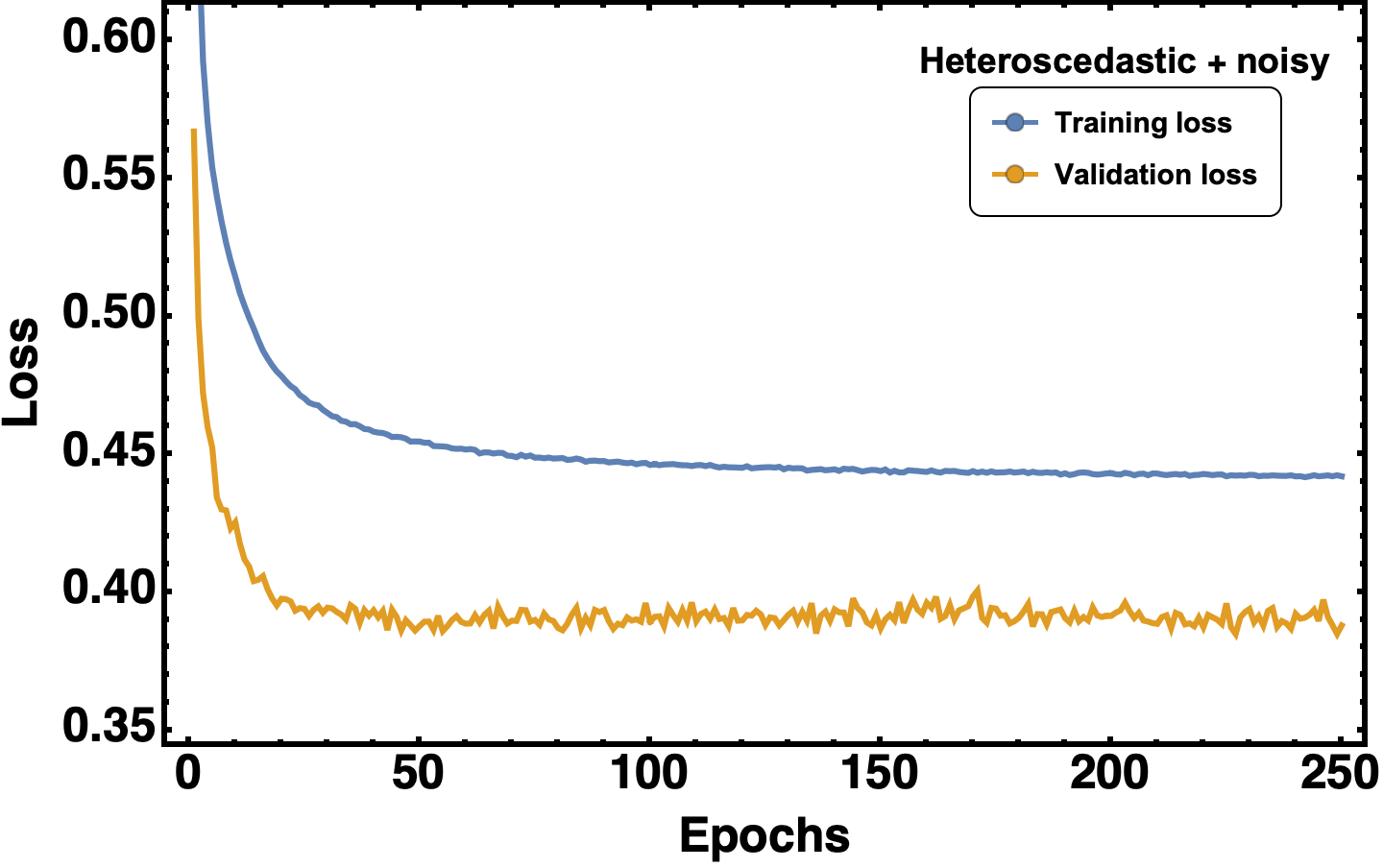} \\
& (b) ESO 444-G084 & \\[5pt]
\includegraphics[width=0.28\linewidth]{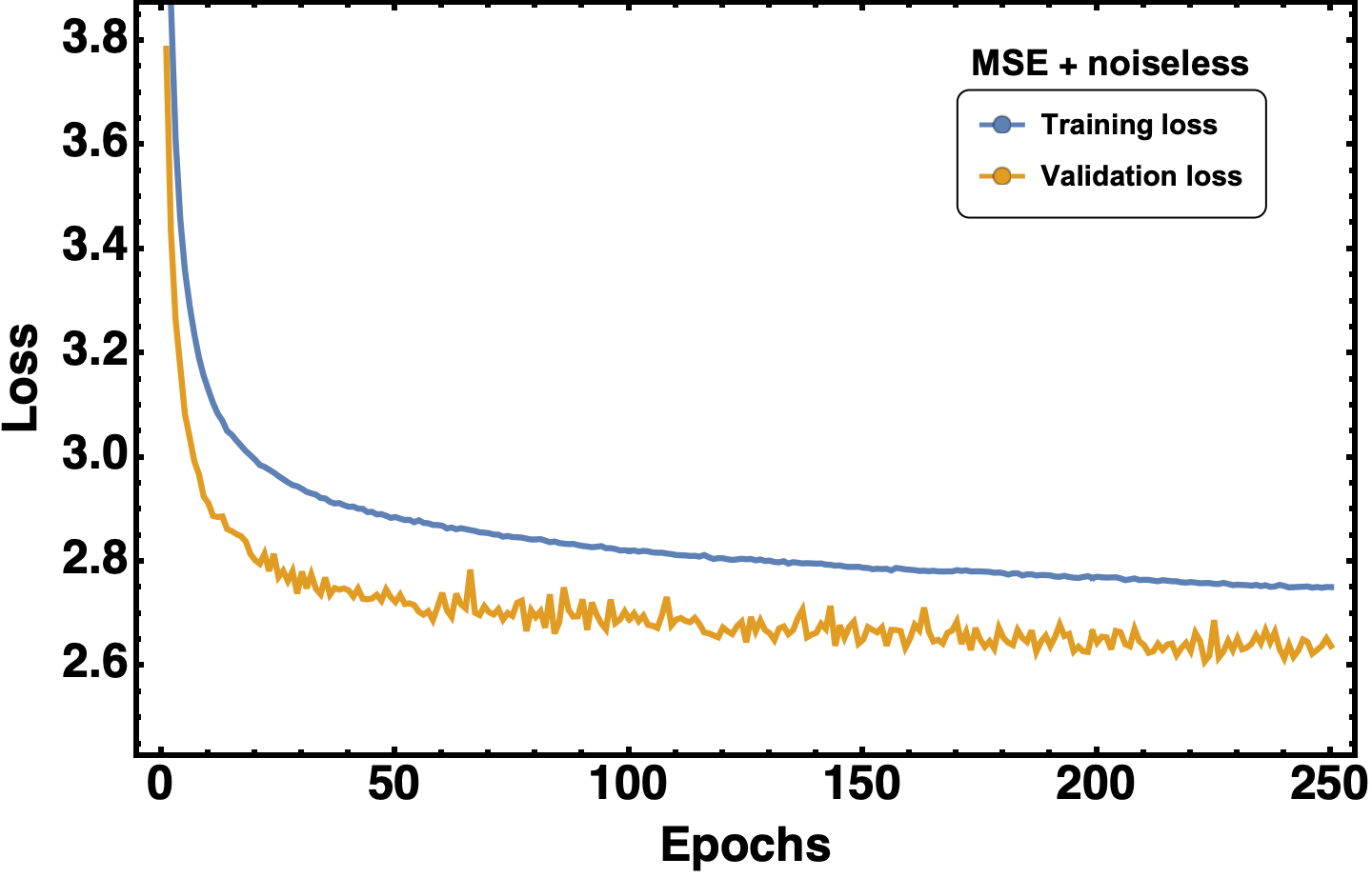} &   \includegraphics[width=0.28\linewidth]{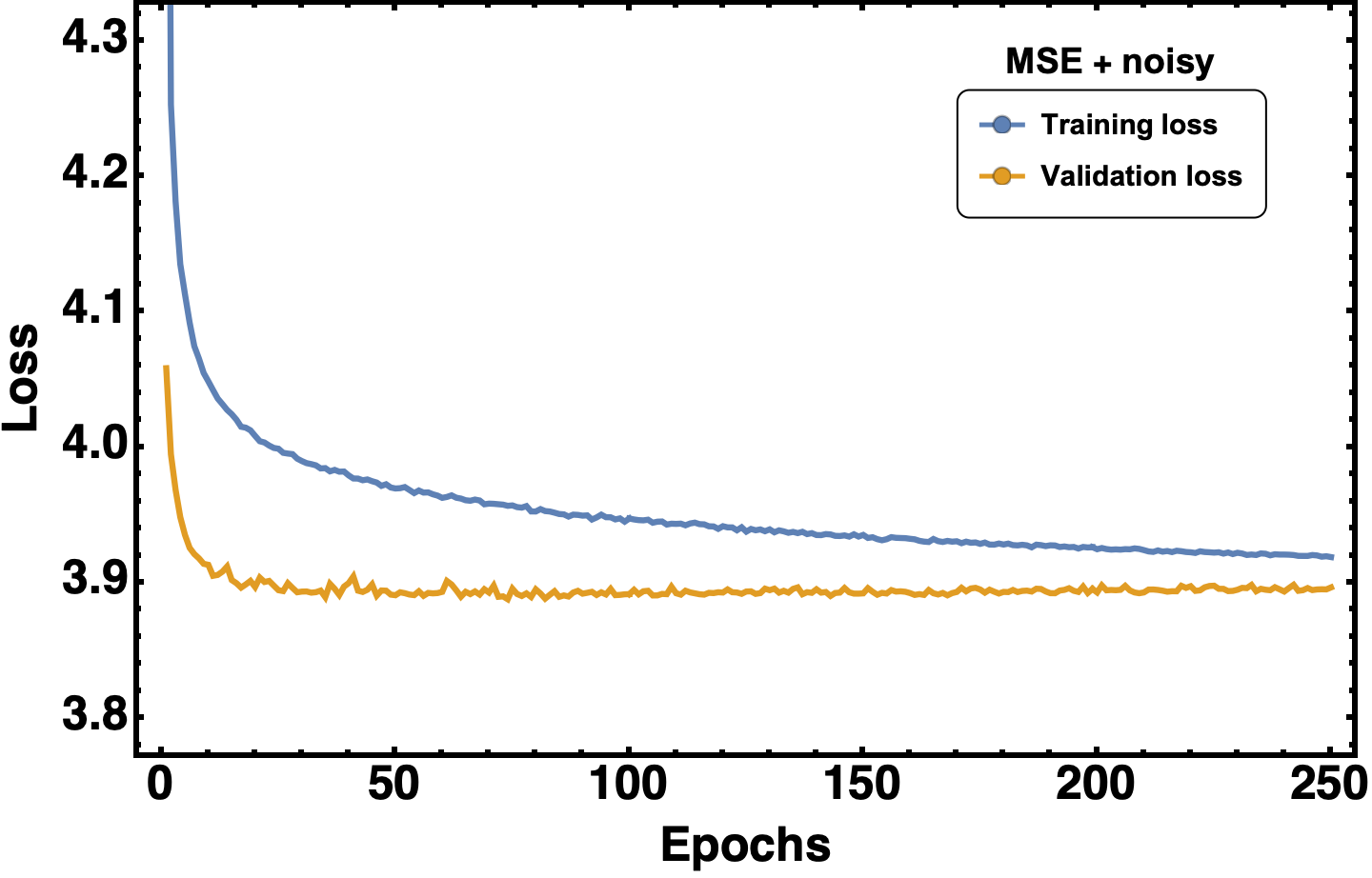} &
\includegraphics[width=0.28\linewidth]{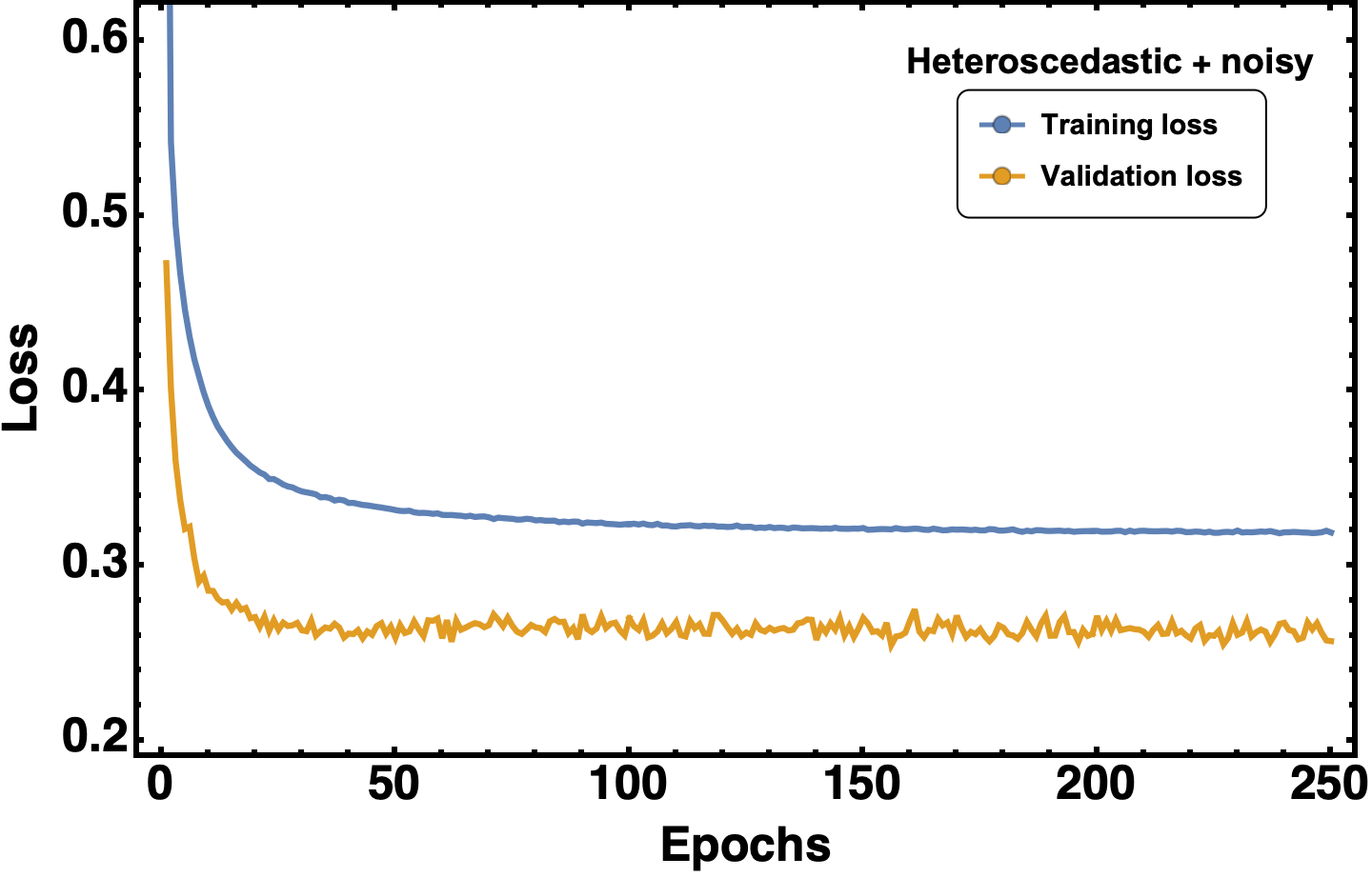} \\
& (c) UGC 5721 & \\[5pt]
\includegraphics[width=0.28\linewidth]{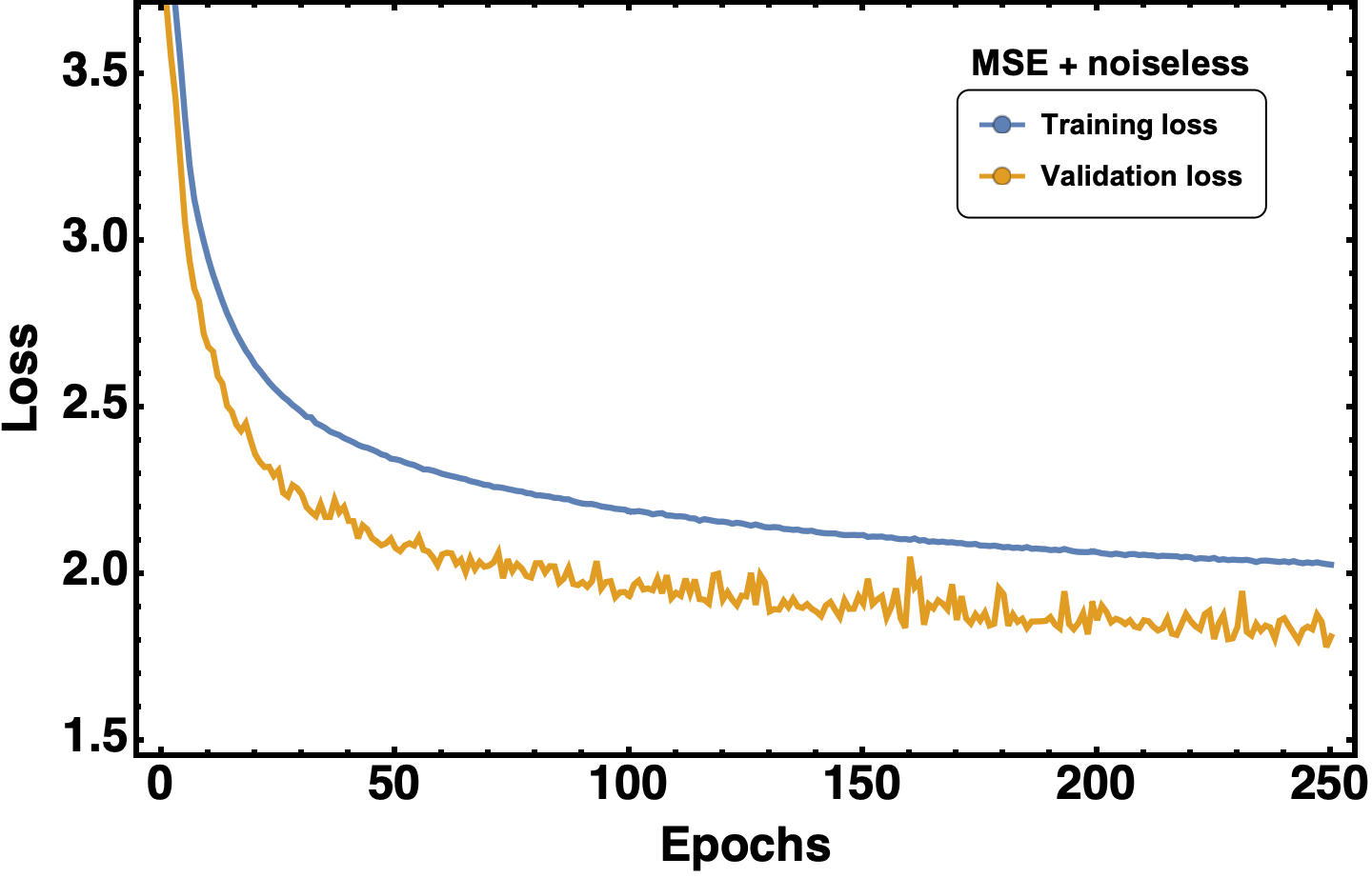} & 
\includegraphics[width=0.28\linewidth]{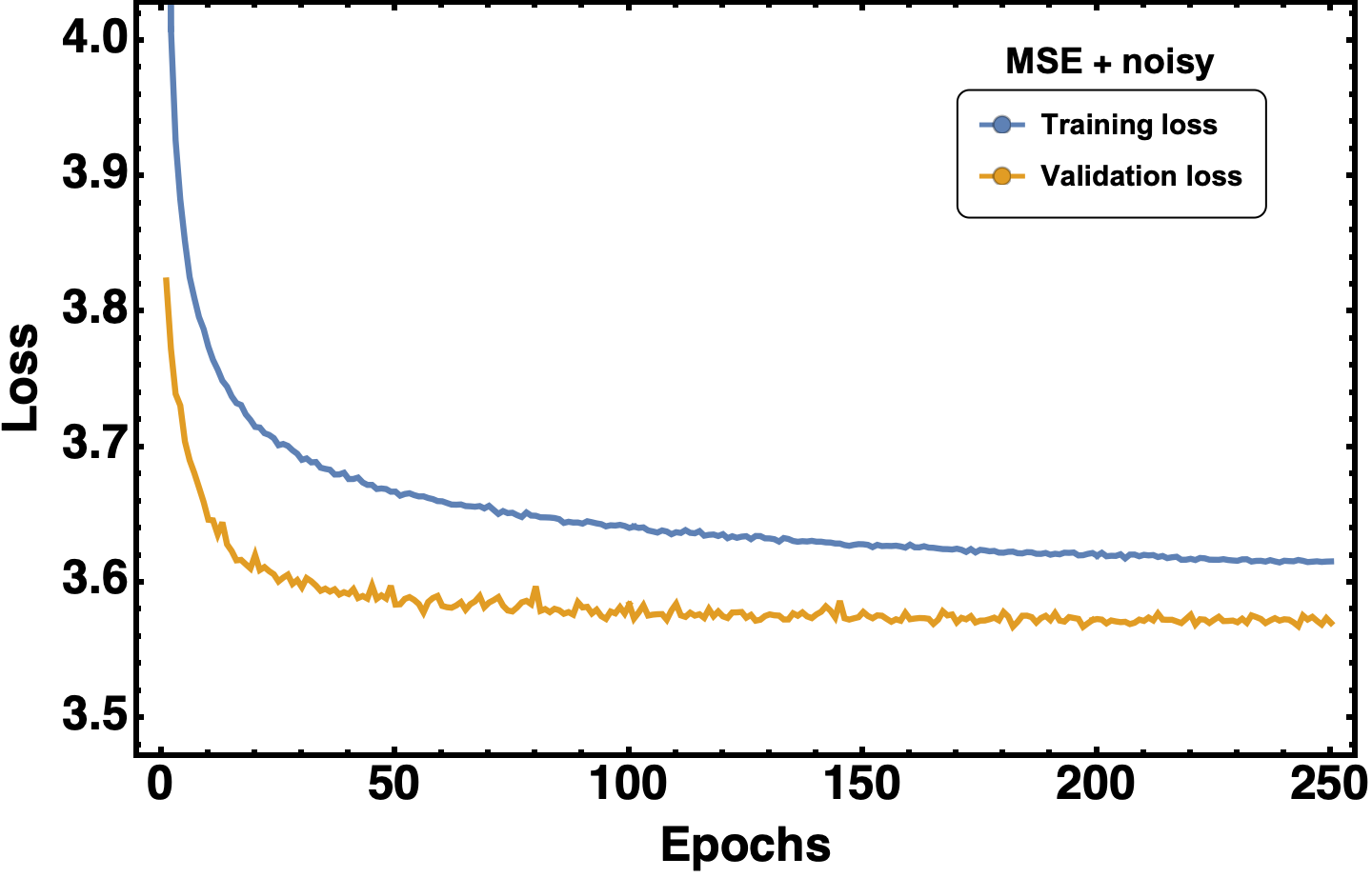} &
\includegraphics[width=0.28\linewidth]{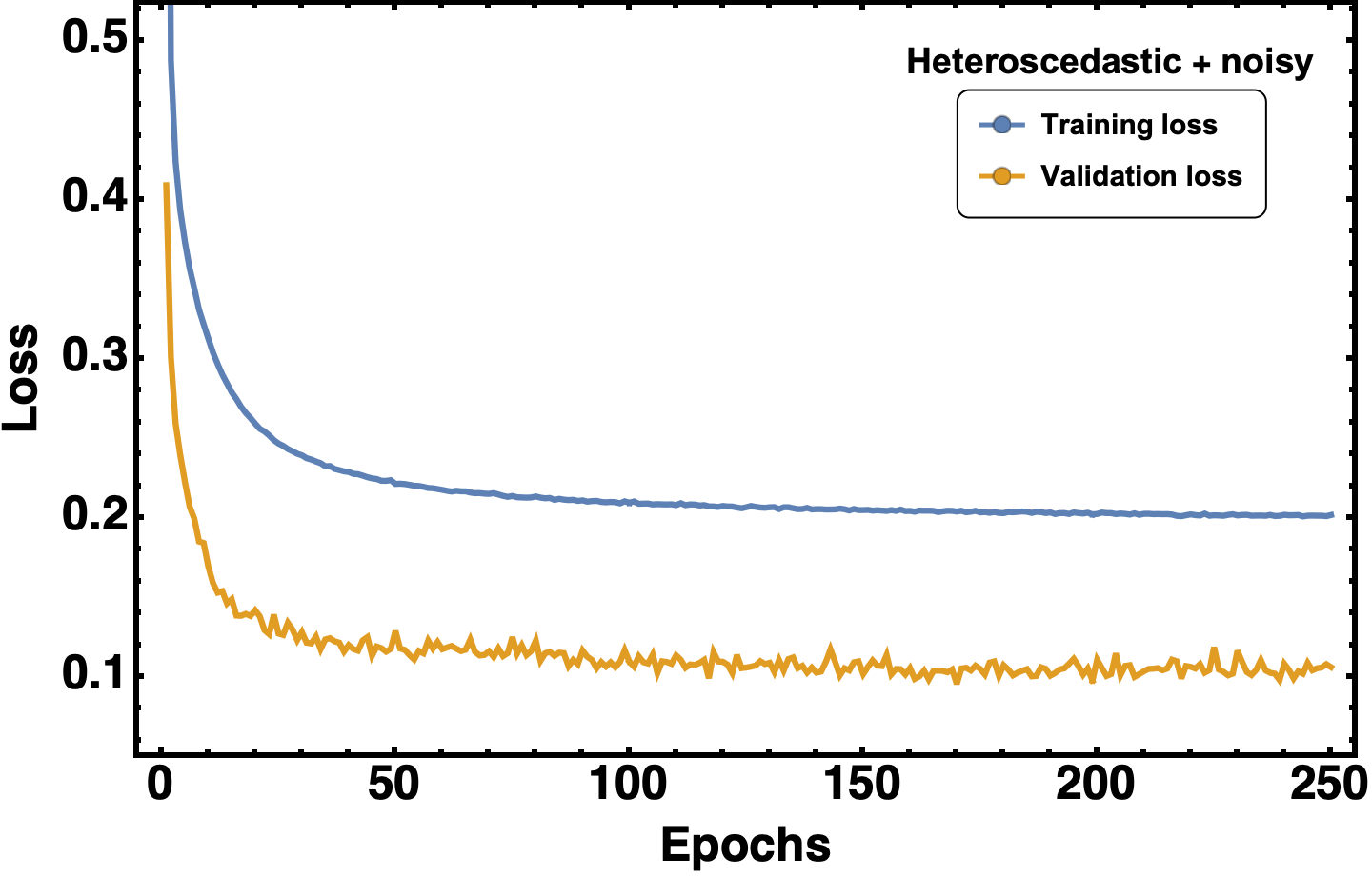} \\
& (a) UGC 5764 & 
\end{tabular}
\caption{\justifying Loss plotted with respect to epochs for the three cases we have considered as mentioned in the text as well as the top right corner of each plot. 
Training loss is denoted by blue while validation loss is denoted by orange for three galaxies from our sample.}
\label{fig:loss_1}
\end{figure*}

\begin{figure*}
\begin{tabular}{ccc}
\includegraphics[width=0.28\linewidth]{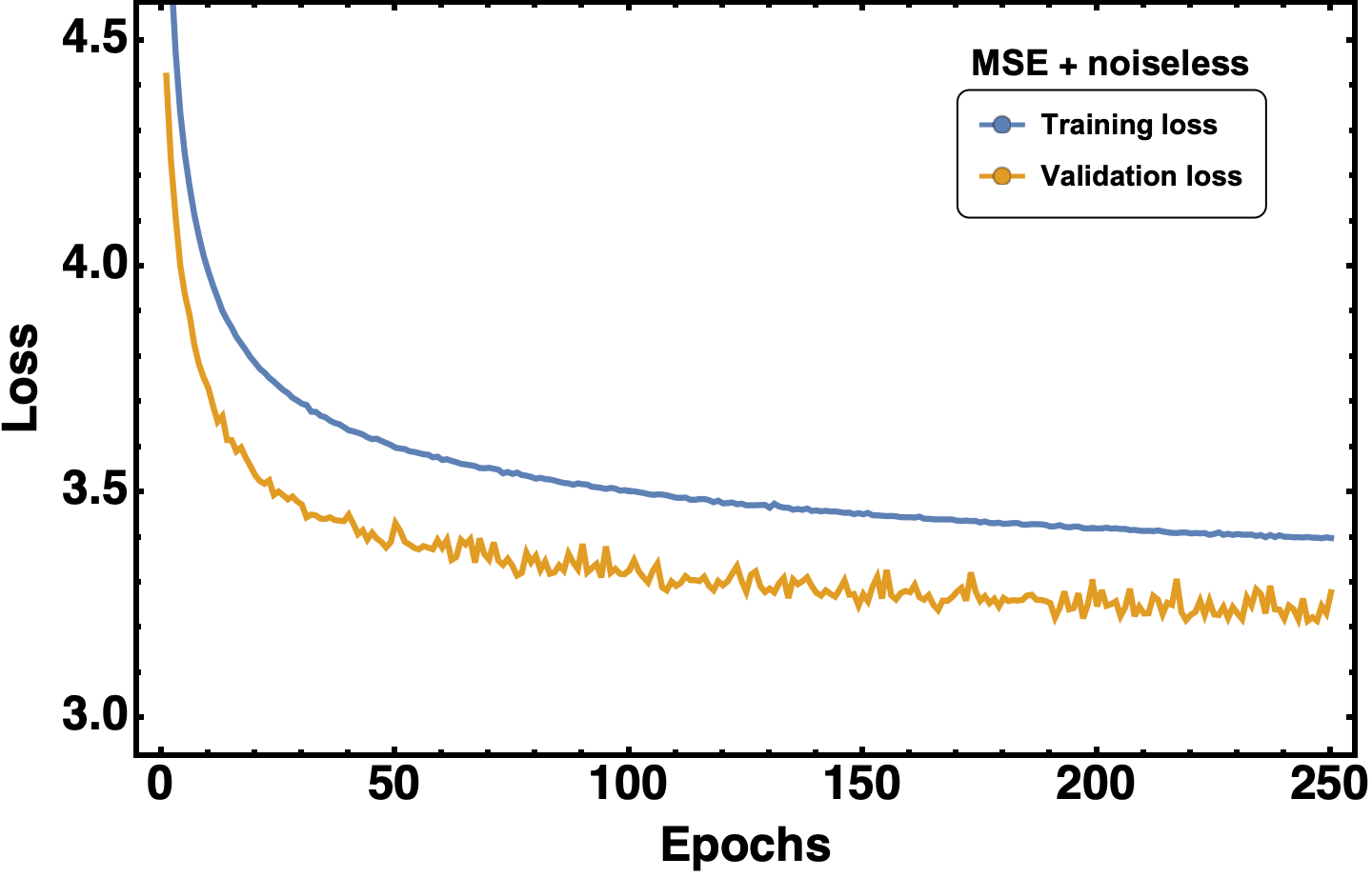} &   \includegraphics[width=0.28\linewidth]{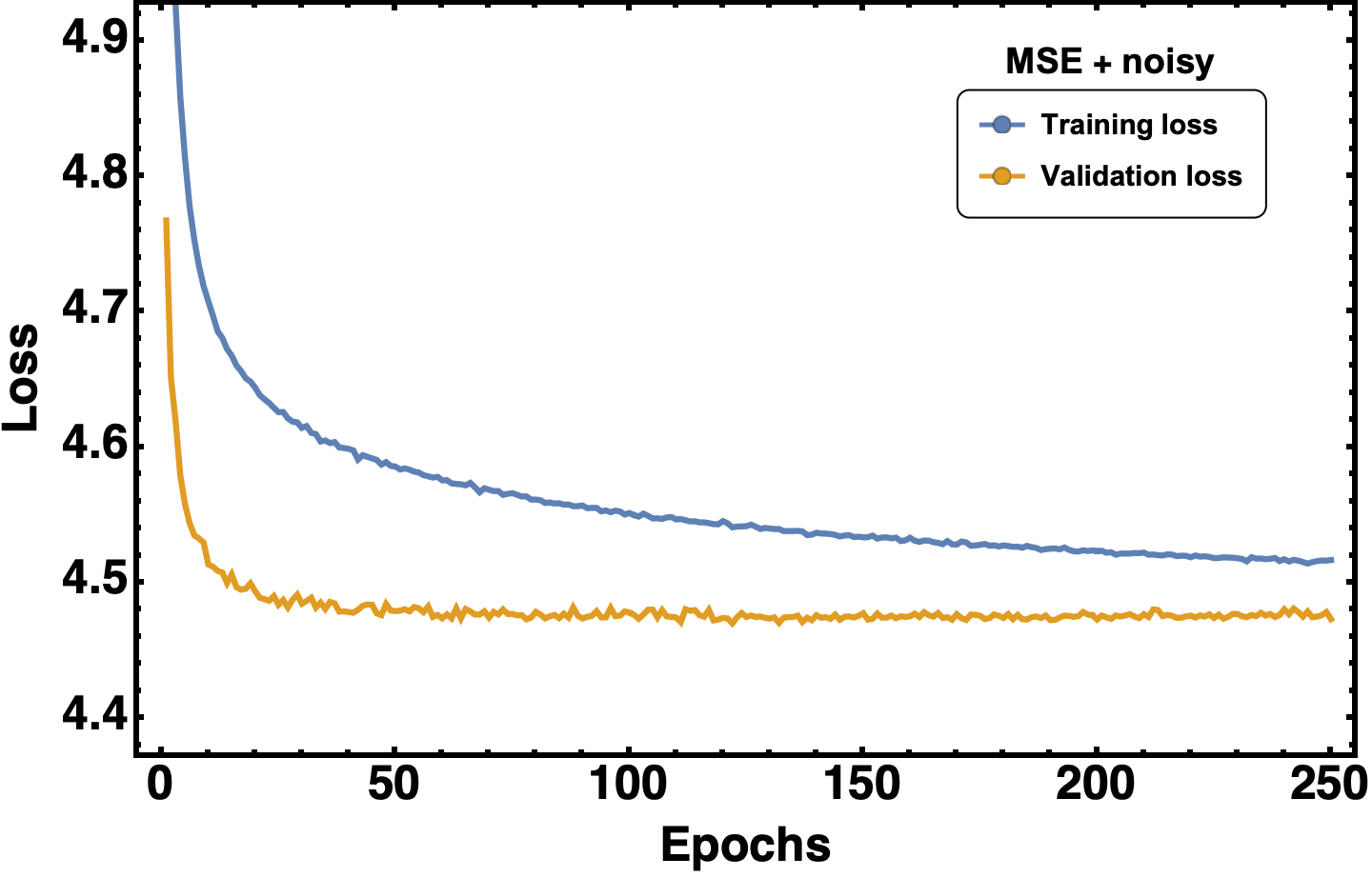} &
\includegraphics[width=0.28\linewidth]{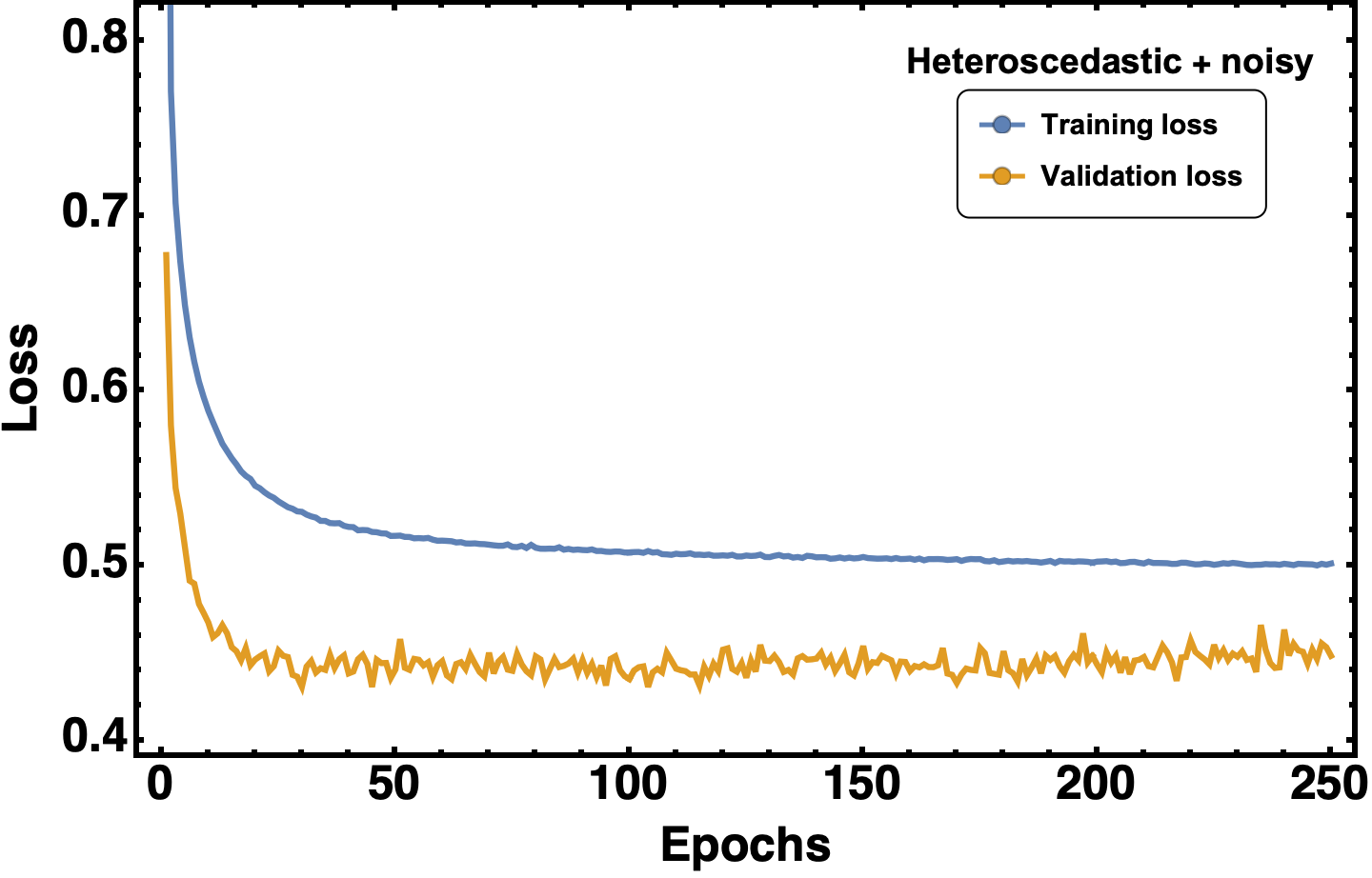} \\
& (b) UGC 7524 & \\[5pt]
\includegraphics[width=0.28\linewidth]{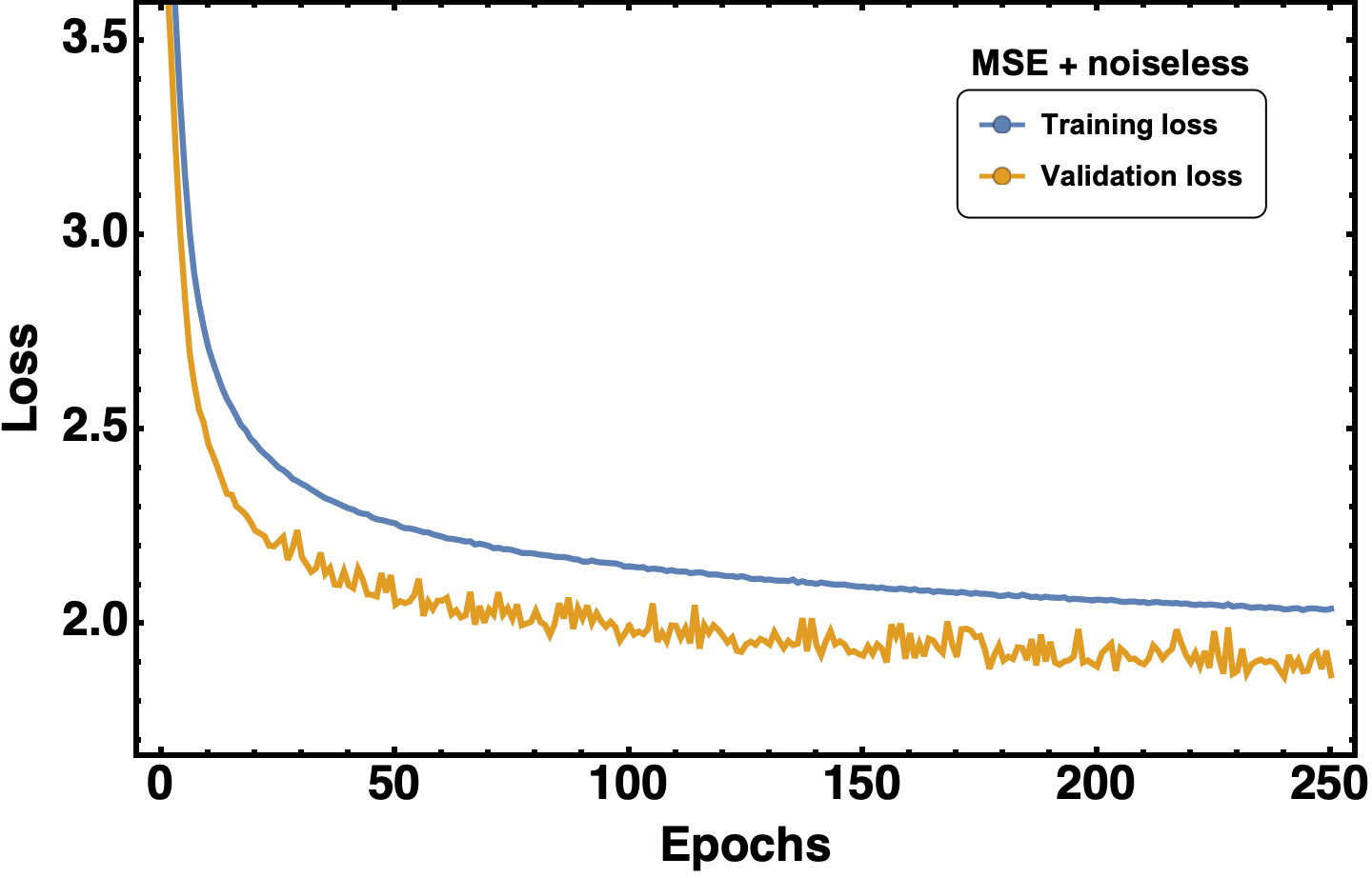} & 
\includegraphics[width=0.28\linewidth]{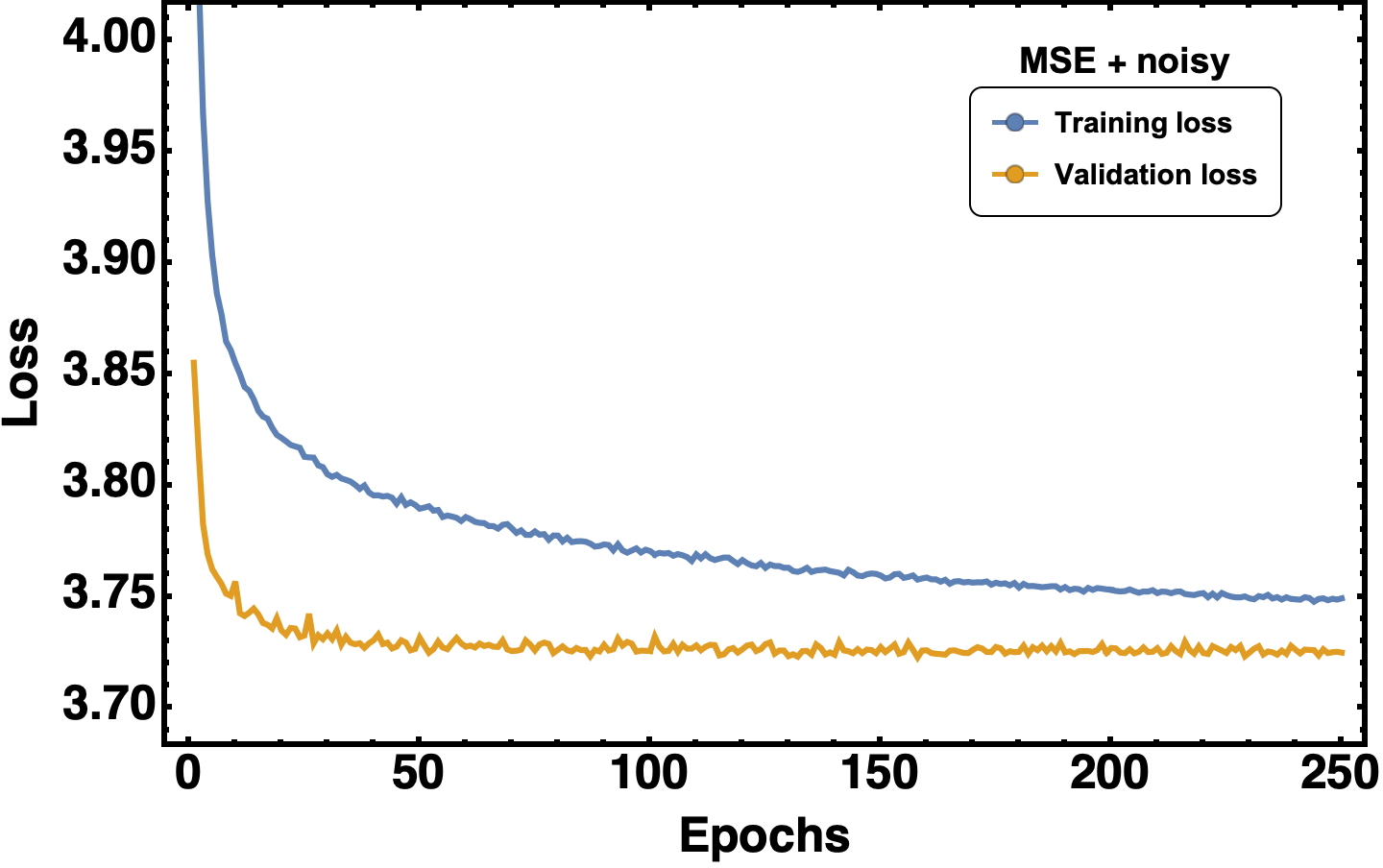} &
\includegraphics[width=0.28\linewidth]{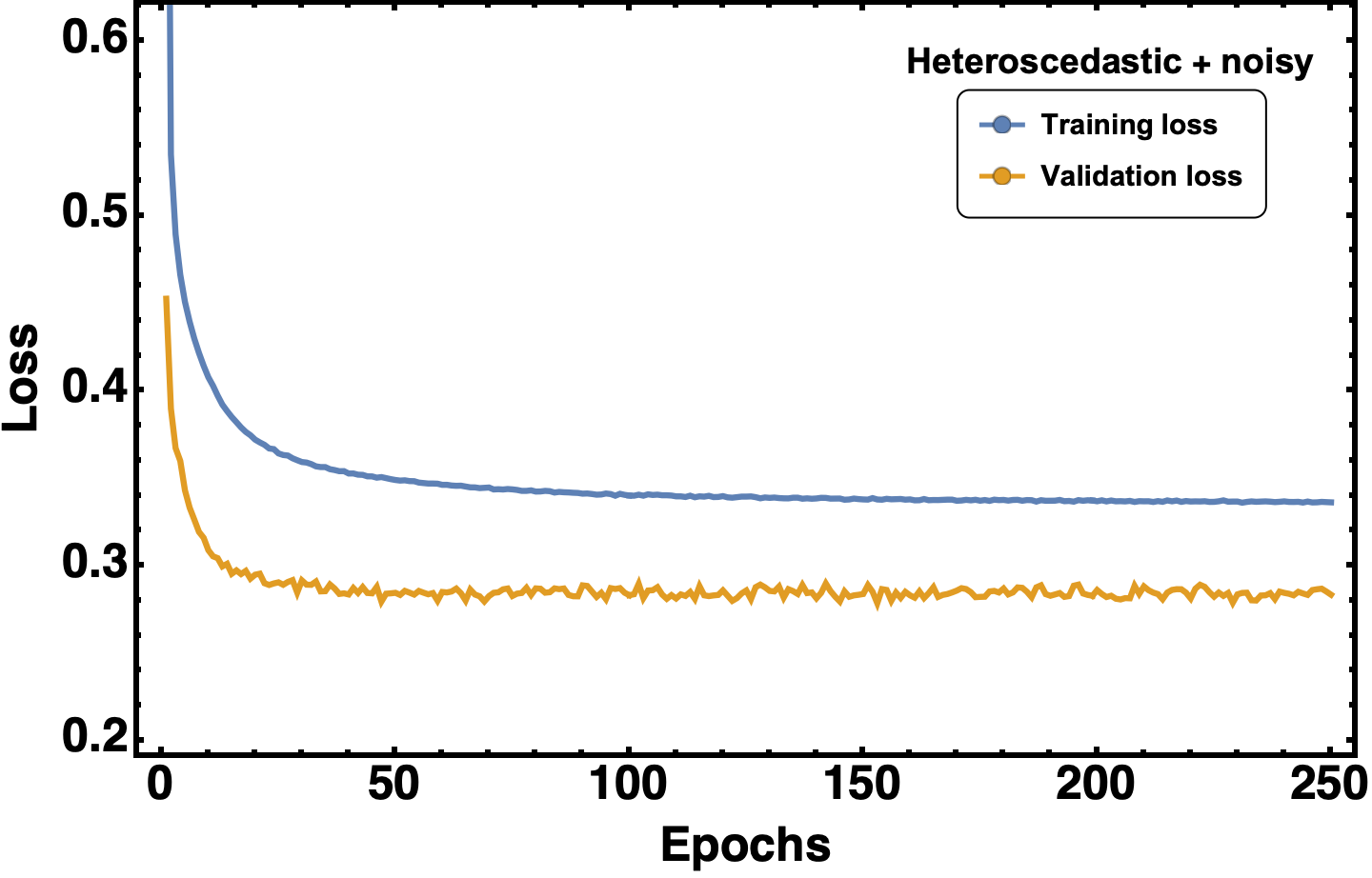} \\
& (c) UGC 7603 & \\[5pt]
\includegraphics[width=0.28\linewidth]{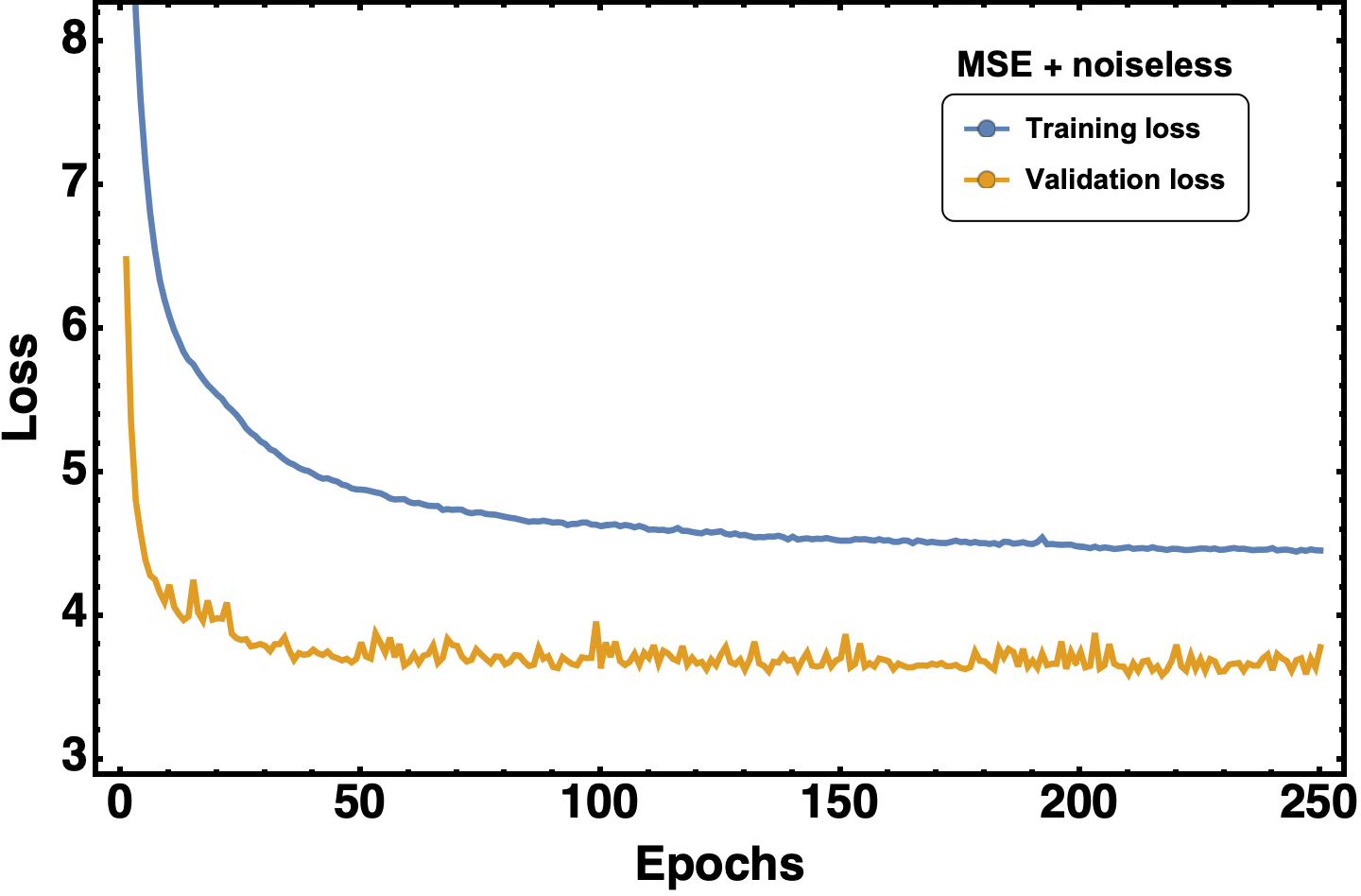} &   \includegraphics[width=0.28\linewidth]{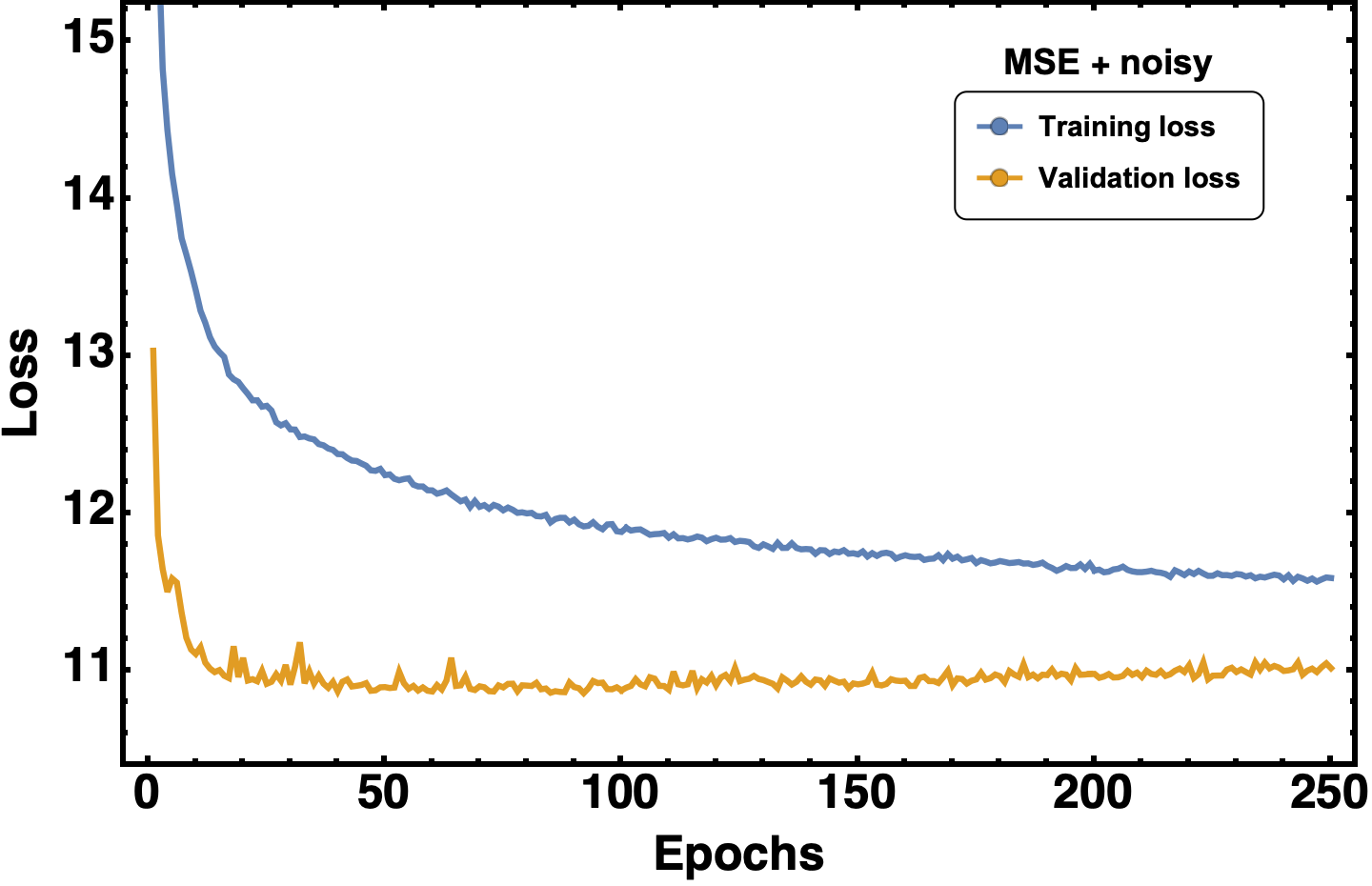} &
\includegraphics[width=0.28\linewidth]{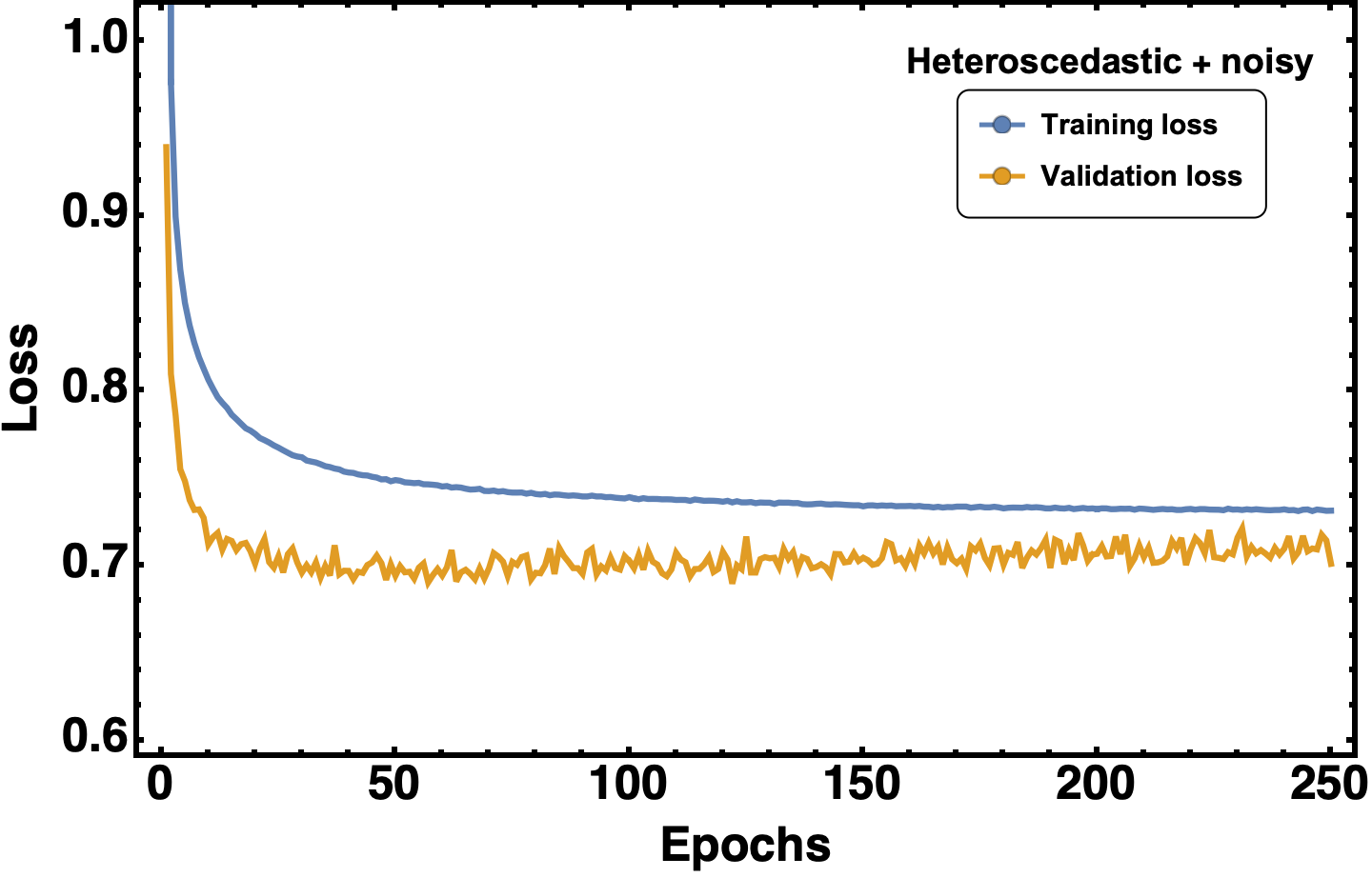} \\
& (d) UGCA 444 & 
\end{tabular}
\caption{Same as figure~\ref{fig:loss_1} for the remaining three galaxies in our sample. }
\label{fig:loss_2}
\end{figure*}

The plots are show in Figs.~\ref{fig:loss_1} and~\ref{fig:loss_2}. Note that for all galaxies, by $250$ epochs, validation loss for every case has stopped changing noticeably. 
Moreover, for the case of noisy training data with MSE loss function, validation loss has started increasing slightly for some galaxies (see the case for UGC 5721 and UGCA 444).
This implies that the neural network is beginning to overfit and further training will lead to worse performance on unseen data. 

\end{document}